\documentclass[11pt,a4paper]{article}
\usepackage{graphicx}

\usepackage{authblk}
\usepackage{comment}
\usepackage{braket}
\usepackage{tikz}
\usepackage[compat=1.1.0]{tikz-feynman} 
\tikzfeynmanset{warn luatex=false}
\usepackage{caption}

\usepackage{color}
\usepackage{todonotes}
\usepackage{booktabs}
\usepackage{array}
\usepackage{multirow,rotating}
\usepackage{amsmath,amssymb,bm,cite,graphicx,wasysym,mathrsfs,dsfont,amsfonts}

\usepackage{slashed,relsize}
\usepackage[colorlinks=true,linkcolor=blue,anchorcolor=black,citecolor=blue,filecolor=cyan,menucolor=red,runcolor=filecolor,urlcolor=blue,bookmarks=true,bookmarksnumbered=true]{hyperref}

\usepackage[font=small,labelfont=bf]{caption}
\usepackage{subcaption}

\usepackage[mathscr]{eucal}

\newcommand\pd{\textsc{v}}
\newcommand\md{\textsc{u}}
\newcommand\td{\textsc{t}}

\newcommand{\ba}{\begin{eqnarray}}
\newcommand{\ea}{\end{eqnarray}}

\title{ \bf Low-virtuality splitting in the Standard Model \vspace{0.4cm}} 
\date{}

\author[1]{Filippo Nardi}
\author[2]{Lorenzo Ricci}
\author[3,4]{Andrea Wulzer}

\affil[1]{\emph{Theoretical Particle Physics Laboratory (LPTP), Institute of Physics, EPFL, Lausanne, Switzerland}}
\affil[2]{\emph{Maryland Center for Fundamental Physics, Department of Physics, University of Maryland, \newline College Park, MD 20742, USA}}
\affil[3]{\emph{ICREA, Instituci\'o Catalana de Recerca i Estudis Avan\c{c}ats, \newline Passeig de Llu\'{\i}s Companys 23, 
08010 Barcelona, Spain}}
\affil[4]{\emph{Institut de F\'{\i}sica d'Altes Energies (IFAE), Campus UAB, 08193 Bellaterra, Barcelona, Spain}}

\begin{document}

\renewcommand{\theequation}{\thesection.\arabic{equation}}
\numberwithin{equation}{section}

\baselineskip=14pt
\maketitle
\begin{abstract}
When the available collision energy is much above the mass of the particles involved, scattering amplitudes feature kinematic configurations that are enhanced by the much lower virtuality of some intermediate particle. Such configurations generally factorise in terms of a hard scattering amplitude with exactly on-shell intermediate particle, times universal factors. In the case of real radiation emission, such factors are splitting amplitudes that describe the creation or the annihilation---for initial or final state splittings---of the low-virtuality particle and the creation of the real radiation particles. We compute at tree-level the amplitudes describing all the splittings that take place in the Standard Model when the collision energy is much above the electroweak scale. Unlike previous results, our splitting amplitudes fully describe the low-virtuality kinematic regime, which includes the region of collinear splitting, of soft emission, and combinations thereof. The splitting amplitudes are compactly represented as little-group tensors in an improved bi-spinor formalism for massive spin-1 particles that automatically incorporates the Goldstone Boson Equivalence Theorem. Simple explicit expressions are obtained using a suitably defined infinite-momentum helicity basis representation of the spinor variables. Our results, combined with the known virtual contributions, could enable systematic predictions of the leading electroweak radiation effects in high-energy scattering processes, with particularly promising phenomenological applications to the physics of future colliders with very high energy such as a muon collider.
\end{abstract}

\thispagestyle{empty}

\newpage

\begingroup
\tableofcontents
\endgroup 

\setcounter{equation}{0}
\setcounter{footnote}{0}
\setcounter{page}{1}

\newpage

\section{Introduction}

\subsubsection*{Motivations}
Processes at multi-TeV or higher energy display novel phenomena driven by the large separation between the collision energy $E$ and the mass scale $m_{\textsc{ew}}\sim100$~GeV associated with the breaking of the Electro-Weak (EW) symmetry. The separation enhances the emission of real and virtual particle quanta. When the enhancement is sufficient to compensate for the small coupling factors, such emissions become an order-one component of the very high energy scattering phenomenology. The enhancement occurs by the same mechanism that underlies the copious emission of hadrons when the collision energy exceeds the QCD scale $\Lambda_{\textsc{qcd}}\sim1$~GeV, or the QED emission of photons---at any energy, since the photon is massless---or of collinear electrons at energies much above the electron mass. Because of the analogy with QED and QCD radiation, we refer to the corresponding phenomena in the EW sector as \emph{EW radiation} effects.

Intriguing manifestations of EW radiation are for instance the emergence of a partonic content of massive vector bosons in elementary colliding particles such as electrons or muons. Or, conversely, the emergence of a rich content of particles inside the scattered partons from final state radiation. For example, the neutrinos produced in the hard scattering will split through gauge interactions into a charged $W$ boson plus an electron with high probability, making the neutrino detectable in the form of a neutrino jet. Our notion of ``EW'' radiation includes effects that are not actually mediated by EW gauge interactions but by other SM couplings. For instance, the top Yukawa mediates peculiar effects like Higgs radiation from top quarks, Higgs splitting into top pairs and more. The Higgs trilinear coupling also plays a role.

There is inherent interest in the physics of the SM at very high energy and in the quantitative theoretical comprehension and modeling of EW radiation. The interest is further boosted by the perspective of studying EW radiation experimentally at future colliders---see e.g. \cite{EuropeanStrategyforParticlePhysicsPreparatoryGroup:2019qin}---with a partonic collision energy in the range from several to 10~TeV. The case for EW radiation is particularly striking in the perspective of a muon collider~\cite{Delahaye:2019omf,Accettura:2023ked,Black:2022cth,AlAli:2021let}. EW radiation will be prominent especially in the high-energy stage at  10~TeV or above, but considerable also at a possible first stage with 3~TeV centre of mass energy. In contrast with possible future hadron colliders probing comparable partonic energies, the reduced background from QCD interactions will enable precise experimental studies of EW radiation, which will have to be accompanied by accurate theoretical predictions. Theoretical studies of the SM in the very high energy regime and of EW radiation effects span several decades~\cite{Dawson:1984gx,Kane:1984bb,Kunszt:1987tk,Chanowitz:1985hj,E_th-original,Gounaris:1986cr,Yao,Bagger,Kilgore:1992re,10.1103/PhysRevLett.69.2619,10.1103/PhysRevD.55.1515,10.1103/PhysRevD.49.4842,Fadin:1999bq,Ciafaloni:2000rp,Ciafaloni:2000df,Melles:2000gw,Manohar:2014vxa,Bauer:2017isx,Bauer:2018xag,Manohar:2018kfx,Fornal:2018znf,Chiu:2007dg,Chiu:2007yn,Chiu:2009ft,Denner:2000jv,Denner:2001gw,Borel:2012by,Wulzer:2013mza,Cuomo:2019siu,Bothmann:2020sxm,Chen:2016wkt,Han:2021kes,Ruiz:2021tdt,Chen:2022msz,Garosi:2023bvq}, but a complete understanding is still to come.

Generically, two complementary approaches can be envisaged for the study of EW radiation. One is to start from asymptotically high energies where the scale separation is very large and progress can be made by analogy with QED and QCD radiation, which is extensively studied in the large scale separation regime. This approach has produced techniques for the resummation of EW radiation effects at different orders in the log expansion~\cite{Fadin:1999bq,Ciafaloni:2000rp,Ciafaloni:2000df,Melles:2000gw,Chiu:2007dg,Chiu:2007yn,Chiu:2009ft,Manohar:2014vxa,Bauer:2017isx,Bauer:2018xag,Manohar:2018kfx,Fornal:2018znf}. The second approach is to start from low energy where EW radiation effects are calculable in fixed-order perturbation theory. Logarithms of the scale separation $E^2/m_{\textsc{ew}}^2$ generically enhance higher-order contributions, but do not invalidate the perturbative expansion within a rather wide range of energies, because the logarithmic growth is slow. This approach might be a valid quantitative description of few-TeV energy processes, and offer at least a qualitative guidance at 10~TeV or higher energy, where resummation of the double-logarithmic effects will be mandatory for quantitatively accurate predictions~\cite{Chen:2022msz}. 

It should be stressed that EW radiation poses radically different challenges than QCD and QED radiation, which stem from the breaking of the EW symmetry. Since the symmetry is broken, individual particles in the EW multiplets are physically distinct and easily distinguishable experimentally. For instance, we can tell apart a longitudinally-polarised $W$ or $Z$ from a Higgs boson, in spite of the fact that they all belong to the same doublet of states. Unlike in QCD, the EW ``color'' is a detectable quantum number and therefore the observables of interest are not color singlets. This violates the assumptions of the KLN theorem and thus entails the non-cancellation of virtual and real EW radiation effects~\cite{Ciafaloni:2000rp,Ciafaloni:2000df}. In light of these differences, the adaptation of QCD and QED calculation methodologies to the EW radiation problem is a delicate process. The development of a robust and systematic fixed-order understanding is arguably a necessary intermediate step. 

The fixed-order approach has produced, in particular, the complete classification of log-enhanced virtual corrections to hard scattering amplitudes up to the two-loops order~\cite{Denner:2000jv,Denner:2001gw}. These results allow us to express the log-enhanced loop contributions to a generic process as a linear combination of the tree-level amplitude of the process under examination and of amplitudes that belong to the same EW symmetry multiplet. The coefficients of the linear combination are  universal and process-independent. This enables to compute analytically and systematically the (virtual) EW radiation effects on fixed-multiplicity exclusive scattering processes, characterised by energetic and well-separated external legs.\footnote{Final states with a fixed number of massive particles, such as massive vector bosons, are theoretically and experimentally well-defined. On the contrary, observables must include the emission of extra massless states like photons or gluons. Our notion of exclusiveness refers to the massive particles content of the final state.} However, most of the interesting effects entail on the contrary the emission of real rather than virtual EW radiation. Additionally, real corrections must be included and combined with virtual effects in several calculations such as semi-inclusive cross-sections~\cite{Chen:2022msz}. The two contributions will not cancel as previously explained. 

\subsubsection*{Tree-level splitting}

At the leading order in perturbation theory, real emission emerges from tree-level diagrams. The present paper deals with a general classification of real radiation effects at tree-level. We derive factorised formulas for generic scattering amplitudes that describe real radiation in the enhanced kinematic configurations. These formulas could be used to model the distribution of the radiation, or integrated over the phase space to compute more inclusive observables. In the latter case, the integral over the enhanced configurations will provide log-enhanced contributions which are the real counterpart of the virtual terms in Refs.~\cite{Denner:2000jv,Denner:2001gw} at the one-loop order.

The kinematic configurations of interest are represented in Figure~\ref{fig:splitting}. The left and right panels represent Final State (FSR) and Initial State (ISR) splitting configurations, respectively. Both topologies correspond to the splitting of a particle $A$ into two particles, $B$ and $C$. In the case of FSR, $B$ and $C$ are on-shell final-state particles while $A$ is virtual, with a virtuality $Q^2\equiv p_A^2-m_A^2$. The $A$ particle is produced in a scattering process of hardness $S=E^2$, with $E$ the collider energy. Specifically, the hard scattering is a one where all the scattered particles---including the $A$ particle---have order $\sqrt{S}=E$ energy and are well-separated in angle. In the case of ISR, $A$ is an on-shell initial-state particle, $B$ is also on-shell while the $C$ particle is virtual, with a virtuality $Q^2\equiv p_C^2-m_C^2$. The $C$ particle undergoes the hard scattering process with a hardness of order $E^2$. In both the FSR and ISR configurations, the enhancement of the amplitude is controlled by the hierarchy between the absolute value of virtuality (which is typically negative, in the case of ISR) and the hardness: $|Q^2|\ll E^2$.

Tree-level splittings has been studied already. Textbook results cover the region $|Q^2|\ll m_{\textsc{ew}}^2$, where the only relevant splittings are those mediated by QED and QCD interactions. Splitting effects due to EW interactions and to other SM vertices like the top quark Yukawa and the Higgs trilinear coupling emerge instead when $Q^2$ is of order 
$m_{\textsc{ew}}^2$ or larger. Namely, when $Q^2$ is comparable or larger than the EW bosons, Higgs and top quark mass. The most famous ISR splitting in the EW theory is the emission of a virtual massive vector boson from a light fermion. The factorised treatment of this splitting leads to Effective Vector boson Approximation (EVA)~\cite{
Dawson:1984gx,Kane:1984bb,Kunszt:1987tk,Borel:2012by}, which provides the leading-order description of the vector boson partonic content of the light fermions. All the ISR and FSR splittings that occur in the SM have been considered as well~\cite{Cuomo:2019siu,Chen:2016wkt}. However, these studies are limited to \emph{collinear} splitting configurations where both the $B$ and the $C$ particles carry a significant fraction of the energy of $A$. The low splitting virtuality $Q^2$ is attained because of the small angle between the $3$-momenta of the particles. Current results do not model instead the soft region, where $Q^2$ is small because one of the particles has small energy. Also the region where the emitted particles are both soft and collinear is not modelled by the current results. This region is particularly relevant because it produces the leading double logarithm contributions to the integrated cross-section. 

\subsubsection*{Structure of the paper}

In the present paper we fill this gap in the literature by studying factorization and computing the splitting amplitudes in the whole low-virtuality phase space. Our formulas encompass at the same time collinear and soft splitting, and the soft-collinear region. Furthermore, we show how the amplitudes can be compactly represented as little-group tensors by an improved bi-spinors formalism for massive spin-1 particles that automatically incorporates the Goldstone Boson Equivalence Theorem. We also obtain simple explicit expressions in the infinite-momentum helicity representation of the spinor variables.

\begin{figure}
    \centering
    \begin{tikzpicture}[scale=1]
	\begin{feynman}
	\vertex[blob,fill=white,minimum size=25pt] (m) at (-1, 0){$E$};
	\vertex (ai) at (2,-1.3){$C$};
    \vertex (af) at (1.6,-2){$B$};
    \vertex (ac) at (1,-1.2);
    \vertex (bc) at (-2,0.5);
    \vertex at (-2.5,0){$X$};
    \vertex (bd) at (-2,-0.5);
    \vertex (f1) at (0.47,0.85);
    \vertex (f3) at (0.57,0.65);
    \vertex (f2) at (0.64,0.44);
    \vertex at (0.84,0.765){$Y$}; 
	\diagram* {
		(ac) -- [momentum=$p_C$] (ai),
    (ac)--[momentum'=$p_B$] (af),
        (m) -- [momentum'=$p_A$,edge label=$A$] (ac),
        (bc) -- (m)--(bd),
        (m) -- (f1),
        (m) -- (f2),
        (m) -- (f3),
	};
	\end{feynman}
    \node at (0,-3) {$Q_F^2 = p_A^2 - m_A^2$};
	\end{tikzpicture}	
  \hspace{1cm}
 \begin{tikzpicture}[scale=1]
	\begin{feynman}
	\vertex[blob,fill=white,minimum size=25pt] (m) at (1.5, 0){$E$};
	\vertex (ai) at (-2,1){$A$};
    \vertex (ac) at (0,1);
    \vertex (af) at (2,2){$B$};
    \vertex (bc) at (0,-1){$X$};
    \vertex (f1) at (2.3,0.2);
    \vertex (f3) at (2.3,0);
    \vertex (f2) at (2.3,-0.2);
    \vertex at (2.8,0){$Y$}; 
	\diagram* {
		(ai) -- [momentum=$p_A$] (ac)--[momentum=$p_B$] (af),
        (ac) -- [ momentum'=$p_C$,edge label=$C$] (m),
        (bc) -- (m),
        (m) -- (f1),
        (m) -- (f2),
        (m) -- (f3),
	};
	\end{feynman}
     \node at (0,-2) {$Q_I^2 = p_C^2 - m_C^2$};
	\end{tikzpicture}	
    \caption{Schematic representation of Final State Radiation (FSR, left) and Initial State Radiation (ISR, right) splittings.  \label{fig:splitting}}
\end{figure}
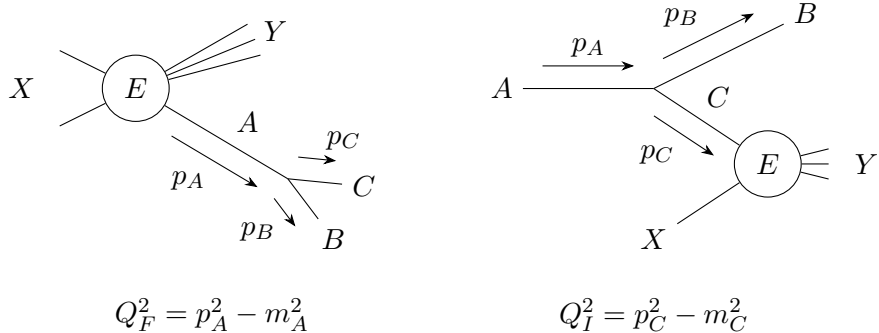

In Section~\ref{sec:rdsa} we define the splitting amplitudes by analysing the low-$Q^2$ expansion of the Feynman diagrams for the complete collision process, i.e., in Figure~\ref{fig:splitting}, for the generic $X\to BC Y$ or $AX\to BY$ scatterings. The leading term is a linear combination of the amplitudes for the hard scattering sub-process that does not involve the $B$ and $C$ (or $A$ and $B$, for ISR) particles and where the $A$ (or $C$) particle is exactly on-shell. The splitting amplitudes are defined as the coefficients of this linear combination. 

Among the possible diagrams of the complete scattering process, the leading contribution at low $Q^2$ comes from the ones with a topology like in Figure~\ref{fig:splitting}. Namely, from diagrams where the low-virtuality momentum---i.e., $p_A=p_B+p_C$ and $p_C=p_A-p_B$ for FSR and ISR, respectively---flows into a propagator line. We call these diagrams ``resonant'' because they feature a $1/Q^2$ pole from the virtual particle propagator. Other diagrams do not have propagators that are enhanced in comparison with the dimensional analysis scaling of $1/E^2$. Therefore, they are naively expected not to experience an enhancement in the splitting configuration $|Q^2|\ll E^2$, and be negligible in comparison with the resonant diagrams. However, it was pointed out in Ref.~\cite{Borel:2012by}---see also~\cite{Kleiss:1986xp,Kunszt:1987tk} for earlier discussions---that this naive expectation based on power-counting is generically violated in theories with massive particles of spin 1 such as the SM. Non-resonant diagrams can be ignored only in specifically-designed formulations of the theory~\cite{Wulzer:2013mza,Cuomo:2019siu}, which entail Feynman rules that are different from the habitual $R_\xi$ or Unitary gauge rules. Since these findings play a major methodological role in our calculation, and impact the final result, they are summarised more extensively below.

The basic issue is that power-counting is not manifest in the canonical diagrammatic formulation of massive gauge theories. The high-energy behavior of individual diagrams is generically enhanced by powers of $E/m_{\textsc{ew}}$---with $m_{\textsc{ew}}$ the mass of the vector bosons---with respect to the behaviour of the physical scattering amplitude. Cancellations---the so-called gauge cancellations---occur between the leading powers of energy of the different diagrams that contribute to the physical amplitude. It can thus happen that non-resonant diagrams feature an enhancement of order  $E^2/m_{\textsc{ew}}^2$, relative to the scaling with energy of the physical scattering amplitude. This enhancement is comparable or larger than the $E^2/Q^2$ enhancement of the resonant diagrams, in the regime $|Q^2|\gtrsim m_{\textsc{ew}}^2$ that is relevant for EW radiation. Therefore, non-resonant diagrams are not negligible. Furthermore, the occurrence of gauge cancellations prevents the resonant diagrams to factorise as the product of a splitting amplitude times the on-shell amplitude for the hard scattering~\cite{Kleiss:1986xp,Borel:2012by}. Factorization eventually holds true, but it emerges from a conspiracy between resonant and non-resonant diagrams that can be only established on a case-by-case basis. An explicit example in the context of the EVA is discussed in Ref.~\cite{Borel:2012by}. 

We avoid this problem by employing an improved diagrammatic formulation of the theory that exploits Goldstone Equivalence (GE), which was developed in Refs.~\cite{Wulzer:2013mza,Cuomo:2019siu} (see also \cite{Veltman:1989ud}). It employs (see Appendix~\ref{App:EqFromalism}) the regular $R_\xi$ gauge Feynman rules, but it describes external spin-1 particles using a double line that represents the combined contribution of gauge and Goldstone bosons external fields (see e.g.~Figure~\ref{fig:EquivalentFR}). The formalism is designed in such a way that polarisation vectors for longitudinal (i.e., zero helicity) states have a regular high-energy (and low-mass) limit, unlike the regular polarisation vector that grows like $E/m_{\textsc{ew}}$. In the regular formalism, the anomalous energy behaviour is due to the energy growth of the longitudinal polarisation. Since this is avoided, power-counting is manifest at the level of individual Feynman diagrams in the GE formalism. No gauge cancellation occurs and consequently only the resonant diagrams are enhanced in the low $Q^2$ limit and need to be retained. Clearly, these diagrams are different from the resonant diagrams in the regular formalism, because the Feynman rules are different. We will see that these differences leave an imprint on the results.

We find convenient to express the splitting amplitudes as tensors with indices in the little-group of the external states, in terms of bi-spinors~\cite{Dittmaier:1998nn,Arkani-Hamed:2017jhn,Feige:2013zla,Franken:2019wqr,Lai:2023upa}. This requires an adaptation of the bi-spinors notation that accounts for the modified Feynman rule and longitudinal polarisation vector which we encounter in the GE formalism. We show in Appendix~\ref{sec:wf} that the GE rules are easily and naturally incorporated in the bi-spinors notation by allowing for more general amplitude tensors that are not symmetric under the exchange of the little-group indices of the spin-1 particle. 

Section~\ref{sec:aimh} is devoted to the explicit evaluation of the splitting amplitudes and to the comparison with the partial results that are available in the literature. Specifically, we recover the collinear splitting amplitudes~\cite{Cuomo:2019siu,Chen:2016wkt} and the eikonal description of the soft radiation emission. Compact explicit formulas for the splitting amplitudes are obtained in a specific representation of the bi-spinor variables that enjoys simple transformation rules under rotations and Lorentz boosts performed around the direction of motion of the $A$ particle. These bi-spinors are a minor generalisation of the ``infinite-momentum frame helicity'' spinors defined by Soper in Ref.~\cite{LCHSoper}, based on Weinberg's idea of studying particles' dynamics in an infinitely boosted Lorentz frame~\cite{Weinberg:1966jm}. Such ``Soper--Weinberg'' (SW) spinors are constructed in Appendix~\ref{app:spi} and used in Section~\ref{sec:aimh} for the explicit evaluation of the splitting amplitudes. The relation with regular helicity amplitudes that employ Jacob-Wick (JW) spinors~\cite{Jacob:1959at} is also worked out.

Our conclusions and an outlook to future work are reported in Section~\ref{sec:conc}. Appendix~\ref{sec:salist} lists our results for the amplitudes of all splitting processes that occur in the SM theory.

\section{Resonant diagrams  and splitting amplitudes}\label{sec:rdsa}

The resonant diagrammatic contributions to the scattering processes under examination are schematically represented in Figure~\ref{fig:splitting} in the case of FSR (left panel) and ISR (right panel) emission. For FSR, we have in mind a generic $X\to BCY$ process, where $B$ and $C$ are some specific SM particles while $X$ and $Y$ are unspecified  multi- or single-particle states. Two conditions need to be fulfilled in order for the $X\to BCY$ scattering amplitude to experience a low-virtuality enhancement. First, a third SM particle $A$ must exist featuring a $3$-point vertex with $B$ and $C$, as in the figure. This vertex produces an FSR splitting $A\to BC$. The second condition for enhancement is the existence of the $X\to A Y$ scattering process, with $A$ on-shell. In the kinematic regime of interest, the hardness of this process is much larger than the virtuality of the $A$ particle. We thus refer to $X\to AY$ as the hard scattering. The situation is similar in the case of ISR. The generic process is $AX\to B Y$, and the conditions for enhancement are once again the existence of the $A\to BC$ splitting and the existence of the hard scattering, which is the process $CX\to Y$ in this case. The enhancement is controlled by the virtuality of the $C$ particle.

\subsection{Kinematics}\label{sec:kin}

The splitting kinematics is described by the $4$-momenta $p_A$, $p_B$ and $p_C$ of the particles involved, subject to $4$-momentum conservation
\begin{equation}
    p_B+p_C-p_A=0\,.
\end{equation}
In the case of FSR, $p_B$ and $p_C$ are physical on-shell momenta with $p_{B,C}^2=m_{B,C}^2$, while $p_A=p_B+p_C$ is off-shell. The virtuality parameter $Q^2$ is defined as $Q^2\equiv Q_{\textsc{f}}^2 = p_A^2-m_A^2$, where $m_A$ is the physical mass of the $A$ particle. For FSR we have instead $p_{A,B}^2=m_{A,B}^2$, $p_C=p_A-p_B$, and the virtuality is defined as $Q^2\equiv Q_{\textsc{i}}^2=p_C^2-m_C^2$. 

For both FSR and ISR splittings, the kinematics is best represented using the following (light-cone) coordinates. Be $e_3={\hat{p}}_A$ the direction of motion of the $A$ particle and ${e}_{i}$ two additional norm-one vectors that complete a right-handed Cartesian system of coordinates $(e_{1},e_{2},e_3)$. We define a basis for the 4-vectors as
\begin{equation}\label{NulFrame}
    n^\mu_\pd=\frac12\{1,\,e_3\}^\mu\,,\quad
    {n}_\md^\mu=\frac12\{1,\,-e_3\}^\mu\,,\quad n^\mu_{\td,i}=\{0,e_{i}\}^\mu\,,
\end{equation}
and we decompose a generic $4$-momentum in the lab frame as
\begin{equation}\label{eq:coord}
    k^\mu=k_\pd n_\pd^\mu+k_\md{n}_\md^\mu+\sum_i k_{\td,i} n^\mu_{\td,i}\,.
\end{equation}
It is useful to define the complex combination ${\bf{k_{\td}}}=k_{\td,1}+i\,k_{\td,2}$ and describe the momentum in our coordinates as a vector made of two real and one complex variable as 
\begin{equation}
    k=(k_\pd,k_\md,{\bf{k_\td}})\,.
\end{equation}
The Lorentz product between vectors reads, in this notation
\begin{equation}
    k\cdot {{q}}=\frac{k_{\pd}q_{\md}+k_{\md}q_{\pd}}2-\frac{{\bf{k_{\textsc{t}}}}{\bf{q^*_{{\textsc{t}}}}}+{\bf{k^*_{{\textsc{t}}}}}{\bf{q_{\textsc{t}}}}}2\,,\qquad
    k^2=k_{\pd}k_{\md}-|{\bf{k_{\textsc{t}}}}|^2\,.
\end{equation}

The $n_\pd$ ($n_\md$) vectors can be interpreted as the 4-momentum of a massless particle with energy $1/2$ moving (anti-)parallel to the $A$ particle. If the $A$ particle moves along the $z$ axis with positive velocity, the Cartesian frame $(e_{1},e_{2},e_3)$ can be taken to coincide with the lab coordinates $(e_x,e_y,e_z)$, and we recover the regular light-cone coordinates where $k_\pd=E+ k_z$, $k_\md=E- k_z$ and ${\bf{k_{\textsc{t}}}}$ is the momentum in the $x$-$y$ plane. The coordinate system for generic $A$ particle momentum is related by a rotation---see later eq.~(\ref{eq:rot})---to the regular light-cone system.

The $A$, $B$ and $C$ momentum components in our coordinates can be parametrised as 
\begin{eqnarray}
    &&p_A=\left(p_{\pd},\, \frac{p_A^2}{p_{\pd}},\,{\bf{0}}\right)\,,\nonumber\\
    &&p_B=\left((1-x)\,p_{\pd}\,,\frac{m_B^2+|{\bf{p}_{\textsc{t}}}|^2}{(1-x)\,p_{\pd}},{\bf{p}_{\textsc{t}}}\right)\,,\\
    &&p_C=\left(x\,p_{\pd},\,\frac{p_C^2+|{\bf{p}_{\textsc{t}}}|^2}{x\,p_{\pd}},\,-{\bf{p}_{\textsc{t}}}\right)\,,\nonumber
\end{eqnarray}
using two real and one complex variable, $p_{\pd}$, $x$ and ${\bf{p}_{\textsc{t}}}$. The parametrisation simultaneously accounts for FSR and ISR splitting. In the former case, $p_C^2=m_C^2$, while $p_A^2\neq m_A^2$ is determined by the conservation of the momentum along the $n_{\md}$ direction. Explicitly
\begin{equation}\label{eq:FSRVIRT}
    Q_{\textsc{f}}^2=p_A^2-m_A^2=\frac{|{\bf{p}_{\textsc{t}}}|^2-x(1-x)m_A^2+x
    \,m_B^2+(1-x)m_C^2
    }{x(1-x)}
    \,.
\end{equation}
In the case of ISR, $p_A^2=m_A^2$ and momentum conservation determines $p_C^2\neq m_C^2$
\begin{equation}\label{eq:ISRVIRT}
    Q_{\textsc{i}}^2=p_C^2-m_C^2=\frac{-|{\bf{p}_{\textsc{t}}}|^2+x(1-x)m_A^2-x
    \,m_B^2-(1-x)m_C^2
    }{1-x}\,.
\end{equation}
The $p_B$ momentum is instead on-shell in both configurations: $p_B^2=m_B^2$. The transverse momentum of $A$ vanishes by definition in our frame. Momentum conservation along the transverse direction is imposed explicitly, leading to a single transverse momentum parameter ${\bf{p}_{\textsc{t}}}$. Momentum conservation is also imposed along the $n_{\pd}$ direction. The variable $x$ is defined as the fraction of $n_{\pd}$ momentum that the $C$ particle carries away from the mother particle $A$ in the splitting, and a $1-x$ fraction is carried away by the $B$ particles.

The parameter $p_{\pd}$ is necessarily positive both in the FSR and in the ISR configurations, because it is the $n_{\pd}$ component of a 4-vector ($p_A$) which is time-like and has positive temporal component.~\footnote{It is easy to see that $p_{\pd}$ is in fact larger than $\sqrt{p_A^2}$ because the $p_A$ 3-momentum is parallel and points in the same direction as $e_3$.} The parameter $x$ is smaller than 1 because $(1-x)\,p_{\pd}$ is the $n_{\pd}$ component of $p_B$, which is a physical on-shell momentum in both splitting configurations. In the case of FSR, we also have that $x>0$ because the $C$ particle is physical, therefore
\begin{equation}
    x \overset{\textrm{FSR}}{\in}(0,1)\,.
\end{equation}
In the case of ISR instead, $x$ could be negative because the $p_C$ momentum could have negative $n_{\pd}$ component, a priori. However, the kinematic regime that is relevant for the study of factorisation is characterised by a large scattering scale for the hard process $CX\to{Y}$, which is initiated by the $C$ particle colliding with the  $X$ initial particle. A large collision centre of mass energy can be attained only if $x$ is positive and also sufficiently large. We thus restrict our analysis to the $x$ range
\begin{equation}\label{eq:risr}
    x \overset{\textrm{ISR}}{\in}(x_{\textrm{min}},1)\,,
\end{equation}
where $x_{\textrm{min}}$ is not much smaller than 1.

We finally report, for future use, the explicit form of the rotation that connects our coordinates with the lab coordinates. We employ the spinor representation of the Lorentz group transformations, with the notations defined in Appendix~\ref{sec:2cs}, and we represent the momentum by a $2\times2$ matrix as in eq.~(\ref{eq:mmnot}). In this notation, the relevant rotation reads
\begin{equation}\label{eq:rot}
    {\mathcal{R}}=e^{-i\varphi_A \mathscr{J}^3} e^{-i\theta_A \mathscr{J}^2} e^{i\varphi_A \mathscr{J}^3}= \begin{bmatrix}
        \cos\frac{\theta_A}2 & -e^{-i\varphi_A}\sin\frac{\theta_A}2\\        e^{i\varphi_A}\sin\frac{\theta_A}2 &\cos\frac{\theta_A}2
    \end{bmatrix}\,,
\end{equation}
where $\theta_A$ and $\varphi_A$ are the polar and azimuthal angles of the particle $A$ in the lab frame. The rotation transforms the lab coordinates axes $(e_x,e_y,e_z)$ into $(e_{1},e_{2},e_3)$. Hence, it enables to express the 4-momentum in the lab frame~(\ref{eq:coord}) as 
\begin{equation}\label{eq:momlab}
    k=k^\mu\sigma_\mu={\mathcal{R}}\cdot
   \begin{bmatrix}
       k_\md & -{\mathbf{k^*_\td}} \\
        -{\mathbf{k_\td}} & k_\pd 
    \end{bmatrix}\cdot{\mathcal{R}}^\dagger\,.
\end{equation}

\subsection{Low-virtuality expansion and factorization}\label{sec:lwefac}

Denoting as ``$E$'' the total energy of the complete scattering process in the centre of mass frame, the kinematic regime that is relevant for factorization is the one where the hard scattering scale is also of order $E$, while the virtuality is small. An order-$E$ scale for the hard scattering necessarily requires $p_{\pd}\sim{E}$ and, in the case of ISR, a momentum fraction $x$ which is not too much smaller than 1 as previously discussed. A small virtuality $Q_{\textsc{f}}^2$~(\ref{eq:FSRVIRT}) or $Q_{\textsc{i}}^2$~(\ref{eq:ISRVIRT}) can instead be attained by a variety of different scalings for the other variables ${\mathbf{p_\td}}$ and $x$. For instance, low virtuality is obtained in the collinear scaling where ${{\bf{p_{\textsc{t}}}}}\ll E$ and $x$ is generic and far from the extremes $x=0$ and $x=1$. In the soft region where $x$ is close to one, namely $x-1=\epsilon\ll1$, small virtuality requires (see Section~\ref{sec:SL} for a more careful discussion) ${{\bf{p_{\textsc{t}}}}}\lesssim\sqrt{\epsilon} \, E$, and similarly when $x$ is close to zero in the case of FSR. In fact, a small (exactly vanishing) virtuality can be also obtained when $x$ is far from the extremes if the particle $A$ is heavier than the sum of the masses of the $B$ and of the $C$ particles: $m_A>m_B+m_C$. This configuration corresponds to the on-shell decay of the $A$ particle to $B$ and $C$. 

The low-virtuality expansion that we perform in the present section describes all these different configurations at once, because it relies exclusively on the scale separation $Q^2\ll E^2$ making no assumption on the scaling of ${{\bf{p_{\textsc{t}}}}}$ and $x$ by which the low virtuality is attained. 

In order to proceed, we define \emph{on-shell} splitting 4-momenta: 
\begin{eqnarray}\label{eq:onmom}
    &&\bar{p}_A=\left(p_{\pd},\, \frac{m_A^2}{p_{\pd}},\,{\bf{0}}\right)\,,\nonumber\\
    &&\bar{p}_B=\left((1-x)\,p_{\pd}\,,\frac{m_B^2+|{\bf{p}_{\textsc{t}}}|^2}{(1-x)\,p_{\pd}},{\bf{p}_{\textsc{t}}}\right)=p_B\,,\\
    &&\bar{p}_C=\left(x\,p_{\pd},\,\frac{m_C^2+|{\bf{p}_{\textsc{t}}}|^2}{x\,p_{\pd}},\,-{\bf{p}_{\textsc{t}}}\right)
    \,.
    \nonumber
\end{eqnarray}
The on-shell momenta do not obey momentum conservation, but rather
\begin{equation}\label{eq:momnc}
    \bar{p}_B+\bar{p}_C-\bar{p}_A = 
    \Delta p
    \equiv\frac{|{\bf{p}_{\textsc{t}}}|^2-x(1-x)m_A^2+x
    \,m_B^2+(1-x)m_C^2
    }{x(1-x)\,p_{\pd}}\,n_{\md}\,.
\end{equation}

For those particles that are physically on-shell in the splitting, the on-shell momenta coincide with the corresponding $p$ momenta. For the off-shell particles instead
\begin{equation}
    \bar{p}_A\overset{\textrm{FSR}}{=}
    p_A-\Delta p=
    p_A-\frac{Q_{\textsc{f}}^2}{p_{\pd}}n_{\md}\,,
    \qquad
    \bar{p}_C\overset{\textrm{ISR}}{=}
    p_C+\Delta p=
    p_C-\frac{Q_{\textsc{i}}^2}{x\,p_{\pd}}n_{\md}\,,
    \label{eq:offonmom}
\end{equation}
in the case of FSR and ISR, respectively. The momentum shift $\Delta p$ is proportional to the virtuality $Q_{{\textsc{i}},{\textsc{f}}}^2$ in both cases, enabling us to take $\Delta p=0$ at the leading order in the low-virtuality expansion. Notice that $\Delta p$ features a factor of $1/x$ for ISR splitting which is absent for FSR. This does not invalidate the expansion because we restricted $x$ to the interval in eq.~(\ref{eq:risr})---which excludes zero---for ISR splittings.

\begin{figure}
    \centering
     \begin{tikzpicture}[scale=1]
	\begin{feynman}
	\vertex[blob,fill=white,minimum size=25pt] (m) at (-1, 0){$\mathcal{A}_{\textrm{H}}^b$};
	\vertex (ai) at (2.48,-1.41){$C$};
    \vertex (af) at (1.89,-2.41){$B$};
    \vertex[blob,fill=white,minimum size=25pt] (ac) at (1,-1.2){$\mathcal{A}_{\textrm{S}}^a$};
    \vertex (bc) at (-2,0.5);
    \vertex at (-2.5,0){$X$};
    \vertex (bd) at (-2,-0.5);
    \vertex (f1) at (0.47,0.85);
    \vertex (f3) at (0.57,0.65);
    \vertex (f2) at (0.64,0.44);
    \vertex at (0.84,0.765){$Y$}; 
	\diagram* {
		(ac) -- [momentum=$p_C$] (ai),
    (ac)--[momentum'=$p_B$] (af),
        (m) -- [fermion, momentum'=$p_A$,edge label=$A$] (ac),
        (bc) -- (m)--(bd),
        (m) -- (f1),
        (m) -- (f2),
        (m) -- (f3),
	};
	\end{feynman}
    \node at (-1,-1.7) {$\frac{i}{Q_F^2} (\slashed{p}_A + m_A)_{ab}$};
	\end{tikzpicture}	
    \hspace{1.5cm}
 \begin{tikzpicture}[scale=1]
	\begin{feynman}
	\vertex[blob,fill=white,minimum size=25pt] (m) at (-1, 0){$\mathcal{A}_{\textrm{H}}^N$};
	\vertex (ai) at (2.48,-1.41){$C$};
    \vertex (af) at (1.89,-2.41){$B$};
    \vertex[blob,fill=white,minimum size=25pt] (ac) at (1,-1.2){$\mathcal{A}_{\textrm{S}}^M$};
    \vertex (bc) at (-2,0.5);
    \vertex at (-2.5,0){$X$};
    \vertex (bd) at (-2,-0.5);
    \vertex (f1) at (0.47,0.85);
    \vertex (f3) at (0.57,0.65);
    \vertex (f2) at (0.64,0.44);
    \vertex at (0.84,0.765){$Y$}; 
	\diagram* {
		(ac) -- [momentum=$p_C$] (ai),
    (ac)--[momentum'=$p_B$] (af),
        (m) -- [photon,momentum'=$p_A$,edge label=$A$] (ac),
        (m) -- [dashed] (ac),
        (bc) -- (m)--(bd),
        (m) -- (f1),
        (m) -- (f2),
        (m) -- (f3),
	};
	\end{feynman}
     \node at (-1,-1.7) {$G_{MN}(p_A)$};
	\end{tikzpicture}	
    \caption{The resonant diagrams for ISR splitting when the intermediate state is a Dirac (left panel) or a vector (right panel). The $5\times5$ propagator $G_{MN}$ is reported in eq.~(\ref{eq:Gprop}).
    \label{fig:Splitting-Dirac-Vector}}
\end{figure}
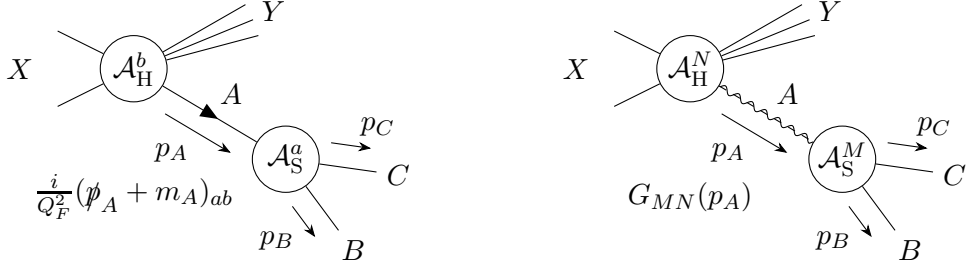

The Feynman amplitude of the resonant diagrams consists of the virtual particle propagator connecting the two portions of the diagram that describe the hard process and the splitting. The hard diagrams are the amplitude of the hard scattering process, where the virtual particle leg ($A$ or $C$, for FSR or ISR) is amputated. We denote as $\mathcal{A}_{\textrm{H}}$ this (partially) amputated amplitude.\footnote{The amplitude is amputated only partially, on the $A$ or $C$ particle leg. It includes instead the wave-function factors for the other external legs of the hard process, which are present in the original Feynman amplitude.} The splitting diagrams consist of the relevant 3-points vertex, contracted with the wave-function factors for the on-shell particles ($B$ and $C$, or $A$ and $B$) involved in the splitting. They form the amputated splitting amplitude $\mathcal{A}_{\textrm{S}}$. The resonant diagrams for FSR splitting are schematised in Figure~\ref{fig:Splitting-Dirac-Vector} in the case of a Dirac (left panel) or a vector (right panel) intermediate state.

The momenta that flow in the hard scattering are of order $E$ and are well-separated in angle. Therefore, applying the momentum shift $\Delta{p}$ in eq.~(\ref{eq:offonmom}) is a small perturbation of the hard process kinematics and the virtual particle momenta can be put on-shell in the hard amputated amplitude up to small corrections. Since $\Delta{p}\sim Q^2/E$, the corrections are proportional to $Q^2$ times---by dimensional analysis---a coefficient of order $1/E^2$. Namely
\begin{equation}\label{eq:off}
    {\mathcal{A}}_{\textrm{H}}={\bar{\mathcal{A}}}_{\textrm{H}}(1+Q^2{\mathcal{O}}(1/E^2))\,, 
\end{equation}
where $\bar{\mathcal{A}}_{\textrm{H}}$ is the on-shell amputated amplitude. 

If the intermediate particle is a scalar, $\bar{\mathcal{A}}_{\textrm{H}}$ is equal to the scattering amplitude ${\mathcal{M}}_{\textrm{H}}$ of the hard process. Otherwise, recovering ${\mathcal{M}}_{\textrm{H}}$ requires contracting $\bar{\mathcal{A}}_{\textrm{H}}$ with a wave-function factor for particle $A$ or $C$. This factor emerges from the decomposition of the numerator of the virtual particle propagator, as we discuss in details in Sections~\ref{sec:VD} and~\ref{sec:VV} for, respectively, Dirac and vector intermediate states. The decomposition takes the schematic form of $\sum_h u_h{\bar{u}}_h$, with ``$u$'' the wave functions factors with on-shell $A(C)$ particles of momenta ${\bar{p}}_{A(C)}$ and helicity $h$. Operating this decomposition up to small low-virtuality correction is simple for a fermion propagator, and more involved in the case of the massive vector.

Eventually, the scattering amplitude ${\mathcal{M}}$ of the complete scattering process ($X\to BCY$, or $AX\to BY$) at the leading order in the virtuality expansion is a linear combination of on-shell hard amplitudes. The splitting amplitudes ${\mathcal{M}}_{\textrm{S}}$ are defined as the coefficients of this combination. The factorization formula reads
\begin{eqnarray}
    {\mathcal{M}}(X\to BCY)\overset{\textrm{FSR}}{=}\frac1{Q_{\textsc{f}}^2} 
    \sum\limits_{h_A}
    {\mathcal{M}}_{\textrm{H}}(X\to A(h_A)Y){\mathcal{M}}_{\textrm{S}}(A^*(h_A)\to B C)+{\Delta\mathcal{M}}_{\textrm{off}}+
    {\Delta\mathcal{M}}_{\textrm{pro}}\,,\nonumber\\
    {\mathcal{M}}(AX\to BY)\overset{\textrm{ISR}}{=}\frac1{Q_{\textsc{i}}^2}
    \sum\limits_{h_C}
    {\mathcal{M}}_{\textrm{S}}(A\to B C^*(h_C)) {\mathcal{M}}_{\textrm{H}}(C(h_C)X\to Y)+{\Delta\mathcal{M}}_{\textrm{off}}+{\Delta\mathcal{M}}_{\textrm{pro}}\,.
\label{eq:facform}
\end{eqnarray}
The sum runs over the helicity of the intermediate ($A$ or $C$) particle. If relevant (like for virtual $Z$ or photon ISR emission from fermions), an additional summation should be considered over the different species of intermediate particles that contribute to the same physical process. 

In eq.~(\ref{eq:facform}) the terms ${\Delta\mathcal{M}}_{\textrm{off}}$ and ${\Delta\mathcal{M}}_{\textrm{pro}}$ represent correction to the factorised approximation that emerge, respectively, from taking the on-shell limit $\Delta p\to0$ of the virtual particle momenta in the hard amplitude, and from the expansion of the propagator. The estimate of the corrections follows closely the analysis performed in~\cite{Borel:2012by,Cuomo:2019siu} for the collinear limit. Because of eq.~(\ref{eq:off}), the former off-shell term features a factor of $Q^2$, which gets cancelled by the denominator of the propagator, times a coefficient that scales with the appropriate power of the hard scale $E$. The off-shell correction ${\Delta\mathcal{M}}_{\textrm{off}}$ is thus independent of $Q^2$ in the low-virtuality limit and its scaling with $E$ is dictated by dimensional analysis, given the energy dimension of the amplitude ${\mathcal{M}}$. Its contribution is generically small and comparable with the contribution of the non-resonant diagrams. The propagator corrections ${\Delta\mathcal{M}}_{\textrm{pro}}$ are also small and will be discussed later in this section. 

In eq.~(\ref{eq:facform}) we label with `` $^*$ '' the virtual particle in the ${\mathcal{M}}_{\textrm{S}}$ amplitude in order to distinguish FSR from ISR splitting, because it is not obvious a priori that the splitting amplitudes are the same in the two cases. In the next subsections we focus on FSR, postponing to Section~\ref{sec:ISRvsFSR} the discussion of ISR splitting. We will see that the ISR amplitudes are equal to the FSR amplitudes and therefore no distinction actually needs to be made in the factorization formulas for the two splitting configurations.

\subsection{Scalar intermediate particle}\label{sec:VS}

It is particularly simple to prove eq.~(\ref{eq:facform}) and to compute the splitting amplitude if the intermediate particle is a scalar. There is no sum over the helicity and no propagator correction (i.e., ${\Delta\mathcal{M}}_{\textrm{pro}}=0$). The only correction term, ${\Delta\mathcal{M}}_{\textrm{off}}$, comes from the on-shell limit $\Delta p\to0$ in the hard amplitude and is small as previously discussed. 

The splitting amplitudes ${\mathcal{M}}_{\textrm{S}}$ is just the Feynman rule (listed in Appendix~\ref{Sec:FeynmanRules}) of the corresponding vertex (times $-i$), multiplied by the corresponding wave function factor for the external real particles involved in the splitting. For instance, the amplitude for the final state splitting of a virtual Higgs boson to fermion anti-fermion (e.g., top quarks) pair reads
\begin{equation}\label{eq:hff0}
{\mathcal{M}}_{\textrm{S}}(h^*\to f(h_B)\bar{f}(h_C))=
-\frac{y_f}{\sqrt{2}}{\overline{u}}_{h_B}({\bar{p}}_B)v_{h_C}({\bar{p}}_C)\,,
\end{equation}
with $y_f$ the Yukawa coupling, $u$ and $v$ the Dirac 4-spinors for the fermion and anti-fermion with helicities $h_B$ and $h_C$. 

Rather than employing Dirac 4-spinors as in eq.~(\ref{eq:hff0}), the splitting amplitudes are most conveniently expressed using bi-spinors~\cite{Dittmaier:1998nn,Arkani-Hamed:2017jhn,Feige:2013zla,Franken:2019wqr,Lai:2023upa}. Our conventions are defined in Appendices~\ref{app:spi} and~\ref{sec:wf}. The amplitude is represented by a tensor with 2-dimensional Latin capital indices (e.g., $I$, running over $I=+,-$) in the \mbox{SU$(2)$} little-group of each external particle. The indices for the different particles $A$, $B$ and $C$ are carried by the angular or squared bracket bi-spinors associated with their on-shell momenta ${\bar{p}}_A$, ${\bar{p}}_B$ and ${\bar{p}}_C$. The amplitudes for particles of a given helicity $h$ can be obtained from the amplitude tensor by contracting with constant tensors $\tau(h)$ that we dub ``little-group wave functions''. For particles of spin $1/2$, the little-group wave functions are 1-index tensors reported in eqs.~(\ref{eq:Dirwf}) for incoming ($\tau_I$) and outgoing (${\bar{\tau}}^I$) external particles. Contracting with these tensors simply identifies the $I=+$ component of the tensor as the $h=+1/2$ helicity amplitude, while $I=-$ is the $h=+1/2$ helicity amplitude. In this notation, the $h^*\to f\bar{f}$ amplitude~(\ref{eq:hff0}) is represented by
\begin{equation}\label{eq:hff}
{\mathcal{M}}_{\textrm{S}}(h^*(A)\to f_{I}(B)\bar{f}_{\,J}(C))=
-\frac{y_f}{\sqrt{2}}
\bigg(
\langle B_I C_J\rangle+
[ B_I C_J]
\bigg)
\,.
\end{equation}

\begin{figure}
    \centering
 \begin{tikzpicture}[scale=1.3]
	\begin{feynman}
	\vertex (m) at (0, 0);
	\vertex (a) at (-0.8,0){};
    \vertex (b) at (0.87,0.5){${\bar{e}}_{M , IJ}$};
    \vertex (c) at (0.87,-0.5){${\bar{e}}_{M , KL}$};
	\diagram* {
		(a) -- [scalar,edge label = $h$] (m),
        (m)--[photon] (b),
        (m)--[scalar] (b),
       (m)--[photon] (c),
       (m)--[scalar](c),
	};
	\end{feynman}
	\end{tikzpicture}
 \begin{tikzpicture}[scale=1.3]
	\begin{feynman}
	\vertex (m) at (0, 0);
	\vertex (a) at (-0.8,0){$=\,$};
    \vertex (b) at (0.87,0.5){${\bar{e}}_{\mu , IJ}$};
    \vertex (c) at (0.87,-0.5){${\bar{e}}_{\mu , KL}$};
	\diagram* {
		(a) -- [scalar,edge label = $h$] (m),
        (m)--[photon] (b),
       (m)--[photon] (c),
	};
	\end{feynman}
	\end{tikzpicture}
 \begin{tikzpicture}[scale=1.3]
	\begin{feynman}
	\vertex (m) at (0, 0);
	\vertex (a) at (-0.8,0){$+\,$};
    \vertex (b) at (0.87,0.5){${\bar{e}}_{\pi , IJ}$};
    \vertex (c) at (0.87,-0.5){${\bar{e}}_{\mu , KL}$};
	\diagram* {
		(a) -- [scalar,edge label = $h$] (m),
        (m)--[dashed] (b),
       (m)--[photon] (c),
	};
	\end{feynman}
	\end{tikzpicture}
 \begin{tikzpicture}[scale=1.3]
	\begin{feynman}
	\vertex (m) at (0, 0);
	\vertex (a) at (-0.8,0){$+\,$};
    \vertex (b) at (0.87,0.5){${\bar{e}}_{\mu , IJ}$};
    \vertex (c) at (0.87,-0.5){${\bar{e}}_{\pi , KL}$};
	\diagram* {
		(a) -- [scalar,edge label = $h$] (m),
        (m)--[photon] (b),
       (m)--[dashed] (c),
	};
	\end{feynman}
	\end{tikzpicture}
 \begin{tikzpicture}[scale=1.3]
	\begin{feynman}
	\vertex (m) at (0, 0);
	\vertex (a) at (-0.8,0){$+\,$};
    \vertex (b) at (0.87,0.5){${\bar{e}}_{\pi , IJ}$};
    \vertex (c) at (0.87,-0.5){${\bar{e}}_{\pi , KL}$};
	\diagram* {
		(a) -- [scalar,edge label = $h$] (m),
        (m)--[dashed] (b),
       (m)--[dashed] (c),
	};
	\end{feynman}
	\end{tikzpicture}
    \caption{Feynman diagrams for Higgs splitting to a pair of massive bosons in the GE formalism.\label{Splitting-hVV}}
\end{figure}
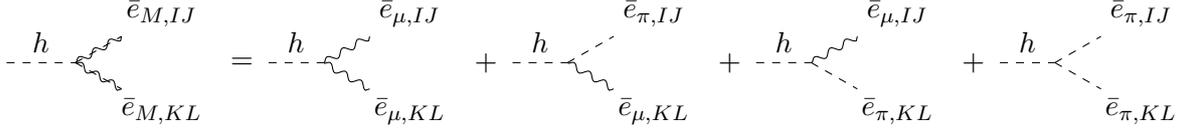

The calculation of the splitting amplitudes that involve spin-1 particles is slightly less standard because, as explained in the Introduction and in Appendix~\ref{App:EqFromalism}, the Goldstone-Equivalent (GE) Feynman rules have to be employed. Consider for illustration the FSR splitting of a Higgs boson to a pair of massive vector bosons. Each vector boson is represented by a double external line that describes an amputated leg of the five-components gauge and Goldstone field $\Phi^M=(V^\mu,\pi)$. The free index $M$ of the leg is contracted with a five-components polarisation vector that we denote as $\bar{\mathcal{E}}_M^h$---see eq.~(\ref{eq:pvGEAP}) and Figure~\ref{fig:EquivalentFR}---for final-state particles. By splitting the 5-vector into Lorentz 4-vector and scalar component we obtain a total of 4 diagrams for the $h^*\to VV$ splitting, like in Figure~\ref{Splitting-hVV}. 

Also for external spin-1 particles we employ little-group tensors to represent the amplitude \cite{Dittmaier:1998nn,Arkani-Hamed:2017jhn,Feige:2013zla,Franken:2019wqr,Lai:2023upa}. Our formalism, described in Appendix~\ref{sec:WFVP}, is an extension of the standard notation in which the GE Feynman rules are incorporated by employing amplitude tensors that are \emph{not symmetric} under the exchange of the little-group indices of the particle. The amplitude tensors are obtained by using eq.~(\ref{eq:GEPVTEout}) (see also eq.~(\ref{eq:GEPVTE}), for incoming particles) to express the GE polarisation vectors $\bar{\mathcal{E}}_M^h$ as an helicity-independent 2-index tensor ${\bar{e}}_{IJ,M}$ contracted with the little-group wave functions ${\bar\tau}^{IJ}(h)$. We thus obtain the amplitude tensor by contracting the $\mu$ index of the gauge external legs with the $M=\mu$ component of ${\bar{e}}$ and by multiplying the Goldstone external leg by the $M=\pi$ component, like in Figure~\ref{Splitting-hVV}. Helicity amplitudes are obtained from the amplitude tensor by contracting with the little-group wave functions in eqs.~(\ref{eq:VectoWF}) and~(\ref{eq:VectoWFout}). The transverse $h=\pm1$ helicity amplitudes are the $\pm\pm$ components of the tensor and the longitudinal amplitude is the $+-$ component.

The $h^*\to VV$ splitting amplitude tensor contains four terms, which are in direct correspondence with the diagrams in Figure~\ref{Splitting-hVV}. Using the Feynman rules for the Higgs-gauge-gauge, Higgs-gauge-Goldstone and Higgs-Goldstone-Goldstone vertices  in Appendix~\ref{Sec:FeynmanRules}, we obtain
\begin{eqnarray}\label{eq:hvv}
&&{\mathcal{M}}_{\textrm{S}}(h^*(A)\to V_{IJ}(B)V_{KL}(C))=
\frac{G}{m_V}
\bigg\{ \langle  B_J C_L\rangle [C_K B_I]\\
&&
-\frac12 \epsilon_{IJ} (\langle C_L B^P \rangle [B_P C_K]-m_V^2\epsilon_{KL})
-\frac12 \epsilon_{KL} (\langle  B_J C^P\rangle [C_P B_I]-m_V^2\epsilon_{IJ})
+\frac{m_h^2}4 \epsilon_{IJ} \epsilon_{KL} 
\bigg\}
\,.\nonumber
\end{eqnarray}
Inverse powers of the mass of the vector, $m_V$, arise in the formula from ${\bar{e}}_{IJ}^\mu$ (see eq.~(\ref{eq:VectoSpin1}) and from our parametrisation of the Golstone bosons vertices. Notice the presence of the anti-symmetric $\epsilon_{IJ}$ and $\epsilon_{KL}$ tensors.  The $m_V^2$ terms on the second line emerge from the square of the $\bar{p}_B$ and $\bar{p}_C$ momenta, which are on-shell. The $m_h^2$ term on the last line comes instead from the Feynman rule for the Higgs vertex with two Goldstones.

The GE Feynman rules are constructed such as to provide the same physical scattering amplitudes as the standard formalism. In light of this, one might ask if a naive calculation of the splitting amplitudes based on the standard Feynman rules would produce the same results. In the standard formalism, only the first (Higgs-gauge-gauge) diagram of Figure~\ref{Splitting-hVV} contributes, and the amplitude tensor is provided by the first term of eq.~(\ref{eq:hvv}). The tensor has to be contracted with the symmetric tensors ${\mathcal{S}}^{1,h}$ reported in eq.~(\ref{eq:stdpolten}) in order to extract the helicity amplitudes. If the vector bosons are transverse ($h=\pm1$), it is easy to check that one recovers the correct splitting amplitude evaluated with the GE rules, as it must be the case since the GE Feynman rules are different from the standard ones only for $h=0$ (longitudinal) states. If instead both vectors in the final state are longitudinal, the naive amplitude computed with the regular Feynman rules differs from the true amplitude obtained from eq.~(\ref{eq:hvv}) by an amount
\begin{equation}\label{eq:diff}
{\mathcal{M}}_{\textrm{S}}^{\textrm{naive}}-{\mathcal{M}}_{\textrm{S}}=
\frac{c}{4m_V}[({\bar{p}}_B+{\bar{p}}_C)^2-m_h^2]=
\frac{c}{4m_V}(p_A^2-m_h^2)=
\frac{c}{4m_V}Q^2_{\textsc{f}}\,.
\end{equation}
The naive splitting amplitude is equal to the complete one only at the kinematic point $Q^2_{\textsc{f}}=0$ where---if allowed by the masses---an on-shell Higgs boson decays to the two vectors.

The mismatch between the naive splitting amplitude and the correct one can be understood as follows. If the Higgs is on-shell, the amplitude is a physical decay amplitude, and hence the GE and the regular Feynman rules must give the same result. This is guaranteed for all physical amplitudes due to the validity of Ward identities like the one in eq.~(\ref{eq:WI}) for the amputated connected correlators of one or several $\Phi^M$ fields, on physical on-shell external states. If one of the states---the Higgs field, in the case at hand---is instead off-shell, the Ward identity is violated by calculable extra terms from which it is possible to recover eq.~(\ref{eq:diff}).

In the Introduction we emphasised that, using the standard formalism, the non-resonant diagrams can contribute sizably to the low-virtuality splitting process due to the anomalous energy-growth~\cite{Borel:2012by,Wulzer:2013mza,Cuomo:2019siu}. Avoiding the contribution from non-resonant diagrams is precisely the reason for employing the GE Feynman rules. On the other hand, the scattering amplitude of the complete physical process including both the resonant and the non-resonant diagrams must be the same in the two formalisms. The spurious contributions from the non-resonant diagrams must thus be accompanied and cancelled by spurious contributions from the resonant diagrams. The emergence of such spurious contribution is what we are observing in eq.~(\ref{eq:diff}) as a correction to the splitting amplitude. 

\subsection{Fermion intermediate particle}\label{sec:VD}

If the virtual particle has spin, establishing factorization requires a decomposition of the numerator of the propagator in terms of on-shell wave-function factors. The decomposition holds up to corrections, which produce the propagator correction term ${\Delta\mathcal{M}}_{\textrm{pro}}$ in eq.~(\ref{eq:facform}). The structure of the propagator indices contraction with the amputated hard and splitting amplitudes $\mathcal{A}_{\textrm{H}}$ and $\mathcal{A}_{\textrm{S}}$ is displayed in Figure~\ref{fig:Splitting-Dirac-Vector} in the case of a Dirac (left panel) and a vector (right panel) intermediate particle $A$ that undergoes splitting in the final state. The opposite orientation of the Dirac propagator line has to be considered if the $A$ particle is an anti-fermion. The $\mathcal{A}_{\textrm{H},\textrm{S}}$ amplitudes in the figure are the sum of amputated Feynman diagrams contracted with the wave-function factors for their on-shell external legs. Their free indices (e.g., Dirac indices $a$ and $b$) are the ones of the $A$ particle field.

It is easy to obtain a valid decomposition in the case of a Dirac intermediate particle. Performing the momentum shift in eq.~(\ref{eq:offonmom}), the numerator of the propagator becomes
\begin{equation}\label{eq:propdecD}
    {\slashed{p}}_A+m_A=({\slashed{\bar{p}}}_A+m_A) + \frac{Q_{\textsc{f}}^2}{p_{\pd}}{\slashed{n}}_{\md}=
    \sum\limits_{h_A} u_{h_A}({\bar{p}}_A){\bar{u}}_{h_A} ({\bar{p}}_A)+ \frac{Q_{\textsc{f}}^2}{p_{\pd}}{\slashed{n}}_{\md}\,,
\end{equation}
by the standard completeness relation for the Dirac spinors. The second term in the decomposition, times the amputated amplitudes $\mathcal{A}_{\textrm{H},\textrm{S}}$ and $1/Q_{\textsc{f}}^2$, gives rise to the propagator correction ${\Delta\mathcal{M}}_{\textrm{pro}}$ in the factorization formula~(\ref{eq:facform}). Since the virtuality cancels and $p_\pd$ is of order $E$, the scaling of ${\Delta\mathcal{M}}_{\textrm{pro}}$ with $E$ is dictated by the energy dimension of the amplitude. It scales like the off-shell corrections ${\Delta\mathcal{M}}_{\textrm{off}}$ and it is of the order of the non-resonant diagrams contribution.

The first term in the decomposition~(\ref{eq:propdecD}) contains the wave function factors ${\bar{u}}_{h_A}$ and $u_{h_A}$ to be contracted with $\mathcal{A}_{\textrm{H}}$ and $\mathcal{A}_{\textrm{S}}$, respectively. The hard amputated amplitude can be evaluated on the on-shell ${\bar{p}}_A$ momentum up to small off-shell corrections like in eq.~(\ref{eq:off}). By contracting with ${\bar{u}}$, this produces the hard amplitude term of eq.~(\ref{eq:facform}) 
\begin{equation}
{\mathcal{M}}_{\textrm{H}}(X\to A(h_A)Y)=({\bar{u}}_{h_A})_a\bar{\mathcal{A}}_{\textrm{H}}^a\,,
\end{equation}
with an on-shell fermion particles $A$ of helicity $h_A$ in the final state.

The splitting amplitude ${\mathcal{M}}_{\textrm{S}}$ is the contraction of $\mathcal{A}_{\textrm{S}}$ with $u_{h_A}$. The $\mathcal{A}_{\textrm{S}}$ amplitude already contains the wave-function factors associated with the real $B$ and $C$ particles. Further contracting with $u_{h_a}$ gives  ${\mathcal{M}}_{\textrm{S}}$ equal to the Feynman vertex associated with the splitting, contracted with exactly on-shell wave functions for all the external legs. Notice however that the on-shell momenta ${\bar{p}}_A$, ${\bar{p}}_B$ and ${\bar{p}}_C$ do not obey momentum conservation. 

Consider for illustration the radiation of a Higgs boson from a virtual fermion (e.g., from a top quark). The splitting amplitude reads
\begin{equation}
{\mathcal{M}}_{\textrm{S}}(f^*(h_A) \to f(h_B)h(C))=
-\frac{y_f}{\sqrt{2}}{\overline{u}}_{h_B}({\bar{p}}_B)u_{h_A}({\bar{p}}_A)\,.
\end{equation}
Alternatively, in the tensor notation
\begin{equation}
{\mathcal{M}}_{\textrm{S}}({f^*}^I(A)) \to f_J(B)h(C))=-\frac{y_f}{\sqrt{2}}
\bigg(
\langle B_J A^I\rangle-
[ B_J A^I]
\bigg)
\,.
\end{equation}

\subsection{Vector intermediate particle}\label{sec:VV}

It is slightly harder to work out a suitable decomposition of the propagator in the case of a massive spin-1 intermediate particles. The resonant Feynman diagrams are those where the final state particles $B$ and $C$ are emitted either from a gauge or from a Goldstone boson propagator, because in general if the $BC$ particle pair has a vertex with the gauge field $V^\mu$, it has also have a vertex with the associated Goldstone boson $\pi$.~\footnote{There can also be a vertex with the Higgs boson field. The associated diagrams feature a Higgs propagator and they have been discussed in Section~\ref{sec:VS}.} The contributions from the gauge and from the Goldstone are collected in a single diagram, depicted on the right panel of Figure~\ref{fig:Splitting-Dirac-Vector}, where the double line propagator denotes the 2-points vacuum correlator of the 5-components field $\Phi_M=(V_\mu,\pi)$. This $5\times5$ propagator is 
\begin{equation}\label{eq:Gprop}
    G_{MN}(p_A)= \frac{i}{Q_{\textsc{f}}^2}
    \begin{bmatrix}
    -\eta_{\mu\nu} & \mathbf{0}_{4\times 1}\\
        \mathbf{0}_{1\times 4} & 1 \\
    \end{bmatrix}_{MN}\,,
\end{equation}
with the Feynman choice $\xi=1$ for the $R_\xi$ gauge-fixing parameter.

The scattering amplitude takes the form
\begin{equation}\label{eq:grd}
i\,{\mathcal{M}}=
\mathcal{A}_{\textrm{S}}\left\{
\Phi^M[p_A]
\right\}
G_{MN}[p_A]
\mathcal{A}_{\textrm{H}}\left\{
\Phi^N[-p_A]
\right\}\,,
\end{equation}
where $\mathcal{A}_{\textrm{S,H}}\{\cdot\}$ denote the splitting and the hard amputated amplitudes. The amputated legs are the 5-component fields $\Phi^M$ and $\Phi^N$, and they are reported as an argument in the curly brackets. The momentum flowing in the leg ($+p_A$ or $-p_A$) is oriented towards the diagrams.  

Since $G_{MN}$ is a full-rank matrix, it cannot be decomposed on a basis formed only by the three (for helicity $h=\pm1,0$) polarisation vectors. Like in familiar massless gauge theories, two additional basis vectors are needed, whose contribution to the complete scattering amplitude~(\ref{eq:grd}) vanish because of Ward identities. We will make use of the massive gauge theory Ward identity at tree-level~\cite{Cuomo:2019siu} (see also Appendix~\ref{App:EqFromalism})
\begin{equation}\label{eq:wai}
{\mathcal{K}}_M[k]\,\mathcal{A}\{V^M[k]\}=0\,,\;{\textrm{where}}\;\;
{\mathcal{K}}_M[k]=[i\,k_\mu,m]_M\,,
\end{equation}
with $m$ the mass of the vector. The identity holds for generic off-shell momentum $k$ and for an amplitude $\mathcal{A}$ with arbitrary physical external particle legs. It thus applies both to the splitting and to the hard amplitude.

In order to operate the decomposition it is convenient to start from the completeness relation for the on-shell 5-components polarisations, reported in eq.~(\ref{eq:cr5d}). This equation  can be cast in the form
\begin{eqnarray}\label{eq:dec0}
    \begin{bmatrix}
    -\eta_{\mu\nu} & \mathbf{0}_{4\times 1}\\
        \mathbf{0}_{1\times 4} & 1 \\
    \end{bmatrix}_{MN}-
    \sum_{h=0\pm 1} \mathcal{E}_{M}^h({\bar{p}}_A)\bar{\mathcal{E}}_{N}^h({\bar{p}}_A)= \begin{bmatrix}
      [{{\bar{p}}_{A\mu}} \varepsilon^0_\nu({\bar{p}}_A)+\varepsilon^0_\mu({\bar{p}}_A){{\bar{p}}_{A\nu}}]/{m}& i\,\varepsilon^0_\mu({\bar{p}}_A)\\
       -i\,\varepsilon^0_\mu({\bar{p}}_A) & 0 \\
    \end{bmatrix}_{MN}\nonumber\\
    =
    -\frac{Q_{\textsc{f}}^2}{m\,p_{\pd}}
    \begin{bmatrix}
        n_{\md,\mu}\varepsilon^0_\nu({\bar{p}}_A)+\varepsilon^0_\mu({\bar{p}}_A)n_{\md,\nu}
     & \mathbf{0}_{4\times 1}\\
        \mathbf{0}_{1\times 4} & 0 \\
    \end{bmatrix}_{MN}\\
    -\frac{i}m{\mathcal{K}}_M[p_A]
    \begin{bmatrix}
    \varepsilon^0_\mu({\bar{p}}_A)\\0
    \end{bmatrix}_N
    +\frac{i}m{\mathcal{K}}_N[-p_A]
    {\begin{bmatrix}
    \varepsilon^0_\nu({\bar{p}}_A) \\ 0
    \end{bmatrix}_M}
    \,,\nonumber
\end{eqnarray}
where we used eq.~(\ref{eq:offonmom}) to shift the on-shell ${\bar{p}}_A$ back to the original off-shell momentum $p_A$, and we employed ${\mathcal{K}}[p_A]$ and ${\mathcal{K}}[-p_A]$ to eliminate to eliminate any explicit occurrence of $p_{A}$.

The two terms on the last line of eq.~(\ref{eq:dec0}) cancel by the Ward identity when contracted with $\mathcal{A}_{\textrm{S}}$ and $\mathcal{A}_{\textrm{H}}$, respectively. In the scattering amplitude~(\ref{eq:grd}) we can thus effectively replace the numerator of $G_{MN}$ with~\footnote{A more general---namely, in arbitrary gauge and beyond tree-level---derivation of the decomposition formula~(\ref{eq:eqnum}) is given in Ref.~\cite{Cuomo:2019siu}. The simplified derivation is presented here for a self-contained exposition.}
\begin{equation}\label{eq:eqnum}
        \begin{bmatrix}
    -\eta_{\mu\nu} & \mathbf{0}_{4\times 1}\\
        \mathbf{0}_{1\times 4} & 1 \\  \end{bmatrix}_{MN}\underset{\textrm{Ward}}{\rightarrow}
    \sum_{h=0\pm 1} \mathcal{E}_{M}^h({\bar{p}}_A)\bar{\mathcal{E}}_{N}^h({\bar{p}}_A)-\frac{Q_{\textsc{f}}^2}{m\,p_{\pd}}
    \begin{bmatrix}
        n_{\md,\mu}\varepsilon^0_\nu({\bar{p}}_A)+\varepsilon^0_\mu({\bar{p}}_A)n_{\md,\nu}
     & \mathbf{0}_{4\times 1}\\
        \mathbf{0}_{1\times 4} & 0 \\
    \end{bmatrix}_{MN}\,.
\end{equation}
This formula achieves the desired decomposition and offers the spin-1 analogous of the Dirac propagator decomposition formula in eq.~(\ref {eq:propdecD}). The correction term contains the GE longitudinal polarisation 4-vector $\varepsilon^0$ which---unlike the standard polarisation---scales like $m/E$ in the high-energy limit as discussed in Appendix~\ref{App:EqFromalism}. The $n_\md$ vector is a constant, therefore the corrections scale as the virtuality $Q_{\textsc{f}}^2$, times a factor of order $1/E^2$. The propagator corrections ${\Delta\mathcal{M}}_{\textrm{pro}}$ to the factorization formula are thus found to be small and comparable with the non-resonant diagrams contribution like in the case of fermionic intermediate particle studied in Section~\ref{sec:VD}. 

Using eq.~(\ref{eq:eqnum}), the derivation of the factorization formula~(\ref{eq:facform}) and the calculation of the splitting amplitudes proceeds very similarly to the case of Dirac particles. The on-shell polarisations ${\mathcal{E}}_M$ and ${\bar{\mathcal{E}}}_N$ contract with the splitting and the hard amputated amplitudes, respectively. After the on-shell limit is taken in the hard amplitude, the latter contraction produces the on-shell hard scattering amplitude ${\mathcal{M}}_{\textrm{H}}$. Notice that since this amplitude is physical and on-shell, it does not necessarily need to be computed in the GE formalism using the 5-dimensional description of the outgoing vector. The regular Feynman rules can be employed instead. The splitting amplitude ${\mathcal{M}}_{\textrm{S}}$---which emerges from the contraction of ${\mathcal{E}}_M({\bar{p}}_A)$ with $\mathcal{A}_{\textrm{S}}$---is instead not the scattering amplitude of a physical process. As such, it cannot be equivalently computed using the standard Feynman rules and the usage of the Goldstone Equivalent ones is mandatory and impacts the result similarly to what we discussed at the end of Section~\ref{sec:VS}.

Eq.~(\ref{eq:eqnum}) decomposes the propagator in terms of the 5-dimensional GE polarisation vectors. It is interesting to compare it with a decomposition that employs instead the polarisations $\varepsilon^h_{\textrm{St.}}$ and ${\bar{\varepsilon}}^h_{\textrm{St.}}$ that describe spin-1 particles in the standard formalism. These standard polarisation vectors are 4-dimensional, therefore in order to obtain this alternative decomposition one must first eliminate the contribution from the Goldstone bosons propagation, which corresponds to the 5-5 entry of the $G$ propagator~(\ref{eq:Gprop}). This is possible by using the Ward identity~(\ref{eq:wai}) to eliminate from eq.~(\ref{eq:grd}) the Goldstone field amputated amplitudes $\mathcal{A}_{\textrm{S,H}}\{\pi\}$, substituting them with $(\pm p_{A,\mu}/m) \mathcal{A}_{\textrm{S,H}}\{V^\mu\}$. This substitution effectively produces an additional contribution, equal to $+p_{A\mu}p_{A\nu}/m^2$, to the $4\times4$ upper-left block of the propagator matrix. If $p_A$ was on-shell, $p_A={\bar{p}}_A$ and the $4\times4$ block would now match the right-hand-side of the standard completeness relation in eq.~(\ref{eq:stdcompleteness}). However, the $p_A$ momentum is not on-shell. Its shift from on-shellnes~(\ref{eq:offonmom}) is proportional to $Q_{\textsc{i}}^2$, which is much smaller than $E^2$ but not smaller than $m^2$. Operating the on-shell shift on $+p_{A\mu}p_{A\nu}/m^2$ thus produces large correction terms, of order $Q_{\textsc{i}}^2/m^2$. Therefore, we can operate the decomposition also in terms of the standard polarisation vectors, but only up to correction terms that are not small, unlike the corrections in the GE decomposition formula~(\ref{eq:eqnum}). Both decompositions become exact at the special kinematic point $Q_{\textsc{i}}^2=0$ that corresponds to the pole of the amplitude for on-shell intermediate $A$. At this on-shell point, the two decompositions produce identical factorization formulas because the splitting amplitudes for $Q_{\textsc{i}}^2=0$ can be equivalently computed using the standard polarisation vectors. 

\subsection{Initial versus final state splitting}\label{sec:ISRvsFSR}

Only the case of final state (FSR) splitting has been discussed so far. A fully analogous treatment is possible for ISR splitting. The differences stem from the extra factor of $1/x$ that is present in the shift~(\ref{eq:offonmom}) from the off- to the on-shell momentum of the virtual particle (i.e., of the $C$ particle in the case of ISR). This factor appears in the decomposition of both the Dirac propagator \begin{equation}\label{eq:propdecDI}
    {\slashed{p}}_C+m_C=({\slashed{\bar{p}}}_C+m_C) + \frac{Q_{\textsc{i}}^2}{xp_{\pd}}{\slashed{n}}_{\md}=
    \sum\limits_h u_h({\bar{p}}_C){\bar{u}}_h ({\bar{p}}_C)+ \frac{Q_{\textsc{i}}^2}{xp_{\pd}}{\slashed{n}}_{\md}\,,
\end{equation}
and in the one of the vector propagator
\begin{equation}\label{eq:eqnumI}
        \begin{bmatrix}
    -\eta_{\mu\nu} & \mathbf{0}_{4\times 1}\\
        \mathbf{0}_{1\times 4} & 1 \\  \end{bmatrix}_{MN}\underset{\textrm{Ward}}{\rightarrow}
    \sum_{h=0\pm 1} \mathcal{E}_{M}^h({\bar{p}}_C)\bar{\mathcal{E}}_{N}^h({\bar{p}}_C)-\frac{Q_{\textsc{i}}^2}{x\,m\,p_{\pd}}
    \begin{bmatrix}
        n_{\md,\mu}\varepsilon^0_\nu({\bar{p}}_C)+\varepsilon^0_\mu({\bar{p}}_C)n_{\md,\nu}
     & \mathbf{0}_{4\times 1}\\
        \mathbf{0}_{1\times 4} & 0 \\
    \end{bmatrix}_{MN}\,.
\end{equation}

The extra $1/x$ enhances, for small $x$, both the propagator and the off-shell corrections to factorization in comparison with the FSR estimate of these corrections. However, a small value of $x$ in the ISR configuration reduces the scattering scale of the hard process $CX\to Y$ and therefore it is excluded from our analysis as explained in Section~\ref{sec:kin}. In the relevant region of $x$, defined by a lower threshold $x_{\textrm{min}}$~(\ref{eq:risr}), the scaling of the corrections to the factorised treatment of the ISR splitting is the same as for FSR splitting.

The splitting amplitudes for ISR splitting are given, like for FSR, by the wave-function factors for the particles involved in the splitting---with on-shell momenta ${\bar{p}}_A$, ${\bar{p}}_B$ and ${\bar{p}}_C$---times the 3-point $A\to BC$ Feynman vertex. However, the Feynman vertex, if it depends on the momenta, is evaluated on the true momenta $p_{A,B,C}$ that flow in the vertex. In the case of FSR, the splitting amplitude is thus obtained by substituting in the vertex $p_A={\bar{p}}_B+{\bar{p}}_C$, $p_B={\bar{p}}_B$ and $p_C={\bar{p}}_C$. By operating this substitution we obtain the amplitude tensors listed in Appendix~\ref{app:SA} for all the splitting processes of the SM. These results are thus valid only in the case of FSR.

In the case of ISR, the amplitude tensor should be obtained using instead $p_C={\bar{p}}_A-{\bar{p}}_B$, $p_A={\bar{p}}_A$ and $p_B={\bar{p}}_B$. The difference between the $p_{A,C}$ and ${\bar{p}}_{A,C}$ momenta is small, of order $\Delta p\sim Q^2/p_{\pd}$~(\ref{eq:offonmom}), but this does not imply that the ISR amplitudes are approximately equal to the FSR ones. Dimensional analysis does not prevent the emergence of an order $p_{\pd}/m$ factor of enhancement in front of $\Delta p$. In fact, we checked that the ISR amplitude \emph{tensors} are different from the FSR tensors reported in Appendix~\ref{app:SA}. However, one finds that the $p_{\pd}/m$ enhancement cancels when the amplitude tensors are contracted with the little-group wave functions in order to obtain the helicity amplitudes. The  ISR \emph{helicity} splitting amplitudes are thus equal to the FSR one up to small corrections. Furthermore, the corrections happen to vanish exactly with the SW bi-spinors basis that we will employ in the next section for the explicit evaluation of the splitting amplitudes. 

With this disclaimer, the ISR and FSR splitting amplitudes are effectively equal
\begin{equation}
    {\mathcal{M}}_{\textrm{S}}(A(h_A)\to B(h_B) C^*(h_C)) =
    {\mathcal{M}}_{\textrm{S}}(A^*(h_A)\to B(h_B) C(h_C)) \equiv {\mathcal{M}}_{\textrm{S}}(A(h_A)\to B(h_B) C(h_C))\,.
\end{equation}

\section{Splitting amplitudes computed}\label{sec:aimh}

We turn now to the explicit evaluation of the splitting amplitudes. Very compact expressions are obtained, in Section~\ref{ssec:imha}, by employing an explicit representation of the bi-spinor variables constructed in Appendix~\ref{ssec:sps} as a generalisation of the Soper--Weinberg (SW) bi-spinors~\cite{LCHSoper}. Regular (i.e., Jacob--Wick~\cite{Jacob:1959at}) helicity amplitudes are considered in Section~\ref{ssec:jwha}. In the same section, we also compare our results with the literature by verifying that the collinear splitting amplitudes are recovered in the appropriate limit. In Section~\ref{sec:SL} we study the soft limit and we recover the eikonal approximation for the emission of soft vector particles. 

\subsection{Soper-Weinberg amplitudes}\label{ssec:imha}

With the generalised SW definition of the single-particle states illustrated in Appendix~\ref{ssec:sps}, the bi-spinor of a particle with momentum $k$ reads
\begin{equation}\label{eq:imhsp}
    \ket{k^I}_\alpha=
    \mathcal{R}_\alpha^{\;\;\beta}
    \begin{bmatrix}
        {{m}}/{\sqrt{k_{\pd}}}& - {\bf{k^*_{\textsc{t}}}}/{\sqrt{k_{\pd}}}\\
        0 & {\sqrt{k_{\pd}}}\\
    \end{bmatrix}_{\beta}^{\;\;I}\,,\qquad
   [k_I|_{\dot\alpha} = (|k^I\rangle_{\dot\alpha})^*\,,
\end{equation}
where ${k_{\pd}}$ and ${\bf{k_{\textsc{t}}}}$ are the coordinates defined by eq.~(\ref{eq:coord}), $m$ is the particle mass and $\mathcal{R}$ is the rotation in eq.~(\ref{eq:rot}) that relates our reference frame to the regular light-cone frame. Notice that the explicit form of $\mathcal{R}$ will not be needed, because $\mathcal{R}$ acts as a Lorentz transformation and the splitting amplitudes are Lorentz singlets. The bi-spinor index $I$ assumes the values of $+$ or $-$, and $I=+$ represents the first column of the matrix.

The SW single-particles states enjoy---see Section~\ref{sec:IMH}---simple transformation properties under a boost ($K_3$) or a rotation ($J_3$) performed along the spacial direction, called $\hat{e}_3$, that is employed in the definition of the states. The ${k_{\pd}}$ and ${\bf{k_{\textsc{t}}}}$ coordinates also transform simply: ${k_{\pd}}\to e^{\zeta}k_{\pd}$ under $K_3$, while ${\bf{k_{\textsc{t}}}}$ is invariant; ${\bf{k_{\textsc{t}}}}\to e^{i\phi}{\bf{k_{\textsc{t}}}}$ under $J_3$, while $k_{\pd}$ is invariant. The simple transformation rule of the states is reflected in a simple transformation rule for the spinors~(\ref{eq:imhsp}) under the transformation of the momentum coordinates. It is easy to check that
\begin{eqnarray}
    \ket{k^\pm}_\alpha \overset{{k_{\pd}}\to e^{\zeta}k_{\pd}}{\longrightarrow}
    \left[\mathcal{R}
    \cdot
    \begin{bmatrix}
        e^{-\zeta/2}&0\\
        0 & e^{\zeta/2}\\
    \end{bmatrix}
    \cdot {\mathcal{R}}^{-1}    \right]_\alpha^{\;\;\beta}\ket{k^\pm}_\beta=\left[
    e^{-i\zeta
    \mathscr{K}_3} \right]_\alpha^{\;\;\beta}\ket{k^\pm}_\beta\,,\\
    \ket{k^\pm}_\alpha \overset{
    {\bf{k_{\textsc{t}}}}\to e^{i\phi}{\bf{k_{\textsc{t}}}}}{\longrightarrow}
    e^{\pm i\phi/2}
    \left[\mathcal{R}
    \cdot
    \begin{bmatrix}
        e^{-i\phi/2}&0\\
        0 & e^{i\phi/2}\\
    \end{bmatrix}
    \cdot {\mathcal{R}}^{-1}    \right]_\alpha^{\;\;\beta}\ket{k^\pm}_\beta=
    e^{\pm i\phi/2}\left[
    e^{-i\phi
    \mathscr{J}_3}    \right]_\alpha^{\;\;\beta}
    \ket{k^\pm}_\beta\,,\nonumber
\end{eqnarray}
where $\mathscr{J}_3$ and $\mathscr{K}_3$ denote the $J_3$ and $K_3$ representation matrices in the spinor representation. Under $K_3$, the bi-spinor is invariant up to a Lorentz transformation. The $I=+$ and $I=-$ spinors transform as charge $+1/2$ and $-1/2$ objects under $J_3$, again up to a Lorentz transformation. Since the Lorentz transformation cancels out in the amplitude, which is a Lorentz scalar, these rules entail strong selection rules for the dependence of the amplitudes on the kinematic variables that describe the splitting.

The evaluation of the splitting amplitudes proceeds straightforwardly. The bi-spinors $\ket{A^I}$, $\ket{B^I}$ and $\ket{C^I}$ are obtained by substituting in eq.~(\ref{eq:imhsp}) the ${\bar{p}}_{A,B,C}$ momenta components, parametrised as in eq.~(\ref{eq:onmom}). The rotation $\mathcal{R}$ in eq.~(\ref{eq:imhsp}) identifies the direction along which the SW states are defined. The same direction---i.e., the one of the $A$ particle, as in eq.~(\ref{eq:rot})---is employed for all the three particles involved in the splitting. The amplitudes are tensors in the little-group space, but they are Lorentz scalars constructed by the invariant contractions of the spinors. The $\mathcal{R}$ matrix drops in these contractions, which read
\begin{eqnarray}\label{eq:bsmat}
        &\langle A^I B^J \rangle = \begin{bmatrix}
        0 & -m_A\sqrt{1-x}\\
        m_B/{\sqrt{1-x}}& -{{\bf{p^*_{\textsc{t}}}}}/{\sqrt{1-x}} 
    \end{bmatrix}_{IJ}=[ B_J A_I ]^*\,,&\nonumber\\
    &\langle A^I C^J \rangle = \begin{bmatrix}
        0 & 
        -m_A\sqrt{x}\\
       {m_C}/{\sqrt{x}} & {{\bf{p^*_{\textsc{t}}}}}/{\sqrt{x}} 
    \end{bmatrix}_{IJ}=[ C_J A_I ]^*\,,&\\
    &\langle B^I C^J \rangle = \begin{bmatrix}
        0 & -m_B \sqrt{{x}/{(1-x)}}\, \\
        m_C\sqrt{{(1-x)}/{x}}  & {\bf{p^*_{\textsc{t}}}}/{\sqrt{(1-x)x}}
    \end{bmatrix}_{IJ}=[ C_J B_I ]^*\,.&\nonumber
\end{eqnarray}

The simplicity of the result stems from the selection rules associated with the $K_3$ and $J_3$ symmetries. Among the variables that describe the splitting ($p_{\pd}$, ${\bf{p_{\textsc{t}}}}$ and $x$), only $p_{\pd}$ transforms under $K_3$, like $p_{\pd}\to e^{\zeta}p_{\pd}$. The spinor products are invariant, hence $p_{\pd}$ cannot appear in the result. Notice that $p_{\pd}\sim E$ is the largest energy scale in the low-virtuality splitting configuration. The $K_3$ symmetry forbids the splitting amplitudes to scale with $ p_{\pd}$ and makes them proportional to $\mathbf{p}_T$ or to the masses. 

The only variable that transforms under $J_3$ is ${\bf{p_{\textsc{t}}}}\to e^{i\phi}{\bf{p_{\textsc{t}}}}$. The $(2,2)$---i.e., $(-,-)$---entry of the bi-spinor products matrix has charge $-1$ under $J_3$, hence it is proportional to ${{\bf{p^*_{\textsc{t}}}}}$. The off-diagonal entries have charge zero, while the $(1,1)$ entry should have charge plus and therefore it vanishes because the spinors depend on ${{\bf{p^*_{\textsc{t}}}}}$ and not on ${{\bf{p_{\textsc{t}}}}}$. The splitting variable $x$ is invariant under $K_3$ and $J_3$ because it is defined as the fraction of momentum along the direction $n_{\pd}$. The dependence of the bi-spinor products on $x$ is thus not determined by symmetries.

Interestingly enough, the dependence on the masses $m_{A,B,C}$ of the particles is also dictated by selection rules. The ``$+$'' and ``$-$'' spinors~(\ref{eq:imhsp}) are respectively odd and even under the replacement $m\to-m$. This means, for instance, that the first line of the $\langle A^I B^J \rangle$ bi-spinor product matrix should be odd under the sign flip of the $A$ particle mass, and the first column must be odd under $m_B\to-m_B$, and similarly for the other matrices. This latter selection rule, plus the polynomial dependence of the spinors on ${{\bf{p^*_{\textsc{t}}}}}$ and the masses, and dimensional analysis, determines the dependence of eq.~(\ref{eq:bsmat}) on all the variables apart from $x$. 

The last step for the determination of the helicity amplitudes in the SW basis is to substitute eq.~(\ref{eq:bsmat}) in the amplitude tensors of Appendix~\ref{app:SA}, and to project the little-group indices along the components that correspond to the helicity eigenstates. The results are reported in Appendix~\ref{Sec:SWSplittings} for all splittings that occur in the Standard Model.

\subsection{Jacob--Wick amplitudes}\label{ssec:jwha}

A priori, the splitting amplitudes in the SW representation of the bi-spinors are not useful for calculations. The SW bi-spinors correspond to a definition of the scattering states that is adapted to the direction of the emitter (i.e., the particle $A$) for the specific splitting process under consideration. However, the complete scattering process might receive contributions from multiple splitting topologies. For instance, the process $\mu^+\mu^-\to e^-{\bar{\nu}}_e W^+$ at a muon collider receives contributions from resonant diagrams where the $W$ is emitted from a virtual $\nu_e$, by the $\nu_e\to e^- W^+$ FSR splitting, as well as contributions from the $e^+\to {\bar{\nu}}_e W^+$ splitting, and from the ISR splitting $\mu^+\to W^+{\bar{\nu}}_\mu$. If the $W$ is emitted with a small angular separation from one of the other legs---i.e., in the collinear region---one specific splitting gives the dominant contribution to the amplitude and it can be reasonable to employ a definition of the states that is adapted to the direction of that specific splitting. In the soft region instead, the $W$ is not necessarily close in angle to any other particle and the low-virtuality enhancement is due to the small $W$ energy. In this case, all the different splitting topologies contribute and there is no way to single out a special direction to define SW states.

It is of course straightforward to compute the splitting amplitudes in any explicit representation of the bi-spinors. The most commonly employed basis for the single-particle states is the Jacob--Wick (JW) helicity basis. As reviewed in Appendix~\ref{ssec:sps}, the corresponding bi-spinors are 
\begin{equation}\label{eq:jwsp}
    \ket{k^I_{\textsc{jw}}}_\alpha=
     \begin{bmatrix}
        \sqrt{k^0-|{\bf{k}}|} \cos\frac{\theta}{2} & 
        -\sqrt{k^0+|{\bf{k}}|} \sin\frac{\theta}{2} e^{-i\varphi}
        \\
        \sqrt{k^0-|{\bf{k}}|} \sin\frac{\theta}{2} e^{i\varphi}
         &   \sqrt{k^0+|{\bf{k}}|} \cos\frac{\theta}{2}\\
    \end{bmatrix}_{\alpha}^{\;\;I}\,,
\end{equation}
where $k^0$ is the energy, ${{\bf{k}}}$ the 3-momentum and $(\theta,\phi)$ the polar and azimuthal angles of the particle momentum. By evaluating the amplitude tensors in Appendix~\ref{app:SA} using this explicit representation for the $A$, $B$ and $C$ bi-spinors, and contracting with the little-group wave functions, we could easily obtain the splitting amplitudes in the JW helicity basis. 

The explicit form of the JW splitting amplitudes is complicated. Therefore, in order to study their properties it is convenient to consider their relation with the amplitudes computed on the SW bi-spinors. Bi-spinors in different representations are related by some SU$(2)$ transformation that acts on the little-group indices. The transformation that relates JW with SW spinors
\begin{equation}
 \ket{k_{\textsc{jw}}^I}={\mathcal{U}}^I_{\;\,J}
    \ket{k_{\textsc{sw}}^J}\,,
\end{equation} 
is operated by a matrix ${\mathcal{U}}$ that depends on the particle's momentum and that can be expressed as in eq.~(\ref{eq:WR}), in terms of the bi-spinors in the two representations. The amplitude tensor evaluated with the JW bi-spinors can thus be obtained by rotating with $\mathcal{U}$ the little-group indices of the tensors computed with the SW bi-spinors. Since the matrix $\mathcal{U}$ depends on the momentum, 3 different little-group rotations have to be performed on the indices associated to the $A$, $B$ or $C$ external states. This rotated amplitude tensor can be eventually contracted with the little-group wave functions in order to obtain the helicity amplitudes, or the rotation can be equivalently applied to the little-group wave functions indices like in eq.~(\ref{eq:wftt}). Notice that, as explained in Appendix~\ref{sec:AMPT}, in the case of spin-1 particles this little-group rotation does not correspond to rotating the helicity index in the triplet of SU$(2)$. This is because the splitting amplitudes are not physical amplitudes and they do not obey the Ward identity. The little-group rotation must be then performed on the amplitude tensors and not on the helicity amplitudes.

We show in Section~\ref{sec:rel} that the ${\mathcal{U}}$ transformation simplifies in two limits. One is the massless limit, in which ${\mathcal{U}}$ is diagonal. The other limit is when the particle, either massless or massive, has 3-momentum ${\bf{k}}$ parallel to the direction, $e_3$, that is employed for the definition of SW states. In this case, $\mathcal{U}=\mathds{1}$ because the JW and the SW states definitions coincide. One can also verify directly that the bi-spinors in eqs.~(\ref{eq:jwsp}) and~(\ref{eq:imhsp}) are equal when ${\bf{k}}$ is parallel to $e_3$.\footnote{If ${\bf{k}}$ is parallel to $e_3$, ${\bf{k_{\textsc{t}}}}=0$ and $k_{\pd}=k^0+|\mathbf{k}|$. Furthermore, the $(\theta,\varphi)$ angles are those that define the $e_3$ direction and in turn the $\mathcal{R}$ rotation.}

By this observation we immediately conclude that the JW amplitudes are in fact equal to the SW amplitudes in the case of a \emph{collinear} splitting, where all particles are emitted in the same direction and therefore the ${\mathcal{U}}$ matrices are trivial. The SW splitting amplitudes in Appendix~\ref{Sec:SWSplittings} can thus be directly compared with the collinear splitting amplitudes  computed in~\cite{Cuomo:2019siu,Chen:2016wkt}. More precisely, the configuration studied in~\cite{Cuomo:2019siu,Chen:2016wkt} is the one where the hard scattering scale $E$ is much larger than the EW scale---and so much larger than the mass of the particles involved in the splitting---and much larger than ${{\bf{p_{\textsc{t}}}}}$. The variable $x$ is far from the extreme configurations $x=0$ or $x=1$. Since $p_{\pd}$ is of the order of the hard scale $E$, the momenta of particles $B$ and $C$ in eq.~(\ref{eq:onmom}) align with the direction of the $A$ particle and therefore the splitting is collinear. Furthermore, the splitting particles are ultra-relativistic. Hence, the fraction $x$ of momentum in the $n_{\pd}$ direction that we used to characterise the splitting is equal to the fraction of energy, which is the variable employed in Refs.~\cite{Cuomo:2019siu,Chen:2016wkt}. Consistently, our results for the splitting amplitudes in Appendix~\ref{Sec:SWSplittings} are equal in form to the ones of Refs.~\cite{Cuomo:2019siu,Chen:2016wkt}.

In a generic non-collinear kinematic configuration, JW and SW splitting amplitudes differ by non-trivial ${\mathcal{U}}$ rotations performed on the $B$ and $C$ external legs. The rotation of the $A$ particle is instead always trivial, because the $A$ particle momentum is parallel by construction to $e_3$. The non-trivial rotations are important in particular in the enhanced low-virtuality configuration that corresponds to a \emph{soft} splitting where $x\simeq0$ or $x\simeq1$. We will see in the following section that the amplitudes in the soft limit can be compactly represented by the so-called \emph{eikonal} formula.

\subsection{The soft limit}\label{sec:SL}

We emphasised in Section~\ref{sec:lwefac} that a low-virtuality enhancement can be attained by many different kinematic configurations, several of which---like decay configurations---are extensively studied and not directly relevant for EW radiation. Our motivation is to study the new regime where the virtuality is comparable or larger than the EW scale, $|Q^2|\gtrsim m_{\textsc{ew}}^2$, with much larger hard scale $E^2\gg |Q^2|$. We will show now that, if $|Q^2|\gtrsim m_{\textsc{ew}}^2$, the only enhanced kinematic configurations are the collinear configuration previously described, and the soft configuration. This result follows from the fact that SM particles have a mass that is either comparable or much smaller than $m_{\textsc{ew}}$.

Consider for definiteness the virtuality of an FSR splitting. This was obtained in eq.~(\ref{eq:FSRVIRT}), and can be written as
\begin{equation}\label{eq:Q2}
    Q^2=\frac{|{\bf{p}_{\textsc{t}}}|^2}{x(1-x)}
    -\frac{(1-x)m_A^2-
    \,m_B^2}{1-x}+\frac{m_C^2
    }{x}
    \,.
\end{equation}
For any value of $x$ far from the extremes $x=0$ or $x=1$, the second and the third terms are of order $m_{\textsc{ew}}^2$, if some heavy SM particle is involved in the splitting, or much smaller. The sum of the two terms can be negative, but it is in any case of order $m_{\textsc{ew}}^2$ or less in absolute value. Therefore, it would be impossible to obtain a virtuality $Q^2$ that is comparable or larger than $m_{\textsc{ew}}^2$ if the first term was much above $Q^2$ because the other terms cannot be large and negative in order to cancel the first one. This implies the upper bound $|{\bf{p}_{\textsc{t}}}|^2\lesssim |Q^2|$. 

Let $\lambda$ be the small parameter associated with $|Q^2|$ measured in units of the hard scale $E^2$. Since $|{\bf{p}_{\textsc{t}}}|^2$ is bounded by $|Q^2|$, and the masses are also bounded by $|Q^2|\gtrsim m_{\textsc{ew}}^2$, we can take the limit that corresponds to this configuration by counting the parameters with the rules
\begin{equation}
{{\bf{p_{\textsc{t}}}}}\sim\sqrt{\lambda}\,,\qquad m_{A,B,C}\sim\sqrt{\lambda}\,,\qquad p_\pd\sim E\sim \lambda^0\,.
\end{equation}
This counting corresponds to the collinear configuration with ultra-relativistic splitting particles and small emission angle. 

Note that the bound $|{\bf{p}_{\textsc{t}}}|^2\lesssim |Q^2|$ can be violated if $|Q^2|$ is much smaller than the EW scale $m_{\textsc{ew}}^2$. An arbitrarily small or even vanishing $|Q^2|$ could be obtained in eq.~(\ref{eq:Q2}) by a cancellation between the first (positive) term---which is much larger than $|Q^2|$ if $|{\bf{p}_{\textsc{t}}}|^2$ is large---and the sum of the other terms, which can be negative if $m_A>m_B+m_C$. This enhanced configuration corresponds to the decay of the $A$ particle to $B$ and $C$. However, we have seen that it cannot be realised in the region of interest $|Q^2|\gtrsim m_{\textsc{ew}}^2$ within the SM. 

Let us now take $x$ close to the extremes. We consider for definiteness $x\ll1$, but a fully analogous treatment is possible if $x$ is instead close to 1. The second term in eq.~(\ref{eq:Q2}) is, once again, of order $ m_{\textsc{ew}}^2$ or less in absolute value and therefore the sum of the first and the third term is bounded by $|Q^2|$ if we want $|Q^2|\gtrsim m_{\textsc{ew}}^2$, as previously discussed. The two terms are positive, hence the bound applies to each of them individually and we have
\begin{equation}\label{eq:boundssoft}
    |{\bf{p}_{\textsc{t}}}|^2\lesssim x\,|Q^2|\,,\qquad
    m_C^2\lesssim x\,|Q^2|\,.
\end{equation}
Notice that the $C$ particle is the one that becomes soft---see eq.~(\ref{eq:onmom})---in the limit $x\to0$. The upper bound on its mass is equivalent to a lower bound on $x$: $x\gtrsim m_C^2/|Q^2|$. This bounds acts as a cutoff for the singularity at $x=0$ that would be encountered for massless $C$ particle. 

The soft limit can be thus be taken with the counting rule
\begin{equation}
    {\bf{p}_{\textsc{t}}}\sim \sqrt{x}\,\sqrt\lambda\,,\qquad m_C\sim \sqrt{x}\,\sqrt\lambda\,,\qquad p_\pd\sim\lambda^0\,,
\end{equation}
where $\lambda$ is---like before---the low-virtuality parameter, and the momentum fraction $x$ is treated as an independent small expansion parameter. As a particular case of the two-parameters expansion, one could count $x$ as an order-$\lambda$ parameter. In this combined counting, all the components of the $C$ particle 4-momentum in eq.~(\ref{eq:onmom}) would count as $\lambda$. However, counting powers of $\lambda$ and $x$ separately is more convenient for our purposes.

The masses $m_A$ and $m_B$ of the other particles involved in the splitting must be counted differently depending on whether the soft $C$ particle is one of the heavy SM particles such as the massive $W$ and $Z$ bosons or much lighter such as the massless photon. In the first case, since no SM particle is much heavier that the $W$ and the $Z$, $m_{A}$ and $m_B$ are either comparable or much smaller than $m_C$. Therefore, the upper bound in eq.~(\ref{eq:boundssoft}) applies to $m_A$ and $m_B$ as well, and the counting rule becomes
\begin{equation}\label{eq:scH}
{\bf{p}_{\textsc{t}}}\sim \sqrt{x}\,\sqrt\lambda\,,\qquad m_{A,B,C}\sim \sqrt{x}\,\sqrt\lambda\,,\qquad p_\pd\sim\lambda^0\,.
\end{equation}
If instead the $C$ particle is light, $m_A$ and $m_B$ are in general only bounded by $|Q^2|$, and we have
\begin{equation}\label{eq:scL}
{\bf{p}_{\textsc{t}}}\sim \sqrt{x}\,\sqrt\lambda\,,\qquad m_{C}\sim \sqrt{x}\,\sqrt\lambda\,,\qquad m_{B,C}\sim \sqrt\lambda\,,\qquad p_\pd\sim\lambda^0\,.
\end{equation}

By applying the power-counting rules we will now analyse the soft limit of the splitting amplitude tensors computed on the SW spinors. The scaling with $\lambda$ is trivial and universal, because it is dictated by dimensional analysis: all dimensionful quantities count as $\sqrt{\lambda}$ with the exception of $p_\pd$, however we have seen that the SW amplitude tensors do not depend on $p_\pd$. The amplitude tensor has energy dimension 1, hence it scales like $\sqrt{\lambda}$. 

The rules for counting $x$ powers depends on whether $C$ is a heavy~(\ref{eq:scH}) or a light~(\ref{eq:scL}) SM particle. However, a result that holds true in both cases is that a scaling with a negative power of $x$---specifically, with $1/\sqrt{x}$---is possible only for amplitudes where the soft $C$ particle has spin~1. This can be established by recalling that the amplitude tensors build up as products of the bi-spinor contractions reported in eq.~(\ref{eq:bsmat}), and by noticing that all these contractions are finite in the low-$x$ limit. A divergent behaviour is found only in amplitudes where the soft $C$ particle has spin~1, due to the $1/m_C\sim1/\sqrt{x}$ factor from the normalisation of the $\bar{e}$ polarisation tensor in eq.~(\ref{eq:VectoSpin}).\footnote{If particles $A$ or $B$ are vectors, divergent $1/m_{A,B}\sim1/\sqrt{x}$ factors also arise from the normalisation of the corresponding polarisation tensors. However, these factors are compensated by the scaling with $x$ of the $A$ and $B$ bi-spinors, and they cancel out in the polarisation vectors for physical $A$ and $B$ particle helicity. The amplitudes with physical $A$ and $B$ helicity such as those in eq.~(\ref{eq:generaleikonal}) are thus insensitive to these low-$x$ enhancements.} This result corresponds, in our more general context, to the well-known fact that soft singularities in massless gauge theories arise only from the emission of gauge bosons.

By inspecting the splitting processes where the $C$ particle is a vector, one can show that the leading (order $\sqrt{\lambda}/\sqrt{x}$) contributions to the amplitude are given by the eikonal formula
\begin{equation}\label{eq:generaleikonal}
    {\mathcal{M}}_{\textrm{S}}^{(\textsc{e})}(A(h_A)\to B(h_B)\,V_{IJ}(C))=
    2\,{\mathcal{G}}^C_{AB}\,\delta_{h_B}^{h_A}\,{\bar{p}}_{A,\mu}\, {\bar{e}}^\mu_{IJ}(C)\,.
\end{equation}
In the formula we employ a mixed representation of the external states, where $A$ and $B$ are helicity eigenstates and their little-group indices are contracted with the little-group wave functions. The $C$ particle indices are not contracted and they are carried by the $\bar{e}$ polarisation tensor defined in eq.~(\ref{eq:VectoSpin1}). 

The structure of the eikonal formula~(\ref{eq:generaleikonal}) matches the regular eikonal approximation in massless gauge theories, adapted to the SM in the massless limit (see e.g.~\cite{Fadin:1999bq}). We collect in ${\mathcal{G}}$ the gauge coupling and the non-Abelian charge associated with the $C$ gauge field in the representation of the particles $A$ and $B$. If $A$ and $B$ are fermions, the representation depends on the helicity of the particles because the SM is a chiral gauge theory. If $A$ and $B$ are spin-1 particles and transversely polarised, the ${\mathcal{G}}$ matrices are the SM group generators in the Adjoint representation. If instead $A$ and $B$ are longitudinal vector bosons or the Higgs, the generators are the ones of the Higgs doublet field. This follows from the correspondence between longitudinal vectors and Goldstone bosons in the massless limit of the SM. Note that the eikonal formula is diagonal in the helicities $h_{A,B}$: the soft enhancement is only in helicity-preserving transitions.

It is interesting to illustrate how the eikonal formula~(\ref{eq:generaleikonal}) emerges as the soft limit of the amplitude tensors in some example. Let us first consider the case in which the $C$ particle is one of the massive SM vector bosons $W$ or $Z$. Their masses are of the order of the EW scale, therefore in order to take the soft limit we must apply the counting of eq.~(\ref{eq:scH}) in which $m_A$ and $m_B$ count as small order $\sqrt{x}$ parameters. With this counting, particle $A$ and $B$ are effectively massless in the limit. Furthermore, since ${\bf{p}_{\textsc{t}}}\sim\sqrt{x}$ is also small, the $B$ particle 4-momentum~(\ref{eq:onmom}) approaches the momentum of the $A$ particle. We can thus take the soft limit in the amplitude tensor by replacing the $B$ particle bi-spinor with
\begin{equation}\label{eq:replsL}
    \ket{B^I}\;\rightarrow\; \ket{A^I}\simeq \ket{\hat{A}^I}\,,
\end{equation}
where $\ket{\hat{A}^I}$ denotes the massless limit of the $A$ particle bi-spinor. Only the minus component of $\ket{\hat{A}^I}$ is non-vanishing, and $\ket{\hat{A}^I}$ is related to the momentum as

\begin{equation}\label{eq:mlsA}
   |\hat{A}^{N} \rangle_\alpha[ \hat{A}_{M}|_{\dot{\alpha}}
    =
    {(\bar{p}_A)}_{\alpha \dot{\alpha}}\,\delta_-^N\delta_M^-\,.
\end{equation}

Consider for instance the vector emission from a fermion. The amplitude tensor is reported in eq.~(\ref{eq:Spfvf}). By taking the massless limit for the $A$ and $B$ particles, operating the replacement in eq.~(\ref{eq:replsL}) and using eq.~(\ref{eq:mlsA}), we obtain
\begin{eqnarray}\label{eq:ampfersl}
\mathcal{M}_S(f^N(A)\rightarrow f_M(B)V_{IJ}(C))\simeq\sqrt{2}\,\frac{G_R}{m_C}\langle \hat{A}_M C_J\rangle [C_I \hat{A}^N]-\sqrt{2}\,\frac{G_L}{m_C}[ \hat{A}_M C_I] \langle C_J \hat{A}^N\rangle\,.\nonumber\\
=
2\,G_R \, {\bar{p}}_{A,\mu}\bar{e}^\mu_{IJ}(C)\,\delta_+^N\delta_M^+ 
+2\,G_L\,  {\bar{p}}_{A,\mu}\bar{e}^\mu_{IJ}(C)
\,\delta_-^N\delta_M^- \,.
\end{eqnarray}
This expression matches the eikonal formula~(\ref{eq:generaleikonal}). The coupling factor for $A$ and $B$ fermions with positive helicity ($N=M=+$) is provided by the right-handed coupling, $G_R$, while the left-handed coupling $G_L$ is present when the helicities are negative. There are no helicity-flipping transitions with $M\neq N$. 

It is even simpler to verify the validity of the eikonal formula if the $C$ vector particle is a massless photon or gluon. The parameters counting in eq.~(\ref{eq:scL}) does not allow us to take the massless limit for the $A$ and $B$ particles, however we have that $m_A=m_B$ because massless gauge fields only couple particles with the same mass. The transverse momentum ${\bf{p}_{\textsc{t}}}\sim\sqrt{x}$ is small, therefore $B$ particle 3-momentum becomes equal to the one of the $A$ particle. Since the masses are also equal, the $B$ particle bi-spinor coincides with the one of the $A$ particle, in the limit. Therefore, we can still perform the replacement in eq.~(\ref{eq:replsL}), but we can not take the massless limit. Further simplifications emerge in the amplitude tensors involving the photon or the gluon, due to gauge invariance. For instance, the coupling to fermions is vector-like: $G_R=G_L=G$. The fermion amplitude tensor of eq.~(\ref{eq:Spfvf}) becomes
\begin{equation}
\mathcal{M}_S(f^N(A)\rightarrow f_M(B)V_{IJ}(C))\simeq\sqrt{2}\,\frac{G}{m_C}\big(\langle {A}_M C_J\rangle [C_I {A}^N]-[ {A}_M C_I] \langle C_J {A}^N\rangle\big)\,.
\end{equation}
We can check that it matches the eikonal formula~(\ref{eq:generaleikonal}) by employing the relation
\begin{equation}
    |A^{I} \rangle_\alpha [ A_{J}|_{\dot{\alpha}}-
    |A_{J} \rangle_\alpha [ A^{I}|_{\dot{\alpha}} 
    =({\bar{p}}_A)_{\alpha\dot{\alpha}}\,\delta^I_J\,.
\end{equation}

It should be emphasised that the eikonal approximation holds---as stated in eq.~(\ref{eq:generaleikonal})---for the amplitude tensor with free indices $I$ and $J$ in the little-group of the soft particle $C$, and not only for the components of the tensor that correspond to physical $C$ particle helicities. This is important because it guarantees that the approximation is valid independently of the explicit representation of the bi-spinors, in spite of the fact that the SW spinors are most convenient in order to take the soft limit and to establish the formula. In fact, we stressed in Section~\ref{ssec:jwha} that the splitting amplitude tensors in different bi-spinor representations are related by little-group transformations, but that these transformations do not corresponds to \mbox{SU$(2)$} rotations in the space of physical helicities. However, since it holds for the entire tensor, we could start from eq.~(\ref{eq:generaleikonal}) written with SW bi-spinors and safely rotate it to any other bi-spinors representation such as the JW basis. The rotation acts on the indices of ${\bar{e}}_{IJ}$, which becomes the wave-function tensor in the new basis. Notice that the rotation from the SW to the JW bi-spinors also requires a little-group rotation on the $B$ particle indices. However, this rotation is trivial in the soft limit because the $B$ particle 3-momentum becomes parallel (in fact, equal) to the one of the particle $A$. Therefore, it is not an issue that the validity of eq.~(\ref{eq:generaleikonal}) is restricted to the physical helicity-eigenstate components of the $B$ particle indices.

Our formula~(\ref{eq:generaleikonal}) extends to massive gauge theories the eikonal approximation of the soft splitting amplitudes for massless theories. The result could have been guessed by analogy with the massless eikonal formula, apart from one aspect. The amplitudes for emitting transversly-polarised (the $\pm\pm$ components~(\ref{eq:VectoWF}) of the tensor) or longitudinally-polarised (the $+-$ component) vector bosons are of the same order ($\sqrt\lambda/\sqrt{x}$) in the soft power-counting. Longitudinal vector bosons thus experiences a soft enhancement, like transverse vectors. Notice however that the soft amplitudes for the emission of transverse vectors (which we can compute either by eq.~(\ref{eq:generaleikonal}), or by taking the soft limit of the amplitudes in Appendix~\ref{Sec:SWSplittings}) are proportional to ${\bf{p}_{\textsc{t}}}$ and therefore they feature a double logarithmic singularity and contribute as Sudakov double-logs to the integral of the cross-section over the phase space of the radiation. Longitudinal amplitudes are instead proportional to the mass and thus they are expected to contribute only with single logarithms.

\section{Conclusions}\label{sec:conc}

We have studied the real emission of radiation in the factorisable regime where the virtuality is much smaller than the scale of the hard collision process. We obtained factorisation formulas and splitting amplitudes (Section~\ref{sec:rdsa}) that describe the radiation emission in the entire low-virtuality phase space and for all splittings that occur in the SM. Apart from QED or QCD radiation, and unstable SM particles decays, the formulas encompass a large variety of splitting phenomena that arise at the EW scale, whose investigation is the motivation of the present paper. We also outlined (Section~\ref{sec:aimh}) a strategy for the explicit analytic evaluation of the splitting amplitudes that exploits the simplicity of the amplitude tensors in a suitably designed representation of the bi-spinor variables. 

A natural continuation of this work would be to square the amplitude and to integrate over the phase space of the radiation, targeting analytic factorised expression for the log-enhanced real radiation effects in inclusive observables. We did not attack this problem, however the results of Section~\ref{sec:aimh} should provide all the required ingredients. In this context, the eikonal description (Section~\ref{sec:SL}) of the amplitudes in the soft limit assumes particular significance. The soft region is the most problematic region of integration because it receives contributions from many different splitting configurations, where the soft particle is emitted from any of the external legs of the hard process, and it is sensitive to the interference between different splitting amplitudes. By the eikonal formula it should be possible to subtract the soft enhancement from the splitting amplitudes and to treat the soft terms in analogy with massless gauge theory calculations. 

Our study required some methodological advances, which could be of use also in different contexts. In Appendix~\ref{app:spi} we constructed a basis for the bi-spinors that is particularly suited for the study of the splitting amplitudes. This is a simple generalisation of the Soper--Weinberg construction of single-particle states in the infinite-momentum frame. In Appendix~\ref{sec:wf} we introduced a description of amplitudes as little-group tensors that automatically accounts for the correspondence between ultra-relativistic longitudinal spin-1 particles and Golstone bosons. At variance with the more standard tensorial formalism, our tensors are not symmetric under the exchange of the little-group indices. The advantage that made this formalism essential for our calculation is the absence of gauge cancellation and the smoothness of the high-energy (or, low-mass) limit of individual Feynman diagrams.

A related question that might be subject of future work is whether and how our splitting amplitudes could be re-derived without using Feynman diagrams. One possible approach based on constructive methods, similar to~\cite{Birthwright:2005ak}, could be to try to relate a generic $n+1$ point amplitude with low-virtuality radiation emission to the hard $n$ point amplitude. As far as we can tell, this program has not been carried out not even in massless theories, and furthermore constructive amplitude methods are not yet fully developed in the massive case~\cite{Badger:2005jv,Lai:2023upa,Ema:2024vww}. An alternative direction could be to use dispersion relations in order to relate the residue of the zero-virtuality pole---which is provided by the on-shell physical amplitudes---to the amplitude with real kinematics in the low-virtuality regime.

\section*{Acknowledgements}

We thank Riccardo Rattazzi and Marc Riembau for useful discussions. The work of F. N. is partially supported by the Swiss National Science Foundation under contract 200020-213104 and through the National Center of Competence in Research SwissMAP. The work of L.~R.~is supported by NSF Grant No.~PHY-2210361 and by the Maryland Center for Fundamental Physics. AW acknowledges support from the Departament de Recerca i Universitats de la Generalitat de Catalunya al Grup de Recerca i Universitats from Generalitat de Catalunya to the Grup de Recerca 00649 (Codi: 2021 SGR 00649).

\appendix

\section{Goldstone Equivalence}\label{App:EqFromalism}

Given a certain gauge-invariant Lagrangian such as the one of the SM, a number of ambiguous steps are needed to obtain a practical formalism for calculations using Feynman diagrams. The ambiguities do not affect the results for physical quantities, but they allow for multiple formulations of the same theory that are equivalent in terms of physical results and yet feature different Feynman rules. The choice of the gauge is a typical example of such ambiguities. In this Appendix we summarise the basic aspects of the Goldstone-Equivalent (GE) formulation of massive gauge theories~\cite{Cuomo:2019siu} that is employed throughout the paper. Ref.~\cite{Cuomo:2019siu} reports the complete study of the formalism dealing with all orders in perturbation theory and with the fact that the massive vector bosons are (in the SM) unstable particles rather than Asymptotic States. An on-shell treatment for stable vector bosons is discussed in Ref.~\cite{Wulzer:2013mza}.

The GE formalism employs the regular $R_\xi$ gauge-fixing functional, therefore the Feynman rules for vertices and propagators are just the same ones of the regular diagrammatic formulation.~\footnote{The Feynman choice $\xi=1$ of the gauge-fixing parameter is employed in the paper (see eq.~(\ref{eq:Gprop})). A manifestly gauge-independent derivation of the  results in Section~\ref{sec:VV} is possible, but slightly more involved.} What changes with respect to the regular formalism is the Feynman rule for external massive spin~1 particles, specifically the one for zero-helicity (longitudinal) states. In the regular formalism, a spin-1 particle is described by an external line of the 4-vector gauge field, $V^\mu(x)$. Namely, it is associated with the amputated connected amplitude with external field $V^\mu$---which we denote as ${\mathcal{A}}\{V^\mu\}$---contracted with the standard polarisation 4-vectors. If the particles is incoming and has helicity $h=\pm1,0$, the Feynman amplitude is
\begin{equation}\label{eq:ampstd}
    {\mathcal{M}}^h=
    \varepsilon^h_{\textrm{St.}\mu}(k)
    {\mathcal{A}}\{V^\mu[k]\}\,,
\end{equation}
where $k$ is the particle's momentum flowing in the diagram. The well-known limitation of this formalism is that when the energy $E$ of the particle is much larger than its mass, the longitudinal ($h=0$) polarisation vector scales like
\begin{equation}\label{eq:pvgrowth}
    \varepsilon^{0}_{\text{St.},\mu}(k)\simeq\frac{k_\mu}{m}\sim \frac{E}{m}\,.
\end{equation}
The high-energy scaling with $E$ is problematic because it is stronger than the naive expectation based on dimensional analysis, namely $\varepsilon\sim 1$. This anomalous behaviour invalidates naive power-counting, making the standard formalism not suited for our analysis as explained in the Introduction.

In the GE formalism instead, the diagrammatic description of external spin-1 states involves both the gauge field $V^\mu(x)$ and the associated Goldstone scalar field, $\pi(x)$. In the case of the SM, the Goldstones associated to the neutral $Z$ and to the charged $W$ are specific components of the Higgs doublet. The gauge and the Goldstone are conveniently collected in a single 5-components field
\begin{equation}
    \Phi^M(x)=[V^{\mu}(x),\pi(x)]^M\,.
\end{equation}
A double gauge-plus-scalar line is employed to represent the $\Phi$ field as in Figure~\ref{fig:EquivalentFR}. The scattering amplitude for an incoming spin-1 particle reads 
\begin{equation}\label{eq:ampge}
    {\mathcal{M}}^h=\mathcal{E}_{M}^h(k)
    {\mathcal{A}}\{\Phi^M[k]\}\,,
\end{equation}
where ${\mathcal{A}}\{\Phi^M[k]\}$ is the connected amputated amplitude for the $\Phi$ field, and 
\begin{equation}\label{eq:GEPV}
\mathcal{E}_M^h(k)=[\varepsilon_\mu^h(k),\delta^h_0\varepsilon_\pi]_M\,,
\end{equation}
is a five-components polarisation vector. Notice that the fifth (i.e., the Goldstone) component of $\mathcal{E}$ is different from zero only for $h=0$ helicity states. 

The GE polarisation vectors in eq.~(\ref{eq:GEPV}) are easily related to the standard 4-components polarisations by exploiting the Ward identity that connects---in a massive gauge theory---gauge and Goldstone connected amputated amplitudes. If the $k$ momentum is on-shell, and at the tree-level order, the Ward identity reads~\footnote{More generally, the $\mathcal{K}$ vector features a non-trivial fifth component $\mathcal{K}_\pi(k^2)$. This eventually produces, beyond tree-level, a non-trivial Goldstone wave-function factor $\varepsilon_\pi\neq i$ in eq.~(\ref{eq:GEPV}). Also notice that a more general Ward identity $\mathcal{K}_{1,M_1} \mathcal{K}_{2,M_2}\ldots\mathcal{A}\{V^{M_1}[k_1]V^{M_2}[k_2]\ldots\}=0$ holds for amplitudes with several amputated legs. 
}
\begin{equation}\label{eq:WI}
    \mathcal{K}_M[k] \mathcal{A}\{V^M[k]\}=0\,,\;{\textrm{with}}\;\;
    \mathcal{K}_{M}[k]=[i\,k_\mu,m]_M\,.
\end{equation}
We will thus obtain the exact same scattering amplitudes as in the standard formalism~(\ref{eq:ampstd}) if we take $\mathcal{E}$ equal to standard polarisations---with vanishing fifth component---plus a shift term proportional to $\mathcal{K}$. The right choice of the proportionality factor is the one that cancels the energy growth~(\ref{eq:pvgrowth}) of the longitudinal polarisation, and this choice is
\begin{equation}\label{eq:deflongituialwpv1}
\mathcal{E}_M^h(k)=
[\varepsilon^{h}_{\text{St.},\mu}(k),\,0]_M+\frac{i}{m}\delta^h_0\mathcal{K}_M[k]
=\left[
\varepsilon^{h}_{\text{St.},\mu}-\delta^h_0\frac{k_\mu}m,\,i\,\delta^h_0
\right]_M
\,.
\end{equation}
By comparing with eq.~(\ref{eq:GEPV}), we find
\begin{equation}\label{eq:deflongituialwpv}
\varepsilon^{0}_{\mu}=\varepsilon^{0}_{\text{St.},\mu}-\frac{k_\mu}m\,,\quad
\varepsilon^{\pm}_{\mu}=\varepsilon^{\pm}_{\text{St.},\mu}\,,\quad
\varepsilon_\pi=i\,.
\end{equation}
The GE longitudinal polarisation 4-vector, $\varepsilon^{0}_{\mu}$, does not grow with the energy unlike the standard polarisation $\varepsilon^{0}_{\text{St.},\mu}$. It is easy to show that it instead decreases with energy, namely
\begin{equation}
    \varepsilon^{0}_{\mu}(k)\sim \frac{m}{E}\,.
\end{equation}
The transverse polarisations $\varepsilon^{\pm}$ are equal to the standard vectors and have a constant high-energy scaling. The Goldstone wave-function factor is a constant $\varepsilon_\pi=i$. All the components of the GE polarisation 5-vectors $\mathcal{E}_M^h(k)$ are thus well-behaved with energy as anticipated.

\begin{figure}
\centering
 \begin{tikzpicture}[scale=1]
	\begin{feynman}
	\vertex (a) at (-2.5,0){$\mathcal{E}_{M}^h(k)$};
    \vertex[blob,minimum size=25pt] (b) at (0,0){};
    \vertex (c) at (2,0){$\varepsilon_{\mu}^h(k)$};
    \vertex[blob,minimum size=25pt] (d) at (4.5,0){};
    \vertex (e) at (6.2,0){$\delta^h_0\,\varepsilon_{\pi}$};
    \vertex[blob,minimum size=25pt] (f) at (8.5,0){};
    \diagram* {
		(a) -- [scalar,momentum=$k$,edge label'=$M$] (b),
      (a) -- [photon] (b),
      (c) -- [photon,momentum=$k$,edge label'=$\mu$] (d),
      (e) -- [scalar,momentum=$k$] (f),
	};
	\end{feynman}
    \node at (1,0) {$=$};
    \node at (5.5,0) {$+$};
	\end{tikzpicture}	
    \caption{Feynman rule for initial state vector particles. For helicity $h=0$, the scattering amplitude of the vector with polarization $\varepsilon^0_\mu(k)$ is summed to the one of the corresponding Goldstone boson with wave function $\varepsilon_\pi$. At tree-level, $\varepsilon_\pi = +i$. \label{fig:EquivalentFR}
    }
\end{figure}

We can similarly derive the GE Feynman rule for outgoing particles
\begin{eqnarray}\label{eq:pvGEAP}
    &\bar{\mathcal{E}}_M^h(k)=
    [{\bar{\varepsilon}}^{h}_{\text{St.},\mu}(k),\,0]_M-\frac{i}{m}\delta^h_0\mathcal{K}_M[-k]=(\mathcal{E}_M^h(k))^*=[\bar\varepsilon_\mu^h(k),-i\,\delta^h_0]_M
    \,,&\nonumber\\
    &\bar\varepsilon_\mu^h=\bar\varepsilon_{\text{St.},\mu}^h-\delta^h_0\frac{k_\mu}m=(\varepsilon_{\text{St.},\mu}^h)^*-\delta^h_0\frac{k_\mu}m=(\bar\varepsilon_\mu^h)^*\,.&
\end{eqnarray}
The shift by $\mathcal{K}_M[-k]$ takes into account that the momentum is now outgoing.

If the standard polarisation vectors are normalised canonically and obey the canonical completeness relation
\begin{equation}\label{eq:stdcompleteness}
\bar{\varepsilon}_{\text{St.}}^{h,\mu}(k){\varepsilon}_{\text{St.},\mu}^g(k)=-\delta^{h\,g}\,,\qquad
\sum_{h=0,\pm1}\varepsilon_{\text{St.},\mu}^h(k)\bar{\varepsilon}_{\text{St.},\nu}^h(k)=-\eta_{\mu\nu}+\frac{k_\mu k_\nu}{m^2}\,, 
\end{equation}
the ${\mathcal{E}}$ and $\bar{\mathcal{E}}$ 5-vectors satisfy the following completeness relation
\begin{equation}\label{eq:cr5d}
 \sum_{h=0,\pm 1} \mathcal{E}_{M}^h(p)\bar{\mathcal{E}}_{N}^h(p)= \begin{bmatrix}
         -\eta_{\mu\nu}-{(k_\mu\varepsilon^0_\nu(k)+\varepsilon^0_\mu(k)k_\nu)}/m & -i\,\varepsilon^0_\mu(k)\\
       i\,\varepsilon^0_\mu(k) & 1 \\
    \end{bmatrix}_{MN}\,.
\end{equation}

\section{Single-particle states and bi-spinors}\label{app:spi}

After setting the notation, this appendix reviews the definition of single-particle states by the method of induced representations~\cite{10.2307/1968551, Weinberg:1995mt}, and the direct connection between such definition---encoded in the choice of the standard Lorentz transformation $\Lambda^{\hspace{-1pt}\textsc{st}}_k$---and the explicit representation of the particle's bi-spinors. The Jacob--Wick (JW) and Soper--Weinberg (SW) bi-spinors are derived. Only massive particles are considered in the Appendix and throughout the paper. The massless limit can be taken smoothly in the final results.

\subsection{Two-component spinor notation}\label{sec:2cs}

The Lorentz group SL$(2,\mathbb{C})$ admits two bi-dimensional representations: the spinor representation and its conjugate. The three rotation ($J_a$) and boost ($K_a$) generators, in the spinor representation, are the matrices
\begin{equation}\label{eq:genex}
    \mathscr{J}_a=\frac12\sigma_a\,,\qquad\mathscr{K}_a=\frac{i}2\sigma_a\,,
\end{equation}
where $\sigma$ are the Pauli matrices and $a=(x,y,z)$ runs over the three coordinates axes.  A lower Greek index $\alpha=1,2$ labels the objects that transform in the spinor representation. A generic Lorentz transformation, consisting of a counterclockwise rotation around a vector $\theta^a$ (of angle $|\theta|$) and of a boost with rapidity vector $y^a$, is represented by a $2\times2$ matrix  
\begin{equation}    \Lambda_{\alpha}^{\,\,\beta}=\exp\big[-i\theta^a \mathscr{J}_a+i\eta^a \mathscr{K}_a\big]_{\alpha}^{\,\,\beta}=\exp\big[-(i\theta^a+\eta^a)\frac{\sigma_a}{2}\big]_{\alpha}^{\,\,\beta}\,.
\end{equation}
The conjugate-spinor generators are $\dot{\mathscr{J}}_a=-(\mathscr{J}_a)^*$ and $\dot{\mathscr{K}}_a=-(\mathscr{K}_a)^*$. Objects transforming in this representation are labeled with a lower dotted index $\dot\alpha=1,2$. The invariant tensors with lower indices are the antisymmetric tensors
\begin{equation}
\epsilon_{\alpha\beta},\quad \epsilon_{\dot\alpha\dot\beta},\quad {\textrm{with}}\;\;\epsilon_{12}=-1\,.
\end{equation}
Invariant contractions and upper-index objects are formed using the tensors
\begin{equation}\nonumber
\epsilon^{\alpha\beta},\quad \epsilon^{\dot\alpha\dot\beta},\quad {\textrm{with}}\;\;\epsilon^{12}=+1\,.
\end{equation}
It is also useful to define
\begin{equation}
\sigma^\mu=(\mathds{1},\,\vec\sigma)\,,
\quad \bar\sigma^\mu=(\mathds{1},\,-\vec\sigma)\,.
\end{equation}

The irreducible linear representations of the Lorentz group transform as tensors
\begin{equation}\label{eq:tenoptr}
{\mathcal{T}}_{\alpha_1\ldots\alpha_{n};\,\dot{\alpha_1}\ldots\dot{\alpha}_{\bar{n}}}\to
\Lambda_{\alpha_1}^{\;\,\beta_1}\ldots
\Lambda_{\alpha_n}^{\;\,\beta_n}
{\mathcal{T}}_{\beta_1\ldots\beta_{n};\,\dot{\beta_1}\ldots\dot{\beta}_{\bar{n}}}
(\Lambda^\dagger)^{\dot\beta_1}_{\;\,\dot\alpha_1}\ldots
(\Lambda^\dagger)^{\dot\beta_{\bar{n}}}_{\;\,\dot\alpha_{\bar{n}}}
\,,
\end{equation}
with $n$ un-dotted and $\bar{n}$ dotted indices. The tensor is is fully symmetric under the permutation of dotted or of un-dotted indices and, if $n=\bar{n}$, it also obeys the Hermiticity condition
\begin{equation}
{\mathcal{T}}^*_{\dot\alpha_1\ldots\dot\alpha_{n};\,{\alpha_1}\ldots{\alpha}_{n}}
={\mathcal{T}}_{\alpha_1\ldots\alpha_{n};\,\dot{\alpha_1}\ldots\dot{\alpha}_{\bar{n}}}\,.
\end{equation} 
The 4-vector is the $n=\bar{n}=1$ tensor representation. Therefore, a 4-momentum $k^\mu$ can be represented as a $2\times2$ Hermitian matrix
\begin{equation}\label{eq:mmnot}
    k_{\alpha \dot{\alpha}}=k_\mu\sigma^\mu_{\alpha \dot{\alpha}} = \begin{bmatrix}
       k^0-k^3 & -k^1+i\,k^2 \\
        -k^1-i\,k^2 & k^0+k^3 
\end{bmatrix}_{\alpha\dot\alpha}\,.
\end{equation}
Under a Lorentz transformation $\Lambda$, the momentum matrix transforms as
\begin{equation}\label{eq:momtrans}
k\rightarrow \Lambda\cdot{k}\cdot\Lambda^\dagger\,.
\end{equation}

\subsection{Single particle states and the wave-function of the tensor fields}\label{ssec:sps}

A basis for the states that describe a single particle with non-vanishing mass $m$ and spin $s$ can be constructed as follows. In the particle's rest frame, where the momentum in the matrix notation is equal to $m$ times the identity, the states $\ket{m\mathds{1},h}$ form a spin-$s$ representation of the SU$(2)$ group of rotations. They are labelled by the integer or semi-integer eigenvalue $h=-s,\ldots,+s$ of the angular momentum operator $J_z$ along the $z$-axis.~\footnote{We refer to the $h$ quantum number as the \emph{helicity}, even if it will corresponds to the regular helicity---i.e., the projection of the angular momentum along the direction of motion---only with the Jacob--Wick choice $\Lambda^{\hspace{-1pt}\textsc{st}}_p=\Lambda^{\hspace{-1pt}\textsc{jw}}_p$ in eq.~(\ref{eq:jwst}). With a different standard transformation, $h$ is the eigenvalue of a different operator.} The other rotation operators $J_{x,y}$ act on the $h$ index by the standard SU$(2)$ matrices in the spin-$s$ representation. The state with generic momentum $k$ is obtained by acting on the states at rest with a \emph{standard Lorentz transformation} $\Lambda^{\hspace{-1pt}\textsc{st}}_k$. Namely 
\begin{equation}\label{eq:DefState}
\ket{k,h}=U(\Lambda^{\hspace{-1pt}\textsc{st}}_k)
\ket{m\mathds{1},h}\,,
\end{equation}
where $U$ is the (unitary) representation of the Lorentz group on the space of states.

The standard Lorentz transformation transforms the rest-frame momentum $m\mathds{1}$ into the particle's momentum $k$, i.e., using eq.~(\ref{eq:momtrans})
\begin{equation}\label{eq:stdc}
k=m\,\Lambda^{\hspace{-1pt}\textsc{st}}_k(\Lambda^{\hspace{-1pt}\textsc{st}}_k)^\dagger\,.
\end{equation}
This condition does not determine $\Lambda^{\hspace{-1pt}\textsc{st}}_k$ uniquely, but only up to the right-multiplication with an SU$(2)$ matrix, resulting in different definition of the states in eq.~\eqref{eq:DefState}. These choices are related by an SU$(2)$ rotation $\mathcal{U}$ which is called a \emph{little-group} transformation:
\begin{equation}\label{eq:lg}
\Lambda^{\hspace{-1pt}\textsc{st}^\prime}_k=\Lambda^{\hspace{-1pt}\textsc{st}}_k\cdot{\mathcal{U}}\,.
\end{equation}
If $\Lambda^{\hspace{-1pt}\textsc{st}}_k$ is changed into $\Lambda^{\hspace{-1pt}\textsc{st}^\prime}_k$, the state in eq.~(\ref{eq:DefState}) transforms by the little-group rotation acting on the $h$ index in the spin-$s$ representation.

The notion of \emph{bi-spinors} naturally emerges if we consider the wave-function factor associated with an interpolating field ${\mathcal{T}}(x)$ in an irreducible tensorial representation of the Lorentz group. The wave function is defined as the matrix element 
\begin{equation}
\Psi_{\alpha_1\ldots\alpha_{n};\,\dot{\alpha_1}\ldots\dot{\alpha}_{\bar{n}}}^{k,h}=\bra{0}{\mathcal{T}}_{\alpha_1\ldots\alpha_{n};\,\dot{\alpha_1}\ldots\dot{\alpha}_{\bar{n}}}(0)\ket{k,h}\,.
\end{equation}
If the particle is at rest, the form of $\Psi$ is strongly constrained by the invariance under the rotational symmetry. The un-dotted spinor indices transform in the doublet representation of the SU$(2)$ rotation group, and dotted indices transform in the (equivalent) conjugate-doublet representation, for a total of $n+\bar{n}$ doublet (or conjugate-doublet) indices. The particle state at rest is in the spin-$s$ representation, therefore a non-vanishing invariant $\Psi$ can be obtained only if $n+\bar{n}\geq 2s$. If in particular $n+\bar{n}$ is equal to $2s$, rotational symmetry determines $\Psi$ up to a normalisation constant, because there is a single spin-$s$ representation in the decomposition of the tensor product of $n+\bar{n}=2s$ doublets. The result is conveniently expressed as 
\begin{equation}\label{wfrest}
\Psi_{\alpha_1\ldots\alpha_{n};\,\dot{\alpha_1}\ldots\dot{\alpha}_{\bar{n}}}^{m\mathds{1},h}\propto 
\delta^{I_1}_{\alpha_1}\ldots \delta^{I_n}_{\alpha_n}\,\epsilon^{I_{n+1}J_1}(\delta^{J_1}_{\dot{\alpha}_1})^*\ldots 
\epsilon^{I_{2s}J_{{\bar{n}}}}(\delta^{J_{{\bar{n}}}}_{\dot{\alpha}_{\bar{n}}})^*\,
{\mathcal{S}}^{s,h}_{I_1\ldots I_{2s}}\,,
\end{equation}
where ${\mathcal{S}}^{s,h}$ are the fully-symmetric tensors that projects the tensor product of $2s$ doublets onto the spin-$s$ representation. The need for the anti-symmetric $\epsilon$ tensors in the previous equation emerges from the fact that the dotted indices are in the conjugate-doublet rather than in the doublet representation.

The states with generic momentum $k$ are obtained by the standard transformation $\Lambda^{\hspace{-1pt}\textsc{st}}_k$ acting on the state at rest. Therefore, in order to obtain the wave-function $\Psi^{k,h}$ we just need to Lorentz-transform eq.~(\ref{wfrest}), given the transformation property of the tensor field in eq.~(\ref{eq:tenoptr}). We thus define the angular \emph{bi-spinor} $\ket{k^I}_\alpha$ as the action of $\Lambda^{\hspace{-1pt}\textsc{st}}_k$ on the tensorial structure---i.e., the $\delta$ tensors---that carries the un-dotted indices in eq.~(\ref{wfrest})
\begin{equation}\label{eq:bsdef}
    \ket{k^I}_\alpha=\sqrt{m}\,{\Lambda^{\hspace{-1pt}\textsc{st}}_k}_\alpha^{\;\beta}\delta^I_\beta=\sqrt{m}\,[\Lambda^{\hspace{-1pt}\textsc{st}}_k]_{\alpha}^{\;\;I}\,,
\end{equation}
with a convenient normalisation factor of $\sqrt{m}$. We also define the square bracket bi-spinors as
\begin{equation}\label{eq:bsdef1}
    [\,k_I|_{\dot{\alpha}}=(\ket{k^I}^*)_{\dot\alpha}
    =\sqrt{m}\,[(\Lambda^{\hspace{-1pt}\textsc{st}}_k)^\dagger]_{I\,\dot\alpha}
    \,.
\end{equation}
Using bi-spinors, the wave function reads 
\begin{equation}\label{eq:WFSpinors}
\Psi_{\alpha_1\ldots\alpha_{n};\,\dot{\alpha_1}\ldots\dot{\alpha}_{\bar{n}}}^{k,h}\propto
\ket{k^{I_1}}_{\alpha_1}\ldots
\ket{k^{I_n}}_{\alpha_n}
[\,k^{I_{n+1}}|_{\dot{\alpha}_1}\ldots 
[\,k^{I_{2s}}|_{\dot\alpha_{\bar{n}}}\,
{\mathcal{S}}^{s,h}_{I_1\ldots I_{2s}}\,,
\end{equation}
where the upper-index square brackets bi-spinor is defined in eq.~(\ref{eq:bispvar}).

The Latin index of the bi-spinor is called a little-group index because it transforms as an SU$(2)$ doublet if the little-group transformation~(\ref{eq:lg}) is operated on $\Lambda^{\hspace{-1pt}\textsc{st}}$. The index is taken to run over $I=(+,-)$, and $I=+$ in eq.~(\ref{eq:bsdef}) means the first column of the matrix $\Lambda^{\hspace{-1pt}\textsc{st}}_k$. little-group indices are lowered and raised with the SU$(2)$ invariant tensors 
\begin{equation}
\epsilon^{IJ},\quad \epsilon_{IJ},\quad {\textrm{with}}\;\;
\epsilon^{+-}=-\epsilon_{+-}=1\,.
\end{equation}
Using the $\epsilon$ tensors we define
\begin{equation}\label{eq:bispvar}
    [\,k^I|_{\dot{\alpha}}\equiv\epsilon^{IJ} [\,k_J|_{\dot{\alpha}}\,,
    \qquad
    \ket{k_I}_\alpha\equiv\epsilon_{IJ}\ket{k^J}_\alpha\,.
\end{equation}

A number of useful bi-spinors properties follow from their definition~(\ref{eq:bsdef}) in terms of the standard Lorentz transformation. Since the determinant of the SL$(2,\mathbb{C})$ matrix $\Lambda^{\hspace{-1pt}\textsc{st}}_k$ is 1
\begin{equation}\label{spinorm}
    \epsilon^{\alpha\beta} \ket{k^I}_\alpha \ket{ k^J}_\beta\equiv\langle k^J \, k^I\rangle=m\,\epsilon^{IJ}
    =[k^I\,k^J]\equiv
    \epsilon^{\dot\alpha\dot\beta} [k^I|_{\dot\alpha}     [k^J|_{\dot\beta}
    \,.
\end{equation}
From eq.~(\ref{eq:stdc}) it follows that 
\begin{equation}
    \ket{k^I}_\alpha\,[\,k_I|_{\dot{\alpha}}=k_{\alpha \dot{\alpha}}\,,
\end{equation}
which also implies
\begin{equation}
    \label{eq:momAS}
\ket{k^I}_\alpha\,[\,k^J|_{\dot{\alpha}}
-\ket{k^J}_\alpha\,[\,k^I|_{\dot{\alpha}}
=\epsilon^{IJ}k_{\alpha \dot{\alpha}}\,.
\end{equation}

\subsection{The JW bi-spinors}

The most common definition of the single-particle states basis was given by Jacob and Wick (JW) in~\cite{Jacob:1959at}, using the standard Lorentz transformation
\begin{equation}\label{eq:jwst}
\Lambda^{\hspace{-1pt}\textsc{st}}_k
=\Lambda^{\hspace{-1pt}\textsc{jw}}_k
=e^{-i\varphi J_z} e^{-i\theta J_y} e^{i\varphi J_z}e^{i\eta K_z}\,,
\end{equation}
where $\theta$ and $\varphi$ are the polar and azimuthal angles of the particle and $\eta$ is the absolute rapidity $\tanh\eta=|{\bf{k}}|/\sqrt{|{\bf{k}}|^2+m^2}$. Explicitly, the particle's 4-momentum is
\begin{equation}
k^\mu=m(\cosh\eta,\,\sinh\eta\cos\phi\sin\theta,\,\sinh\eta\sin\phi\sin\theta,\,\sinh\eta\cos\theta)^\mu\,.
\end{equation}

The JW standard transformation first operates a boost along the $z$-axis, which puts the rest-frame state in motion along a direction parallel to the spin. With this first transformation we obtain an eigenstate of the helicity operator, defined as the projection of the angular momentum along the direction of the 3-momentum, with eigenvalue $h$. The subsequent transformations are rotations, which leave the helicity invariant. The JW states are thus eigenstates of the helicity operator and therefore---again because the helicity is rotational invariant---the rotation SU$(2)$ subgroup of the Lorentz group acts as a diagonal phase on these states. On the contrary, the states feature complicated non-diagonal boost transformation properties because the helicity is not boost-invariant.

The JW standard transformation in the spinor representation can be evaluated using the generators in eq.~(\ref{eq:genex}), obtaining by eq.~(\ref{eq:bsdef}) the explicit form of the bi-spinors for the single-particle states in the JW basis
\begin{equation}
    \ket{k^I_{\textsc{jw}}}_\alpha=
    \sqrt{m}\,[\Lambda^{\hspace{-1pt}\textsc{jw}}_k]_{\alpha}^{\;\;I}=
     \begin{bmatrix}
        \sqrt{m}\,e^{-\eta/2} \cos\frac{\theta}{2} & 
        - \sqrt{m}\,e^{+\eta/2} \sin\frac{\theta}{2} e^{-i\varphi}
        \\
        \sqrt{m}\,e^{-\eta/2} \sin\frac{\theta}{2} e^{i\varphi}
         &   \sqrt{m}\,e^{+\eta/2} \cos\frac{\theta}{2}\\
    \end{bmatrix}_{\alpha}^{\;\;I}\,.
\end{equation}
This expression can be written as in eq.~(\ref{eq:jwsp}) by trading the mass and the rapidity for $|{\bf{k}}|$ and the energy $k^0=\sqrt{|{\bf{k}}|^2+m^2}$. The explicit form of the $\Lambda^{\hspace{-1pt}\textsc{jw}}_k$ matrix can be readily checked to respect the consistency condition in eq.~(\ref{eq:stdc}).

\subsection{The SW bi-spinors}\label{sec:IMH}

The Soper--Weinber (SW) definition~\cite{LCHSoper} of the single-particle states is designed in such a way that the states feature extremely simple transformation properties under a rotation performed around a ``special'' direction in space, $e_3$, and under a boost performed along the same direction. We call $J_3$ and $K_3$ the corresponding generators
\begin{equation}
    J_3=e_3^aJ_a\,,\quad K_3=e_3^aK_a\,,
\end{equation}
and we denote as $\Lambda_3$ a finite Lorentz transformation along $J_3$ or $K_3$
\begin{equation}
    \Lambda_3=e^{-i\,\phi J_3+i\,{y} K_3}\,.
\end{equation}
The special direction is taken to coincide with the $z$-axis in Ref~\cite{LCHSoper}, namely $e_3=e_z=(0,0,1)$. 
The generalisation to arbitrary direction is simply obtained via the rotation $R$ rotating the $z$-axis $e_z$ to the $e_3$ direction:
\begin{equation}\label{eq:jkrot}
    J_3=RJ_zR^{-1}\,,\quad K_3=RK_zR^{-1}\quad\Rightarrow\quad 
    \Lambda_3=R\,e^{-i\,\phi J_z+i\,{y} K_z}R^{-1}\,. 
\end{equation}
In the main text, we identify the special direction $e_3$ as the direction of the $A$ particle. The spinor representation of the rotation as a $2\times2$ matrix $\mathcal{R}$ is reported in eq.~(\ref{eq:rot}).

The coordinates that we define in Section~\ref{sec:kin} to describe the 4-momentum---see in particular eq.~(\ref{eq:momlab})---transform very simply under $\Lambda_3$. By recalling that $J_z$ and $K_z$ are diagonal matrices in the spinor representation, one easily finds
\begin{equation}\label{eq:k3trans}
k={\mathcal{R}}\cdot
   \begin{bmatrix}
       k_\md & -{\mathbf{k^*_\td}} \\
        -{\mathbf{k_\td}} & k_\pd 
    \end{bmatrix}\cdot{\mathcal{R}}^\dagger\rightarrow k^\prime=\Lambda_3\cdot k\cdot \Lambda_3^\dagger=
    {\mathcal{R}}\cdot
    \begin{bmatrix}
       e^{-{y}}k_\md & -e^{-i\,\phi}{\mathbf{k^*_\td}} \\
        -e^{i\,\phi}{\mathbf{k_\td}} & e^{{y}}k_\pd 
    \end{bmatrix}
    \cdot{\mathcal{R}}^\dagger\,.
\end{equation}
The $k_\pd$ and $k_\md=(m^2+|{\mathbf{k^*_\td}}|^2)/k_\pd$ coordinates scale under $K_3$ with opposite weights. The complex ${\mathbf{k_\td}}$ coordinate transforms as unit-charge object under $J_3$.

We define the following generators
\begin{equation}
T_1=R(K_x-J_y)R^{-1}\,,\qquad
T_2=R(K_y+J_x)R^{-1}\,,
\end{equation}
which are simply the ``transverse boost'' generators defined in~\cite{LCHSoper}, rotated by $R$. We can also consider the complex combinations of the tranverse boosts
\begin{equation}
    T=T_1-i\,T_2\,,\qquad \overline{T}=T_1+i\,T_2\,.
\end{equation}
The transverse boosts transform simply under $J_3$ and $K_3$. The Lorentz Algebra gives
\begin{align}
    [J_3,T]=-T\,,\quad [K_3,T]=i\,T\,, 
    \;\Rightarrow\;\Lambda_3 T\Lambda_3^{-1}=e^{-{y}+i\phi}\,T\,,
    \nonumber\\
    [J_3,\overline{T}]=+\overline{T}\,,\quad [K_3,\overline{T}]=i\,\overline{T}\,, 
    \;\Rightarrow\;\Lambda_3 \overline{T}\Lambda_3^{-1}=e^{-{y}-i\phi}\,\overline{T}\,.
    \label{eq:trpgen}
\end{align}
The standard Lorentz transformation that defines the SW states is the following
\begin{equation}\label{eq:swtr1}
\displaystyle
\Lambda^{\hspace{-1pt}\textsc{st}}_k
=\Lambda^{\hspace{-1pt}\textsc{sw}}_k
=e^{\frac{i}{2}\,({\bf{v_\td}}T+{\bf{v_\td^*}}\overline{T})}e^{i\,\zeta_\pd\,K_3}\,R\,,
\end{equation}
where ${\bf{v_\td}}={\bf{k_\td}}/k_\pd$ and $\zeta_\pd=\log(k_\pd/m)$. Consider now operating a $\Lambda_3$ transformation $k\to k^\prime$ on the momentum as in eq.~(\ref{eq:k3trans}). The $\zeta$ parameter shifts to $\zeta+y$. The parameter ${\bf{v_\td}}$ becomes $e^{-y+i\phi}{\bf{v_\td}}$, which matches the transformation of the $T$ generator in eq.~(\ref{eq:trpgen}). The $\Lambda^{\hspace{-1pt}\textsc{sw}}_k$ transformation evaluated on the transformed momentum $k^\prime$ can thus be written as
\begin{equation}
\Lambda^{\hspace{-1pt}\textsc{sw}}_{k^\prime}
=\Lambda_3\Lambda^{\hspace{-1pt}\textsc{sw}}_kR^{-1}\Lambda_3^{-1}e^{i\,y\,K_3}R=
\Lambda_3\Lambda^{\hspace{-1pt}\textsc{sw}}_ke^{i\,\phi\,J_z}\,.
\end{equation}
If we then act with $\Lambda_3$ on the SW states, which are defined by eq.~(\ref{eq:DefState}) using the standard SW transformation, we obtain
\begin{equation}
U(\Lambda_3)\ket{k,h}=
U(\Lambda_3\Lambda^{\hspace{-1pt}\textsc{sw}}_k)
\ket{m\mathds{1},h}=
U(\Lambda^{\hspace{-1pt}\textsc{sw}}_{k^\prime}) \, e^{-i\,\phi\,J_z}
\ket{m\mathds{1},h}=
e^{-ih\phi}\ket{k^\prime,h}\,,
\end{equation}
because $J_z$ is diagonal and equal to $h$ on the rest-frame states. Up to the transformation of the momentum, the SW states are simply invariant under boosts in the special direction, and pick up a phase equal to minus the helicity under $J$ rotations. 

For the explicit evaluation of $\Lambda^{\hspace{-1pt}\textsc{sw}}_k$ in the spinor representation it is convenient to use eq.~(\ref{eq:jkrot}) and express $\Lambda^{\hspace{-1pt}\textsc{sw}}_k$ as $R$ times the exponential of the un-rotated transverse boosts and of $K_z$, which are very simple matrices in the spinor representation~(\ref{eq:genex}). We obtain
\begin{equation}
\Lambda^{\hspace{-1pt}\textsc{sw}}_k=\frac1{\sqrt{m}}
    \mathcal{R}\cdot
    \begin{bmatrix}
        {{m}}/{\sqrt{k_{\pd}}}& - {\bf{k^*_{\textsc{t}}}}/{\sqrt{k_{\pd}}}\\
        0 & {\sqrt{k_{\pd}}}\\
    \end{bmatrix}\,,
\end{equation}
and, in turn, we derive the bi-spinors in the SW basis reported in eq.~(\ref{eq:imhsp}). By the explicit form of $\Lambda^{\hspace{-1pt}\textsc{sw}}_k$ we can also verify its consistency as a valid standard transformation that obeys the condition in eq.~(\ref{eq:stdc}). Indeed
\begin{equation}
    m\,\Lambda^{\hspace{-1pt}\textsc{sw}}_k (\Lambda^{\hspace{-1pt}\textsc{sw}}_k)^\dagger= 
    {\mathcal{R}}\cdot
   \begin{bmatrix}
       k_\md & -{\mathbf{k^*_\td}} \\
        -{\mathbf{k_\td}} & k_\pd 
    \end{bmatrix}\cdot{\mathcal{R}}^\dagger\,,   
\end{equation}
which is equal to the momentum $k$ in eq.~(\ref{eq:momlab}).

\subsection{Relation between the JW and the SW bi-spinors}\label{sec:rel}

Different bases for the single-particle states are related by a little-group transformation, as previously discussed. On the bi-spinor index, the transformation acts as a $2\times2$ matrix ${\mathcal{U}}$. The relation between the JW and SW bi-spinors thus takes the form
\begin{equation}\label{eq:LGrot}
    \ket{k_{\textsc{jw}}^I}={\mathcal{U}}^I_{\;\,J}
    \ket{k_{\textsc{sw}}^J}\,.
\end{equation}
Since the bi-spinors are normalised as in eq.~(\ref{spinorm}), we have that 
\begin{equation}\label{eq:WR}
{\mathcal{U}}^I_{\;\,J}=\frac{1}{m}\langle k_{\textsc{sw}\,J}\,k_{\textsc{jw}}^I\rangle\,,
\end{equation}
which can be computed directly from the explicit spinors in the two bases.

For generic momenta, the explicit form of ${\mathcal{U}}$ is not particularly insightful. However, ${\mathcal{U}}$ simplifies in two particular limits. In the ultra-relativistic limit $m\to0$, only one of the two bi-spinors (for $I=-$) is non-zero, both in the JW~(\ref{eq:jwsp}) and in the SW~(\ref{eq:imhsp}) bases. The off-diagonal entries of ${\mathcal{U}}$ therefore must vanish in the limit, while the diagonal entries must be opposite phases as ${\mathcal{U}}\in{\textrm{SU}}(2)$. Indeed, we find
\begin{equation}
    {\mathcal{U}}^I_{\;\,J}=e^{i\,I\,\Delta}\,\delta^{I}_{J}\,,
    \qquad \Delta=\tan^{-1}\bigg(\frac{\sin(\phi-\beta)}{\cot\frac{\alpha}{2}\cot\frac{\theta}{2}+\cos(\phi-\beta)}\bigg)\,,
\end{equation}
where $\alpha$ and $\beta$ are the polar and azimuthal angles of the $e_3$ vector, or equivalently the Euler angles of the $R$ rotation 
\begin{equation}
    {\mathcal{R}}= \begin{bmatrix}
        \cos\frac{\alpha}2 & -e^{-i\beta}\sin\frac{\alpha}2\\        e^{i\beta}\sin\frac{\alpha}2 &\cos\frac{\alpha}2
    \end{bmatrix}\,.
\end{equation}

Another interesting configuration is when the particle's $3$-momentum $k$ is parallel to $e_3$. In this case, the tranverse momentum ${\bf{k_\td}}$ vanishes and the light-cone rapidity $\zeta_\pd$ coincides with the total rapidity $\zeta$. The $\alpha$ and $\beta$ angles that define the rotation are equal to the $\theta$ and $\varphi$ angles, so that $R$ becomes equal to the rotation that appears in the JW standard transformation. By comparing eq.~(\ref{eq:swtr1}) with eq.~(\ref{eq:jwst}), we see that the two standard transformations becomes equal. Consistently, the bi-spinors in the two bases become equal and ${\mathcal{U}}=\mathds{1}$. 

\section{Dirac wave functions and polarisation vectors}\label{sec:wf}

In this Appendix we summarise the relation between the bi-spinors and the wave-function factors associated with the on-shell external legs of Feynman diagrams for spin-$1/2$ particles and anti-particles, and for spin-1 particles. The interpolating fields are the composition of tensorial irreducible representations of the Lorentz group and therefore the wave-functions can be expressed in terms of bi-spinors using eq.~(\ref{eq:WFSpinors}). The 4-components Dirac field that interpolates for spin-$1/2$ particles has two (Weyl) components $(n=1,\bar{n}=0)$ and $(n=0,\bar{n}=1)$. The corresponding Dirac wave-functions are reported in Section~\ref{sec:WFDP}. Spin-1 particles are interpolated by the (possibly complex, for charged $W$ bosons) vector representation $(n=1,\bar{n}=1)$, but the scalar $(n=0,\bar{n}=0)$ is also involved in the GE formalism of Appendix~\ref{App:EqFromalism}. We will see in Section~\ref{sec:WFVP} how to express the GE polarisation vectors in terms of tensors with little-group indices using bi-spinors. In Section~\ref{sec:AMPT} we summarise our results and describe the simple rules to turn Feynman diagrams into a little-group tensor representation of the corresponding amplitude. 

\subsection{Wave functions of Dirac particles}\label{sec:WFDP}

The 4-components wave functions that describe respectively the annihilation and the creation of a Dirac particle are conveniently expressed in terms of the $u$ and $\bar{u}$ Dirac spinors, defined as
\begin{equation}\label{eq:DirtoSpin}
    u^I(k)=\begin{pmatrix}
        \ket{k^I}_\alpha\\
        |k^I]^{\dot{\alpha}}
    \end{pmatrix}\,,\qquad\bar{u}_I(k)=(-\bra{k_I}^\alpha\,\,[k_I|_{\dot{\alpha}})\,.
\end{equation} 
They satisfy normalisation and completeness relations
\begin{equation}
\bar{u}_I(k)u^J(k)=2m\delta^{J}_{I}\,,\qquad u^I(k)\bar{u}_I(k)=\slashed{k}+m\,,
\end{equation}
where $\slashed{k}=k_\mu\gamma^\mu$ and the $\gamma$ matrices are
\begin{equation}
\gamma^\mu=\begin{pmatrix}
0 & \sigma^\mu\\
{\bar\sigma}^\mu &0
\end{pmatrix}\,.
\end{equation}
The wave functions for an incoming (or outgoing) particle with helicity $h=\pm1/2$ particle are obtained by contracting the little-group indices of $u$ (or of $\bar{u}$) with little-group wave-function tensors $\tau(h)$ and $\bar\tau(h)$
\begin{equation}\label{eq:Dirwf}
\begin{split}
    &\tau_{I}(+1/2)=\delta^+_I\,,\quad
    {\bar\tau}^{I}(+1/2)=\delta_+^I\,,\\
    &\tau_{I}(-1/2)=\delta^-_I\,,\quad
    {\bar\tau}^{I}(-1/2)=\delta_-^I\,.\\
\end{split}
\end{equation}
Namely, we have
\begin{equation}
    u_h(k)=\tau_I(h)u^I(k)\,,\qquad
    {\bar{u}}_h(k)={\bar\tau}^I(h){\bar{u}}_I(k)\,,
\end{equation}
for incoming and outgoing fermion-respectively.

For Dirac anti-particles we define instead the $v$ spinors
\begin{equation}
    v_I(k)=\begin{pmatrix}
        -\ket{k_I}_{\alpha}\\
        |k_I]^{\dot{\alpha}} 
    \end{pmatrix}\,,\qquad\bar{v}^I(k)=(\bra{k^I}^\alpha\,\,[k^I|_{\dot{\alpha}})\,,
\end{equation}
that satisfy
\begin{equation}
\bar{v}^I(k)v_J(k)=-2m\delta^{I}_{J}\,,\qquad v_I(k)\bar{v}^I(k)=\slashed{k}-m\,.
\end{equation}
The wave functions for outgoing and incoming anti-fermions are 
\begin{equation}
    v_h(k)={\bar\tau}^I(h)v_I(k)\,,\qquad
    {\bar{v}}_h(k)={\tau}_I(h){\bar{v}}^I(k)\,.
\end{equation}
Notice that the anti-fermion annihilation is described by $\bar{v}$, which has an upper little-group index like the $u$ spinor that describes the annihilation of the fermions. Hence, both incoming particles and anti-particles lead to an upper-index amplitude tensor compatibly with eq.~(\ref{eq:WFSpinors}). A lower little-group index emerges instead for outgoing particles, from taking the conjugate of eq.~(\ref{eq:WFSpinors}). The little-group wave functions $\tau$ and $\bar\tau$ are the same tensors in eq.~(\ref{eq:Dirwf}) for both particles and anti-particles. Finally, notice that our wave functions are related by the charge conjugation relation
\begin{equation}
    v_h(k)=-i\gamma^2(u_h(k))^*\,.
\end{equation}

\subsection{Polarisation vectors}\label{sec:WFVP}

\subsection*{Standard polarisations}

In the standard formalism (see Appendix~\ref{App:EqFromalism}) the interpolating field for particles of spin-1 is the vector field $V_\mu$. We can equivalently express $V_\mu$ as a tensor field $V_{\alpha\dot\alpha}$, and vice versa, by the relations
\begin{equation}    V_{\alpha\dot\alpha}=V_\mu\sigma^\mu_{\alpha\dot\alpha}\,,\quad V_\mu=\frac12 V_{\alpha\dot\alpha}{\bar\sigma}_\mu^{\dot\alpha\alpha}\,.
\end{equation}
The polarisation vectors are proportional to eq.~(\ref{eq:WFSpinors})---with $n={\bar{n}}=1$---in the case of incoming particles, and to its complex conjugate if the particle is outgoing. We can thus express them in terms of the little-group and Lorentz tensors
\begin{equation}\label{eq:VectoSpin}    e^{IJ}_{\alpha\dot{\alpha}}(k)=\frac{\sqrt{2}}{m}\ket{k^I}_{\alpha}[\,k^J|_{\dot{\alpha}}\,,\qquad
{\bar{e}}_{IJ}^{\dot\alpha{\alpha}}(k)=-\frac{\sqrt{2}}{m}
|k_I]^{\dot{\alpha}}\langle{k_J|}^{\alpha}\,,
\end{equation}
or, equivalently, of the little-group tensors with 4-vector index
\begin{equation}\label{eq:VectoSpin1}
    e^{IJ}_\mu(k)=\frac12 e^{IJ}_{\alpha\dot{\alpha}}(k){\bar\sigma}_\mu^{\dot\alpha\alpha}\,,\qquad
    {\bar{e}}_{IJ}^{\mu}(k)=\frac12{\bar{e}}_{IJ}^{\dot\alpha{\alpha}}(k)\sigma^\mu_{\alpha\dot\alpha}\,.
\end{equation}
Notice that 
\begin{equation}
{\bar{e}}_{IJ}^{\dot\alpha{\alpha}}=\varepsilon^{\dot\alpha\dot\beta}\varepsilon^{\alpha\beta}[(e^{IJ})^*]_{\dot\beta\beta}\,,\qquad
{\bar{e}}_{IJ}^\mu=\eta^{\mu\nu}(e^{IJ}_\nu)^*\,.
\end{equation}

Using eq.~(\ref{eq:WFSpinors}) we can readily express the standard spin-1 polarisation vectors---denoted as $\varepsilon^h_{\textrm{St.}}$ in Appendix~\ref{App:EqFromalism}---in terms of the $e$ and ${\bar{e}}$ tensors, up to a multiplicative constant that can be fixed by the normalisation of the polarisations. The canonically-normalised polarisation vectors for incoming and outgoing particles are, respectively
\begin{equation}\label{eq:stdpolspin}
\varepsilon^h_{\textrm{St.},\mu}(k)=
{\mathcal{S}}^{1,h}_{IJ}\,e^{IJ}_{\mu}(k)\,,\qquad
{\bar\varepsilon}^{h,\mu}_{\textrm{St.}}(k)=
[\varepsilon^{h,\mu}_{\textrm{St.}}(k)]^*=
({\mathcal{S}}^{1,h}_{IJ})^*\,{\bar{e}}_{IJ}^{\mu}(k)\,,
\end{equation}
where ${\mathcal{S}}^{1,h}_{IJ}$ are the symmetric SU$(2)$ 2-tensors associated with the spin-1 representation, for $h=\pm1,0$. The explicit components of these tensors are
\begin{equation}\label{eq:stdpolten}
\begin{split}
    &{\mathcal{S}}^{1,-1}_{IJ}=\delta^-_I\delta^-_J\,,\\
    &{\mathcal{S}}^{1,+1}_{IJ}=\delta^+_I\delta^+_J\,,\\
    &{\mathcal{S}}^{1,0}=\frac{1}{\sqrt{2}}(\delta^-_I\delta^+_J+\delta^+_I\delta^-_J)\,.
\end{split}
\end{equation}
The polarisation vectors in eq.~(\ref{eq:stdpolspin}) are canonically normalised and obey the standard completeness relation in eq.~(\ref{eq:stdcompleteness}). This can be verified by employing the relations
\begin{equation}
    e^{IJ}_{\alpha\dot{\alpha}}(k)\,{\bar{e}}_{MN}^{\dot{\alpha}\alpha}(k)=
    -2\delta^I_M\delta^J_N\,,\qquad
    e^{IJ}_{\alpha \dot{\alpha}}(k){\bar{e}}^{\dot{\beta}\beta}_{IJ}(k)=2\delta^{\beta}_{\alpha}\delta^{\dot{\beta}}_{\dot{\alpha}}\,.
\end{equation}

\subsection*{GE polarisations}

The GE polarisation vectors described in Appendix~\ref{App:EqFromalism}, $\mathcal{E}_M^h(k)$, have 5 components rather than $4$, and their fifth component is a wave-function factor for the scalar Goldstone boson field $\pi(x)$. In order to express it as a little-group 2-tensor we will thus also need a Lorentz scalar, on top of the Lorentz vector tensor $e^{IJ}_\mu$ in eq.~(\ref{eq:VectoSpin}). We take this tensor to be proportional to the anti-symmetric tensor
\begin{equation}
e_{\pi}^{IJ}(k)=\frac{i}{\sqrt{2}}\epsilon^{IJ}\,,
\end{equation}
and we define a 5-component object
\begin{equation}
e_{M}^{IJ}(k)=[e^{IJ}_\mu,\,e^{IJ}_\pi]_M\,.
\end{equation}

It is simple to express $\mathcal{E}_M^h$ in terms of the $e^{IJ}_M$ tensor. We notice that, owing to eq.~(\ref{eq:momAS})
\begin{equation}\label{eq:kpol}
k_{\alpha\dot{\alpha}}=\frac{m}{\sqrt{2}}
\epsilon_{IJ}e^{IJ}_{\alpha\dot{\alpha}}(k)\,.
\end{equation}
The shift that relates the GE polarisations with the standard ones~(\ref{eq:deflongituialwpv1}) can thus be written as 
\begin{equation}\label{eq:Kpol}
\frac{i}{m}{\mathcal{K}}_M[k]=\frac1{\sqrt{2}}\epsilon_{IJ}e^{IJ}_{M}(k)\,.
\end{equation}
By combining with eq.~(\ref{eq:stdpolspin}) we obtain
\begin{equation}\label{eq:GEPVTE}
\mathcal{E}_M^h(k)=\left[{\mathcal{S}}^{1,h}_{IJ}
-\delta^h_0\frac1{\sqrt{2}}\epsilon_{IJ}
\right]e^{IJ}_M\equiv\tau_{IJ}(h)e^{IJ}_M\,,
\end{equation}
where the $\tau(0)$ tensor has been defined as the symmetric ${\mathcal{S}}^{1,0}$ minus the anti-symmetric tensor. 

We can thus express the GE polarisation vectors in terms of little-group 2-tensors, but only by employing little-group wave functions that are not symmetric
\begin{equation}\label{eq:VectoWF}
\tau_{IJ}(-1)=\delta^-_I\delta^-_J\,,\qquad\tau_{IJ}(0)=\sqrt{2}\,\delta^+_I\delta^-_J
\,,\qquad\tau_{IJ}(+1)=\delta^+_I\delta^+_J
\,.
\end{equation}

We can proceed similarly for the polarisation 5-vectors of outgoing particles. By defining
\begin{equation}\label{eq:outpspi} 
{\bar{e}}_{\pi,IJ}(k)=\frac{i}{\sqrt{2}}\epsilon_{IJ}\,,\qquad {\bar{e}}_{M,IJ}(k)=[{\bar{e}}_{IJ,\mu},\,{\bar{e}}_{\pi,IJ}]_M\,,
\end{equation}
we have 
\begin{equation}\label{eq:GEPVTEout}
-\frac{i}{m}{\mathcal{K}}_M[-k]=\frac1{\sqrt{2}}\epsilon^{IJ}{\bar{e}}_{IJ,M}(k)\;\;\;\Rightarrow\;\;\;
\bar{\mathcal{E}}_M^h(k)=\bar{\tau}^{IJ}(h){\bar{e}}_{IJ,M}\,,
\end{equation}
where the little-group wave functions for outgoing particles are
\begin{equation}\label{eq:VectoWFout}
    {\bar\tau}^{IJ}(h)=[{\tau}_{IJ}(h)]^*\,.
\end{equation}

The tensor notation offers a simple understanding of the different scaling at high-energy (or at low-mass) of the GE polarisation vectors in comparison with the standard ones. The bi-spinors in eqs.~(\ref{eq:imhsp}) and~(\ref{eq:jwsp}) scale with energy as follows. The upper-index angular bracket bi-spinors scale like $\ket{k^+}\sim m/E^{1/2}$ and $\ket{k^-}\sim E^{1/2}$, while $|k^+]\sim E^{1/2}$ and $|k^-]\sim m/E^{1/2}$ due to the contraction with $\epsilon$ in the definition of upper-index square bracket bi-spinors. Therefore, the components of $e^{IJ}_\mu$ scale like $e^{++}_\mu\sim e^{--}_\mu\sim 1$, $e^{+-}_\mu\sim m/E$ and $e^{-+}_\mu\sim E/m$. The Goldstone component of $e^{IJ}_M$, $e^{IJ}_\pi$, is a constant. The little-group wave functions associated with $h=\pm1$ particles, $\tau_{IJ}(\pm1)$, pick up the diagonal $\pm\pm$ entries of $e^{IJ}_M$, and thus they scale like a constant. The longitudinal wave function $\tau_{IJ}(0)$ picks up the $+-$ entry of $e^{IJ}_M$, which is suppressed as $m/E$ in the vector $M=\mu$ component, and of order 1 for $M=\pi$. The GE longitudinal polarisation vector is thus well-behaved in the high-energy limit. In contrast, the standard $h=0$ little-group wave function in eq.~(\ref{eq:stdpolten}) is symmetric, and therefore it picks up also the $-+$ entry of $e^{IJ}_\mu$, which grows like $E/m$. This is responsible for the energy growth of the standard longitudinal polarisation vector.

\subsection{Amplitude tensors}\label{sec:AMPT}

With the results of the previous sections we can express Feynman diagrams as little-group tensors,  by proceeding as follows. Each external particle leg is associated with that tensor that produces the corresponding wave function or polarisation vector by the contraction with the little-group wave functions $\tau$ and $\bar\tau$ in eqs.~(\ref{eq:Dirwf}), (\ref{eq:VectoWF}) and~(\ref{eq:VectoWFout}). For instance, the external leg of an incoming spin-$1/2$ particle is associated with the $u^I$ Dirac 4-spinor, while ${\bar{u}}_I$ corresponds to an outgoing fermion. The Lorentz index of the tensor, e.g. the Dirac index in the case of $u^I$, contracts with the tensors from the other external states and with the rest of the diagram in order to form a Lorentz singlet. In the final result, all the Lorentz indices are contracted and the free indices of the amplitude tensor are the indices in the little-group of each external particle. The tensor has one upper index for each incoming fermion or anti-fermion, and one lower index for each outgoing fermion. Spin-1 particles bring two indices from the $e^{IJ}$ and ${\bar{e}}_{IJ}$ tensors. Example applications of these rules to construct amplitude tensors are discussed in Section~\ref{sec:rdsa}. See in particular Figure~\ref{Splitting-hVV} for a diagrammatic representation of the amplitude tensor associated with the $h^*\to VV$ splitting.

Helicity amplitudes are obtained from the amplitude tensors by contracting the free indices of each particle with the little-group wave function of the desired helicity. One chooses whether to employ the standard polarisation vectors for spin-1 particles or the GE ones by using either the symmetric wave-functions in eq.~(\ref{eq:stdpolten}) or the non-symmetric ones in eq.~(\ref{eq:VectoWF}). The Ward identity in eq.~(\ref{eq:WI}) ensures that exactly identical physical scattering amplitudes are obtained with the standard and with the GE polarisation vectors. In the tensor notation, this can be seen by noticing that the 5-vector ${\mathcal{K}}_M[\pm k]$ which appears in the Ward identity (for incoming or outgoing particle momentum) is proportional to the contraction of $e_M$ or ${\bar{e}}_M$ with the anti-symmetric $\epsilon$ tensor, as shown in eqs.~(\ref{eq:Kpol}) and~(\ref{eq:GEPVTEout}). Therefore, the Ward identity ensures that the amplitude tensors that correspond to physical scattering processes have an exactly vanishing anti-symmetric component. When dealing with such amplitudes, the non-symmetric GE little-group wave function in eq.~(\ref{eq:VectoWF}) are fully equivalent to the symmetric ones in eq.~(\ref{eq:stdpolten}). 

However, the validity of the Ward identity~(\ref{eq:WI}) is restricted to physical scattering amplitudes. The splitting amplitudes are not physical amplitudes, they do not obey the Ward identity and their representation in terms of tensors is not symmetric. For this reason, the standard and the GE polarisation vectors are not equivalent in the calculation of the splitting amplitudes as verified explicitly in the main text.

Dealing with amplitude tensors that are not symmetric under the exchange of the little-group indices of the spin-1 states can led to confusions related to how the \mbox{SU$(2)$} little-group symmetry is implemented in the formalism. It should thus be stressed that the amplitude tensors do transform, by construction, as regular tensors under the little group. For instance, if we operate the transformation in eq.~(\ref{eq:LGrot}) that corresponds to a change of basis for the bi-spinors, upper indices of the tensor transform as they should with the  ${\mathcal{U}}$ matrix, and lower indices transform with the conjugate ${\mathcal{U}}^*$ matrix. If we operate a Lorentz transformation, the associated \mbox{SU$(2)$} little-group matrix acts like ${\mathcal{U}}$ does. From the viewpoint of Lorentz and little-group transformations of the tensor, it is immaterial if the indices are symmetrised or not.

The situation is different if we consider instead the transformation of the helicity amplitudes. The little-group transformation of the tensor indices is equivalent to the transformation of the little-group wave functions. In the case of incoming spin-1 particles
\begin{equation}\label{eq:wftt}
    \tau_{IJ}(h)\;\rightarrow\;
    \tau^\prime_{IJ}(h)=\tau_{KL}(h)\,
    {\mathcal{U}}^{K}_{\;\;\;I}\,
    {\mathcal{U}}^{L}_{\;\;\;J}\,.
\end{equation}
If the $\tau$'s were symmetric, the \mbox{SU$(2)$}-rotated $\tau^\prime$ would be a linear combination of the $\tau$'s with coefficients provided by the Wigner-${\mathcal{D}}$ matrix representation of ${\mathcal{U}}$ in the spin-one representation. Consequently, the helicity amplitudes would transform as triplets under the little-group. On the contrary, the zero-helicity $\tau(0)$ wave function has also an anti-symmetric component, which is invariant under \mbox{SU$(2)$}. Therefore, the transformed $\tau^\prime$ is given by a rotation of the helicity indices, plus a shift. Namely
\begin{equation}\label{eq:LGWFTLGT}
    \tau^\prime_{IJ}(h)=
    \sum\limits_{g}
    \tau_{IJ}(g)[{\mathcal{D}}^{1}({\mathcal{U}})]_g^{\;h}
    +\frac1{\sqrt{2}}\left\{
    [{\mathcal{D}}^{1}({\mathcal{U}})]_0^{\;h}
    -\delta_0^h
    \right\}\epsilon_{IJ}\,.
\end{equation}
The shift term proportional to $\epsilon$ has no effect on physical amplitudes that obey the Ward identity. On such amplitudes, the little-group (and in turn the Lorentz) transformation is provided by the regular Wigner rotation. The shift term is instead relevant for non-physical amplitudes like the splitting amplitudes.

In our formalism, the little-group and Lorentz symmetries are realised like in theories with massless spin-1 particles: the unwanted shift of the wave-functions is cancelled on the physical scattering amplitudes by the Ward identity. The analogy can be illustrated more clearly by computing the effect of a little-group transformation on the GE polarisation vectors $\mathcal{E}_M^h(k)$. Using their definition~(\ref{eq:GEPVTE}), eq.~(\ref{eq:Kpol}) and eq.~(\ref{eq:LGWFTLGT}), we find
\begin{equation}
    \mathcal{E}_M^h\;\rightarrow\;
        \sum\limits_{g}
    \mathcal{E}_M^g[{\mathcal{D}}^{1}({\mathcal{U}})]_g^{\;h}
    +\frac{i}{m}
    \left\{
    [{\mathcal{D}}^{1}({\mathcal{U}})]_0^{\;h}
    -\delta_0^h
    \right\}
    {\mathcal{K}}_M\,.
\end{equation}
This result is fully analogous to the shift, by an amount that is proportional to the particle's 4-momentum $k_\mu$, of the massless polarisation vectors. In the massive case, one could avoid the shift by employing the regular polarisation vectors with symmetric little-group wave functions. On the other hand, it is not surprising that the polarisation vectors that are most suited to describe the massless (or, high-energy) limit---the GE polarisations---do instead feature a shift like the massless polarisations.

\section{Standard Model splitting amplitudes}\label{sec:salist}
In this appendix, we summarise our results for the SM splitting amplitudes. In Section~\ref{Sec:FeynmanRules} we list the relevant Feynman rules. In Section~\ref{app:SA} we report the general expressions for the splitting amplitudes tensors. In Section~\ref{Sec:SWSplittings} we report the explicit expressions for the helicity amplitudes for SW bi-spinors.

\subsection{Standard Model Feynman rules}\label{Sec:FeynmanRules}
\paragraph{Fermionic vertices}

The Feynman rules between two fermions $f_{a,b}$ of masses $m_{a,b}$ and a (vector or Goldstone) boson of mass $m_V$ are expressed in terms of the general coupling $G_L$, $G_R$, the chirality projectors $P_{L/R} = (1\mp \gamma_5)/2$ and the Yukawa couplings. The latter are diagonal in the Standard Model and related to the fermion masses through the Higgs VEV $m_f=\sqrt{2} v y_f$. The two couplings $G_L$ and $G_R$ read
\begin{align*}
\begin{tabular}{c|cccc} 
\rule{0pt}{1.2em}
$V$ & $W^-~(f^a=d, f^b=u)$ &  $Z~(f^a=f^b)$ &  $\gamma~(f^a=f^b)$ & $G~(f^a=f^b)$  \\
\hline
$G_L$ & $\frac{gV_{ud}}{\sqrt{2}}$ &  $\frac{g}{c_{\rm w}}(T^3-s_{\rm w}^2q_f)$ &  $eq_f$ & $g_st^\alpha_{i_bi_a}$  \\
$G_R$ & $0$ &  $\frac{g}{c_{\rm w}}(-s_{\rm w}^2q_f)$ &  $eq_f$ & $g_st^\alpha_{i_bi_a}$  
\end{tabular}
\end{align*}
where $T^A$ and $t^{\alpha}$ are respectively the color \mbox{SU$(3)$} and the weak \mbox{SU$(2)$} generators. The gauge couplings are $g_s$ and $g$; $e = g s_{\rm w}$ is the QED coupling constant and $q_f$ is the electric charge.

\begin{align*}
      &\hspace{2cm}\begin{tikzpicture}[baseline=(m), scale = 1.4]
     \begin{feynman}
        \vertex (m) at (0,0);
        \vertex (a) at (1,0) {$\pi$};
        \vertex (b) at (-0.5,0.86) {$f_b$};
        \vertex (c) at (-0.5,-0.86) {$f_a$};
        \diagram* {
            (a) -- [scalar] (m),
            (m) -- [fermion] (b),
            (m) -- [anti fermion] (c),
        };
    \end{feynman}
    \node at (3,0) {$=\frac{m_b-m_a}{m_V}(G_L P_L + G_R P_R),$};
    \end{tikzpicture}\\
    &\begin{tikzpicture}[baseline=(m), scale = 1.4]
     \begin{feynman}
        \vertex (m) at (0,0);
        \vertex (a) at (1,0) {$V^{\mu}$};
        \vertex (b) at (-0.5,0.86) {$f_b$};
        \vertex (c) at (-0.5,-0.86) {$f_a$};
        \diagram* {
            (a) -- [photon] (m),
            (m) -- [fermion] (b),
            (m) -- [anti fermion] (c),
        };
    \end{feynman}
    \node at (3,0) {$=i \gamma^{\mu} (G_L P_L +G_R P_R),$};
    \end{tikzpicture}\hspace{1cm}
    \begin{tikzpicture}[baseline=(m), scale = 1.4]
     \begin{feynman}
        \vertex (m) at (0,0);
        \vertex (a) at (1,0) {$h$};
        \vertex (b) at (-0.5,0.86) {$f_b$};
        \vertex (c) at (-0.5,-0.86) {$f_a$};
        \diagram* {
            (a) -- [scalar] (m),
            (m) -- [fermion] (b),
            (m) -- [anti fermion] (c),
        };
    \end{feynman}
    \node at (2,0) {$=- i \delta_{ab}\frac{y_f}{\sqrt{2}}.$};
    \end{tikzpicture}
\end{align*}

\paragraph{Boson-scalar Vertices}
We now collect the Feynman rules between three bosons, with at least one Higgs field. The rules for the Higgs coupling to vectors and Goldstones are parameterized in terms of a coupling $G$
\begin{align*}
    \begin{tabular}{c|cccc} 
\rule{0pt}{1.2em}
$V^a V^b$ & $W^-W^+$ &  $ZZ$ & $\gamma\gamma$ & $GG$\\
\hline
$G$ & $g$ &  ${g}/{c_{\rm w}}$ & $0$ & $0$
\end{tabular}
\end{align*}
where $m_a$ and $m_b$ are the masses of the vector/scalar.

\begin{align*}
        &\begin{tikzpicture}[baseline=(m), scale = 1.4]
    \begin{feynman}
        \vertex (m) at (0,0);
        \vertex (a) at (1,0) {$h$};
        \vertex (b) at (-0.5,0.86) {$h$};
        \vertex (c) at (-0.5,-0.86) {$h$};
        \diagram* {
            (a) -- [scalar] (m),
            (m) -- [scalar] (b),
            (m) -- [scalar] (c),
        };
    \end{feynman}
    \node at (2,0) {$=-i \frac{3}{2} g \frac{m_h^2}{m_W}$,};
    \end{tikzpicture}\hspace{2.5cm}
    \begin{tikzpicture}[baseline=(m), scale = 1.4]
    \begin{feynman}
        \vertex (m) at (0,0);
        \vertex (a) at (1,0) {$h$};
        \vertex (b) at (-0.5,0.86) {$V^{\mu}$};
        \vertex (c) at (-0.5,-0.86) {$V^{\nu}$};
        \diagram* {
            (a) -- [scalar] (m),
            (m) -- [photon] (b),
            (m) -- [photon] (c),
        };
    \end{feynman}
    \node at (2.5,0) {$=i G m_V\, \eta_{\mu \nu}$,};
    \end{tikzpicture}\\
    &\begin{tikzpicture}[baseline=(m), scale = 1.4]
    \begin{feynman}
        \vertex (m) at (0,0);
        \vertex (a) at (1,0) {$h$};
        \vertex (b) at (-0.5,0.86) {$\pi$};
        \vertex (c) at (-0.5,-0.86) {$\pi$};
        \diagram* {
            (a) -- [scalar] (m),
            (m) -- [scalar] (b),
            (m) -- [scalar] (c),
        };
    \end{feynman}
    \node at (2.5,0) {$=-i G \frac{m_h^2}{2 m_V}$,};
    \end{tikzpicture}\hspace{1.5cm}
    \begin{tikzpicture}[baseline=(m), scale = 1.4]
    \begin{feynman}
        \vertex (m) at (0,0);
        \vertex (a) at (1,0) {$h$};
        \vertex (b) at (-0.5,0.86) {$\pi$};
        \vertex (c) at (-0.5,-0.86) {$V^{\mu}$};
        \diagram* {
            (a) -- [scalar,momentum' = $k_1$] (m),
            (m) -- [photon] (c),
            (b) -- [scalar, momentum' =$k_2$] (m),
        };
    \end{feynman}
    \node at (2.5,0) {$=\frac{G}{2} (k_2-k_1)^{\mu}$.};
    \end{tikzpicture}
\end{align*}

\paragraph{Vector and Goldstone Vertices}
We finally report the Feynman rules between three bosons, vectors or Goldstones, of masses $m_a$, $m_b$ and $m_c$. We express them in terms of a coupling  $G_{abc}$ 
\begin{align*}
    \begin{tabular}{c|ccc} 
    \rule{0pt}{1.2em}
    $V^aV^bV^c$ & $W^-W^+\gamma$ &  $W^-W^+Z$ & $G^a G^b G^c$\\
    \hline
    $G_{abc}$ & $e$ &  $g c_{\rm w}$ & $-ig_sf_{a b c}$ 
    \end{tabular}
\end{align*}
where $f_{abc}$ are the \mbox{SU$(3)$} structure constants.

\begin{align*}
      &\hspace{1.75cm}\begin{tikzpicture}[baseline=(m), scale = 1.4]
    \begin{feynman}
        \vertex (m) at (0,0);
        \vertex (a) at (1,0) {$V^{\rho}_c$};
        \vertex (b) at (-0.5,0.86) {$V^{\mu}_a$};
        \vertex (c) at (-0.5,-0.86) {$V^{\nu}_b$};
        \diagram* {
            (a) -- [photon,momentum = $k_c$] (m),
            (b) -- [photon,momentum = $k_a$] (m),
            (c) -- [photon,momentum = $k_b$] (m),
        };
    \end{feynman}
    \node at (5,0) {$=i G_{abc}[\eta_{\mu \nu}(k_a -k_b)_{\rho}+ \eta_{\nu \rho}(k_b - k_c)_{\mu} +\eta_{\rho \mu}(k_c - k_a)_{\nu}]$,};
    \end{tikzpicture}\\
     &\begin{tikzpicture}[baseline=(m), scale = 1.4]
    \begin{feynman}
        \vertex (m) at (0,0);
        \vertex (a) at (1,0) {$\pi_c$};
        \vertex (b) at (-0.5,0.86) {$V^{\mu}_a$};
        \vertex (c) at (-0.5,-0.86) {$V^{\nu}_b$};
        \diagram* {
            (a) -- [scalar] (m),
            (m) -- [photon] (b),
            (m) -- [photon] (c),
        };
    \end{feynman}
    \node at (2.5,0) {$=G_{abc}\,\eta_{\mu \nu} \frac{m_b^2 - m_a^2}{m_c}$,};
    \end{tikzpicture}\hspace{0.5cm}
    \begin{tikzpicture}[baseline=(m), scale = 1.4]
    \begin{feynman}
        \vertex (m) at (0,0);
        \vertex (a) at (1,0) {$V^{\mu}_c$};
        \vertex (b) at (-0.5,0.86) {$\pi_a$};
        \vertex (c) at (-0.5,-0.86) {$\pi_b$};
        \diagram* {
            (a) -- [photon] (m),
            (b) -- [scalar,momentum =$k_a$] (m),
            (c) -- [scalar,momentum =$k_b$] (m),
        };
    \end{feynman}
    \node at (3.5,0) {$=-\frac{i}{2} G_{abc} \frac{m_a^2 + m_b^2 - m_c^2}{m_a m_b} (k_a -k_b)_{\mu}$.};
    \end{tikzpicture}
\end{align*}

\subsection{Amplitude tensors}\label{app:SA}

\subsubsection*{Higgs splitting into two Higgs particles}

\begin{align}
\mathcal{M}_{\textrm{S}}(h(A)\rightarrow h(B)h(C))=-\frac{3}{2}g\frac{m_h^2}{m_W}\,.
\end{align}

\subsubsection*{Higgs splitting into two fermions}
\begin{align}
{\mathcal{M}}_{\textrm{S}}(h(A)\to f_{I}(B)\bar{f}_{\,J}(C))=
-\frac{y_f}{\sqrt{2}}
\bigg(
\langle B_I C_J\rangle+
[ B_I C_J]
\bigg)\,.
\end{align}
\subsubsection*{Fermion splitting into a Higgs and a fermion}

\begin{align}
{\mathcal{M}}_{\textrm{S}}(f^{I}(A)\to f_{\,J}(B) h(C))=
-\frac{y_f}{\sqrt{2}}
\bigg(
\langle A^I B_J\rangle-
[ A^I B_J]
\bigg),.
\end{align}

\subsubsection*{Higgs splitting into two vectors}
\begin{align}
\begin{aligned}
                \mathcal{M}_{\textrm{S}}(h(A) \rightarrow V_{IJ}(B)V_{MN}(C))= &\frac{G}{m_V}
                \bigg\{ \langle B_J C_N \rangle [C_M B_I] +\frac{1}{2} \epsilon_{IJ} (\langle C_N B^K \rangle [B_K C_M]-\frac{1}{2}m_V^2\epsilon_{MN}) \\
                &+\frac{1}{2} \epsilon_{MN} (\langle B_J C^K \rangle [C_K B_I]-\frac{1}{2}m_V^2\epsilon_{IJ}) +\frac{m_h^2}{4} \epsilon_{IJ} \epsilon_{MN} 
                \bigg\}\,.
            \end{aligned}
\end{align}
\subsubsection*{Vector splitting into a Higgs and a vector}

\begin{equation}
\begin{split}
        \mathcal{M}_S(V^{IJ}(A)\rightarrow h(B)V_{MN}(C))=&\frac{G}{m_V}
                \bigg\{-\langle C_NA^I\rangle[A^JC_M]+\frac{1}{4}m_h^2\,\epsilon^{IJ}\epsilon_{MN}\\
                &+\frac{1}{4}\epsilon_{MN}[A^J|p_B-p_C|A^I\rangle-\frac{1}{4}\epsilon^{IJ}\langle C_N |2p_B+p_C |C_M]\bigg\}\,.
\end{split}
\end{equation}

\subsubsection*{Vector splitting into two fermions}

\begin{equation}\label{eq:SPlittingVff}
\begin{split}
    \mathcal{M}_S(V^{IJ}(A)&\rightarrow f_M(B)\bar{f}_N(C))=\\
    &-\sqrt{2}\frac{G_R}{m_A}\bigg(\langle B_M A^I\rangle [A^J C_N]+\frac{1}{2}m_C\epsilon^{IJ}\langle B_M C_N\rangle - \frac{1}{2}m_B\epsilon^{IJ}[ B_M C_N]\bigg)+\\
    &-\sqrt{2}\frac{G_L}{m_A}\bigg([ B_M A^J] \langle A^I C_N\rangle+ \frac{1}{2}m_C\epsilon^{IJ}[ B_M C_N]-\frac{1}{2}m_B\epsilon^{IJ}\langle B_M C_N\rangle \bigg)\,.
\end{split}
\end{equation}

\subsubsection*{Fermion splitting into a fermion and a vector}

\begin{equation}\label{eq:Spfvf}
\begin{split}
    \mathcal{M}_S(f^N(A)&\rightarrow f_M(B)V_{IJ}(C))=\\
    &+\sqrt{2}\frac{G_R}{m_C}\bigg(\langle B_M C_J\rangle [C_I A^N]+\frac{1}{2}m_A\epsilon_{IJ}\langle B_M A^N\rangle + \frac{1}{2}m_B\epsilon_{IJ}[ B_M A^N]\bigg)+\\
    &-\sqrt{2}\frac{G_L}{m_C}\bigg([ B_M C_I] \langle C_J A^N\rangle+ \frac{1}{2}m_B\epsilon_{IJ}\langle B_M A^N\rangle+\frac{1}{2}m_A\epsilon_{IJ}[ B_M A^N] \bigg)\,.
\end{split}
\end{equation}

\subsubsection*{Vector splitting into two vectors}

\begin{equation}
\begin{split}
    &\mathcal{M}_{\textrm{S}}(V^{IJ}(A))\to V_{KL}(B)V_{MN}(C))=\frac{G_{ABC}}{\sqrt{2}\,m_A m_B m_C}\,\bigg\{\\ 
    &+\langle C_N B_L \rangle [B_K C_M](\langle A^I |p_C-p_B|A^J]+\epsilon^{IJ}(m_C^2-m_B^2))+\frac{1}{4}\epsilon_{MN}\epsilon_{KL}(m_B^2+m_C^2-m_A^2)\langle A^I |p_C-p_B| A^J]\\
    &+2\langle B_L A^I \rangle [A^J B_K](\langle C_N |p_B|C_M]+\frac{1}{2}\epsilon_{MN}(m_A^2-m_B^2-m_C^2))+\frac{1}{2}\epsilon^{IJ}\epsilon_{KL}(m_C^2-m_A^2-m_B^2)\langle C_N |p_B| C_M] \\
    &-2\langle C_N A^I \rangle [A^J C_M](\langle B_L |p_C|B_K]+\frac{1}{2}\epsilon_{KL}(m_A^2-m_B^2-m_C^2))-\frac{1}{2}\epsilon^{IJ}\epsilon_{MN}(m_B^2-m_A^2-m_C^2)\langle B_L |p_C| B_K]\\ 
    &+\frac{1}{4}\epsilon^{IJ}\epsilon_{MN}\epsilon_{KL}((m_A^2+m_B^2-m_C^2)m_C^2-(m_A^2+m_C^2-m_B^2)m_B^2)\bigg\}.
\end{split}
\end{equation}

\subsection{Splitting amplitudes in the SW basis}\label{Sec:SWSplittings}
We now list the SM splitting amplitudes in the Soper-Weinberg basis. We report a minimal set of splitting amplitudes, the missing ones can be obtained according to the two following rules:
\begin{itemize}
    \item the kinematic configuration where the particles $B$ and $C$ are exchanged (see eq.~(\ref{eq:onmom})) is obtained via the replacement
    \begin{align*}
        x\rightarrow 1-x\,, && \mathbf{p}_T\rightarrow -\mathbf{p}_T\,, && m_B \rightarrow m_C\,, &&m_C \rightarrow m_B\,.
    \end{align*}
    \item splitting amplitudes involving anti-particles are obtained form the ones involving particles through CP conjugation via the following relation
    \begin{equation}
    \mathcal{M}_S(\bar{A}(-h_A)\rightarrow \bar{B}(-h_B) \, \bar{C}(-h_C))=\prod_{k=A,B,C} (-1)^{j_k-h_k}\mathcal{M}_S(A(h_A)\rightarrow B(h_B) C(h_C))\bigg|_{\mathbf{p}_T\rightarrow \mathbf{p}_T^*}\,,
    \end{equation}
    where $j_k$ and $h_k$ are, respectively, the spin and the helicity of the particle $k$ (see Appendix~B of \cite{Cuomo:2019siu} for more details).
\end{itemize}

\subsubsection*{Scalar-Fermions splitting}
    
\begin{minipage}[c]{0.4\linewidth}
    \begin{tabular}{|c|c|c|}
    \hline
        \rule{0pt}{11pt} $\rightarrow$  &  $f^+(B)+h(C)$ & $f^-(B)+h(C)$ \\
        \hline
        \rule{0pt}{15pt} $f^+(A)$ & $-\frac{y_f}{\sqrt{2}}\frac{(1-x)m_A+m_B}{\sqrt{1-x}}$ &  $\frac{y_f}{\sqrt{2}}\frac{\mathbf{p}_T}{\sqrt{1-x}}$   \\
        \rule{0pt}{15pt} $f^-(A)$ & $-\frac{y_f}{\sqrt{2}}\frac{\mathbf{p}_T^*}{\sqrt{1-x}} $  & $-\frac{y_f}{\sqrt{2}}\frac{(1-x)m_A+m_B}{\sqrt{1-x}}$ \\
        \hline
    \end{tabular}
\end{minipage}\hspace{2cm}\begin{minipage}[c]{0.4\linewidth}
    \begin{tabular}{|c|c|c|}
    \hline
        \rule{0pt}{11pt} $\rightarrow$  &  $f^-(B)+\bar{f}^+(C)$ &  $f^-(B)+\bar{f}^-(C)$\\
        \hline
        \rule{0pt}{15pt} $h(A)$ &   $-\frac{y_f}{\sqrt{2}}\frac{m_B x -m_C(1-x)}{\sqrt{x(1-x)}}$ & $\frac{y_f}{\sqrt{2}}\frac{\mathbf{p}_T}{\sqrt{x(1-x)}}$ \\
        \hline
    \end{tabular}
\end{minipage}

\subsubsection*{Vector-Fermions splitting}
\begin{minipage}[c]{0.5\textwidth}
    \centering
    \begin{tabular}{|c|c|c|c|c|}
    \hline
        \rule{0pt}{11pt} $\rightarrow$  & $f^+(B) + \bar{f}^+(C)$ &  $f^+(B) + \bar{f}^-(C)$ & $f^-(B)+ \bar{f}^+(C)$ & $f^-(B)+ \bar{f}^-(C)$ \\
        \hline
        \rule{0pt}{15pt} $V^+(A)$ & $-\frac{\sqrt{2}(G_Lm_Bx+G_Rm_C(1-x))}{\sqrt{x(1-x)}} $ & $-\sqrt{2}G_R\,\mathbf{p}_T\,\sqrt{\frac{1-x}{x}}$ & $\sqrt{2}G_L\mathbf{p}_T\sqrt{\frac{x}{1-x}}$ & 0 \\
        \rule{0pt}{15pt} $V^-(A)$ & 0 & $-\sqrt{2}G_R\,\mathbf{p}_T^*\,\sqrt{\frac{x}{1-x}}$  & $\sqrt{2}G_L\,\mathbf{p}_T^*\,\sqrt{\frac{1-x}{x}}$ & $-\frac{\sqrt{2}(G_Rm_Bx+G_Lm_C(1-x))}{\sqrt{x(1-x)}} $ \\ 
        \hline
    \end{tabular}
    \label{Tab:SplitVpmtoFermionfermion}
\end{minipage}

\vspace{0.5cm}

\hspace{-0.6cm}\begin{minipage}[c]{0.5\textwidth}
    \centering
    \begin{tabular}{|c|c|c|}
    \hline
         \rule{0pt}{11pt}$\rightarrow$  & $f^+(B) + \bar{f}^+(C)$ &  $f^+(B) + \bar{f}^-(C)$  \\
         \hline
        \rule{0pt}{15pt} $V^0(A)$ & $\big(\frac{m_B}{m_A}G_L-\frac{m_C}{m_A}G_R\big)\frac{\mathbf{p}_T^*}{\sqrt{x(1-x)}}$ & $\frac{(1-x)m_C^2+xm_B^2-2m_A^2x(1-x)}{m_A\sqrt{x(1-x)}}G_R-\frac{m_B m_C}{m_A\sqrt{x(1-x)}}G_L$ \\ \hline
    \end{tabular}
    \label{tab:my_label}
\end{minipage}

\vspace{0.5cm}

\hspace{-0.6cm}\begin{minipage}[c]{0.5\textwidth}
\centering
    \begin{tabular}{|c|c|c|}
    \hline
        \rule{0pt}{11pt} $\rightarrow$  &  $f^+(B)+V^+(C)$ & $f^-(B)+V^+(C)$ \\
        \hline
        \rule{0pt}{15pt} $f^+(A)$ & $\sqrt{2} G_R\frac{\mathbf{p}_T^*}{x\sqrt{1-x}}$ & $\sqrt{2}G_R \frac{m_B}{\sqrt{1-x}}-\sqrt{2}G_L m_A \sqrt{1-x}$  \\
        \rule{0pt}{15pt} $f^-(A)$ & 0  & $\sqrt{2}G_L \mathbf{p}_T^* \frac{\sqrt{1-x}}{x}$  \\
        \hline
    \end{tabular}
\end{minipage}

\vspace{0.5cm}

\hspace{-0.6cm}\begin{minipage}[c]{0.5\textwidth}
    \begin{tabular}{|c|c|c|}
    \hline
        \rule{0pt}{11pt} $\rightarrow$  &  $f^+(B)+V^-(C)$ & $f^-(B)+V^-(C)$\\
        \hline
        \rule{0pt}{15pt} $f^+(A)$  & $-\sqrt{2}G_R \mathbf{p}_T \frac{\sqrt{1-x}}{x}$  & $0$  \\
        \rule{0pt}{15pt} $f^-(A)$ & $\sqrt{2}G_L \frac{m_B}{\sqrt{1-x}}-\sqrt{2}G_R m_A \sqrt{1-x}$ & $-\sqrt{2} G_L\frac{\mathbf{p}_T}{x\sqrt{1-x}}$ \\
        \hline
    \end{tabular}
\end{minipage}
\vspace{0.5cm}

\hspace{-0.6cm}\begin{minipage}[c]{0.5\textwidth}
    \begin{tabular}{|c|c|c|}
    \hline
        \rule{0pt}{11pt} $\rightarrow$  &  $f^+(B)+V^0(C)$ & $f^-(B)+V^0(C)$  \\
        \hline
        \rule{0pt}{15pt} $f^+(A)$ & $G_R\frac{m_A^2 x(1-x)-m_B^2 x - 2 m_C^2 (1-x)}{m_C x \sqrt{1-x}}+G_L\frac{x m_A m_B }{m_C\sqrt{1-x}}$ & $\frac{\mathbf{p}_T}{\sqrt{1-x}}\big(G_R \frac{m_B}{m_C}-G_L \frac{m_A}{m_C}\big)$ \\
        \rule{0pt}{15pt} $f^-(A)$ & $\frac{\mathbf{p}_T^*}{\sqrt{1-x}}\big(G_R \frac{m_A}{m_C}-G_L \frac{m_B}{m_C}\big)$ & $G_L\frac{m_A^2 x(1-x)-m_B^2 x - 2 m_C^2 (1-x)}{m_C x \sqrt{1-x}}+G_R\frac{x m_A m_B }{m_C\sqrt{1-x}}$  \\
        \hline
    \end{tabular}
\end{minipage}

\subsubsection*{Scalar-Vectors splitting}

\begin{minipage}[c]{0.5\textwidth}
    \begin{tabular}{|c|c|c|c|}
    \hline
        \rule{0pt}{11pt} $\rightarrow$  &  $V^+(B)+V^+(C)$ & $V^+(B)+V^-(C)$ & $V^+(B)+V^0(C)$  \\
        \hline
        \rule{0pt}{15pt} $h(A)$ & 0  & $G m_V$ &  $-G\frac{\mathbf{p}_T^*}{\sqrt{2}(1-x)}$ \\

        \hline
    \end{tabular}
\end{minipage}

\vspace{0.5cm}

\hspace{-0.6cm}\begin{minipage}[c]{0.5\textwidth}
    \begin{tabular}{|c|c|c|}
    \hline
        \rule{0pt}{11pt} $\rightarrow$  &  $V^0(B)+V^0(C)$ & $h(B)+h(C)$   \\
        \hline
        \rule{0pt}{15pt} $h(A)$ & $\frac{G}{2}\big(\frac{m_h^2}{m_V}-m_V\frac{1+x}{1-x}-m_V\frac{2-x}{x}\big)$ &  $-\frac{3}{2}gm_V$ \\

        \hline
    \end{tabular}
\end{minipage}

\vspace{0.5cm}

\hspace{-0.6cm}\begin{minipage}[c]{0.5\textwidth}
    \begin{tabular}{|c|c|c|c|}
    \hline
        \rule{0pt}{11pt} $\rightarrow$  &  $h(B)+V^+(C)$ & $h(B)+V^-(C)$ & $h(B)+V^0(C)$  \\
        \hline
        \rule{0pt}{15pt} $V^+(A)$ & $- G m_V$  & 0 &$- \frac{G}{\sqrt{2}}\mathbf{p}_T$ \\

        \hline
    \end{tabular}
\end{minipage}

\vspace{0.5cm}

\hspace{-0.6cm}\begin{minipage}[c]{0.5\textwidth}
    \begin{tabular}{|c|c|c|c|}
    \hline
        \rule{0pt}{11pt} $\rightarrow$  &  $h(B)+V^+(C)$ & $h(B)+V^-(C)$ & $h(B)+V^0(C)$  \\
        \hline
        \rule{0pt}{15pt} $V^0(A)$ & $\frac{G}{\sqrt{2}}\frac{\mathbf{p}_T^*}{\sqrt{x}}$  & $-\frac{G}{\sqrt{2}}\frac{\mathbf{p}_T}{\sqrt{x}}$ &   $\frac{G}{2}\big(-\frac{m_h^2}{m_V}+m_V(1-2x)-m_V\frac{2-x}{x}\big)$\\

        \hline
    \end{tabular}
\end{minipage}
\subsubsection*{Vectors splitting}

\begin{minipage}[c]{0.5\textwidth}
    \begin{tabular}{|c|c|c|c|c|}
    \hline
        \rule{0pt}{11pt} $\rightarrow$  &  $V^+(B)+V^+(C)$ & $V^+(B)+V^-(C)$ & $V^-(B)+V^+(C)$ & $V^-(B)+V^-(C)$  \\
        \hline
        \rule{0pt}{15pt} $V^+(A)$ &  $-\sqrt{2}G_{ABC}\frac{\mathbf{p}_T^*}{x(1-x)}$ & $\sqrt{2}G_{ABC}\,\mathbf{p}_T\frac{1-x}{x}$ & $\sqrt{2}G_{ABC}\,\mathbf{p}_T\frac{x}{1-x}$ & 0 \\
        \rule{0pt}{15pt} $V^0(A)$ & 0 & $G_{ABC}\frac{m_A^2(1-2x)-m_B^2+m_C^2}{m_A}$ & $G_{ABC}\frac{m_A^2(1-2x)-m_B^2+m_C^2}{m_A}$ & 0\\
        \hline
    \end{tabular}
\end{minipage}

\vspace{0.5cm}

\hspace{-0.6cm}\begin{minipage}[c]{0.5\textwidth}
    \begin{tabular}{|c|c|c|}
    \hline
        \rule{0pt}{11pt} $\rightarrow$  &  $V^+(B)+V^0(C)$  & $V^-(B)+V^0(C)$  \\
        \hline
        \rule{0pt}{15pt} $V^+(A)$ & $G_{ABC}\big(m_C \frac{2-x}{x}+\frac{m_B^2-m_A^2}{m_C}\big)$ & 0 \\
        \rule{0pt}{15pt} $V^0(A)$ & $G_{ABC} \frac{m_B^2-m_A^2-m_C^2}{\sqrt{2} m_A m_C}\frac{\mathbf{p}_T^*}{1-x}$ & $-G_{ABC} \frac{m_B^2-m_A^2-m_C^2}{\sqrt{2} m_A m_C}\frac{\mathbf{p}_T}{1-x}$ \\
        \hline
    \end{tabular}
\end{minipage}

\vspace{0.5cm}

\hspace{-0.6cm}\begin{minipage}[c]{0.5\textwidth}
    \begin{tabular}{|c|c|}
    \hline
        \rule{0pt}{11pt} $\rightarrow$  &  $V^0(B)+V^0(C)$  \\
        \hline
        \rule{0pt}{15pt} $V^+(A)$  &$G_{ABC} \frac{m_A^2-m_B^2-m_C^2}{\sqrt{2} m_B m_C}\mathbf{p}_T$\\
        \rule{0pt}{15pt} $V^0(A)$ & $
            \frac{G_{ABC}}{2} \big(\frac{m_A(m_B^2+m_C^2-m_A^2)}{m_B m_C}(1-2x) -\frac{m_B(m_A^2+m_C^2-m_B^2)}{m_A m_C}\frac{1+x}{1-x}+\frac{m_C(m_A^2+m_B^2-m_C^2)}{m_A m_B}\frac{2-x}{x}\big)$\\
        \hline
    \end{tabular}
\end{minipage}

\bibliographystyle{JHEP}
\bibliography{biblio}

\end{document}